\documentclass[apj]{emulateapj}

\usepackage{apjfonts}
\usepackage{epsfig}
\usepackage{epstopdf}
\usepackage{natbib}
\usepackage{amssymb,latexsym}
\usepackage{graphicx}
\usepackage{appendix}
\newcommand{\be}{\begin{equation}}
\newcommand{\ee}{\end{equation}}

\newcommand{\vpeak}{\mathrm{v}_{\rm{peak}}}
\newcommand{\vacc}{\mathrm{v}_{\rm{acc}}}
\newcommand{\vnpeak}{\mathrm{v}_{\rm{0,peak}}}
\newcommand{\mucut}{{\rm{\mu}}_{\rm{cut}}}
\newcommand{\mvir}{\hbox{$M_{\rm vir}$}}

\newcommand{\LCDM}{$\Lambda$CDM }
\newcommand{\sigate}{ \sigma_{8}}
\newcommand{\Msun}{{\rm M}_\odot}
\newcommand{\hinv}{h^{-1}}
\newcommand{\mpc}{\rm{Mpc}}
\newcommand{\hmpc}{\hinv\mpc}
\newcommand{\ov}{ \Omega_{\rm \Lambda} }
\newcommand{\om}{ \Omega_{\rm m} }
\newcommand{\vmax}{\mathrm{v}_{\rm max}}

\newcommand{\gtsima}{$\; \buildrel > \over \sim \;$}
\newcommand{\gsim}{\lower.5ex\hbox{\gtsima}}

\usepackage[usenames]{color}

\newcommand{\hMpc}{\ h^{-1}\ \rm Mpc}

\newcommand{\Rvir}{R_\mathrm{vir}}

\newcommand{\macc}{\mathrm{M}_{\rm{acc}}}
\newcommand{\mpeak}{\mathrm{M}_{\rm{peak}}}
\newcommand{\mnpeak}{\mathrm{M}_{\rm{0,peak}}}
\newcommand{\mhost}{\mathrm{M}_{\rm{host}}}
\newcommand{\mnow}{\mathrm{M}_{0}}

\shortauthors{Reddick et al}
\shorttitle{The Galaxy--Halo Connection in the Local Universe}

\begin{document}
\title{The Connection between Galaxies and Dark Matter Structures in the Local Universe}
\author{Rachel M. Reddick$^1$, Risa H. Wechsler$^1$, Jeremy L. Tinker$^2$, Peter S. Behroozi$^1$}
\affil{$^1$Kavli Institute for Particle Astrophysics and Cosmology;
 Physics Department, Stanford University, Stanford, CA, 94305\\
 SLAC National Accelerator Laboratory, Menlo Park, CA, 94025\\
 rmredd, rwechsler@stanford.edu\\
$^2$Physics Department, New York University, New York, NY}

%\date{\today}

\begin{abstract}
We provide new constraints on the connection between galaxies in the local Universe, identified by the Sloan Digital Sky Survey (SDSS), and dark matter halos and their constituent substructures in the $\Lambda$CDM model using WMAP7 cosmological parameters. Predictions for the abundance and clustering properties of dark matter halos, and the relationship between dark matter hosts and substructures, are based on a high-resolution cosmological simulation, the Bolshoi simulation.  We associate galaxies with dark matter halos and subhalos using subhalo abundance matching, and perform a comprehensive analysis which investigates the underlying assumptions of this technique including (a) which halo property is most closely associated with galaxy stellar masses and luminosities, (b) how much scatter is in this relationship, and (c) how much subhalos can be stripped before their galaxies are destroyed.  The models are jointly constrained by new measurements of the projected two-point galaxy clustering and the observed conditional stellar mass function of galaxies in groups.  We find that an abundance matching model that associates galaxies with the peak circular velocity of their halos is in good agreement with the data, when scatter of $0.20 \pm 0.03$ dex in stellar mass at a given peak velocity is included.  This confirms the theoretical expectation that the stellar mass of galaxies is tightly correlated with the potential wells of their dark matter halos before they are impacted by larger structures.  The data put tight constraints on the satellite fraction of galaxies as a function of galaxy stellar mass and on the scatter between halo and galaxy properties, and rule out several alternative abundance matching models that have been considered. This will yield important constraints for galaxy formation models, and also provides encouraging indications that the galaxy--halo connection can be modeled with sufficient fidelity for future precision studies of the dark Universe.
\end{abstract}

\maketitle
%\tableofcontents
\section{Introduction}

The connection between galaxies and their dark matter halos is the fundamental link between predictions of a given cosmological model and models of galaxy formation.  Galaxies form in the gravitational potential wells of dark matter halos, and our modern understanding of galaxy formation therefore depends on an understanding of dark matter. 
Dark matter halos are virialized structures that began as high density peaks in the early Universe and grew and collapsed through self-gravity.  Halos grow by accreting additional material from the smooth density field as well as nearby smaller halos.  The galaxies within them grow in tandem with their respective halos.  Accreted halos (or subhalos) generally also contain galaxies. These subhalos (and the galaxies they contain) are stripped by the tidal forces of the (host) halo that have accreted them and are eventually destroyed.  The halo that accreted the subhalo gains this mass, and stellar mass of the disrupted galaxy either accretes onto another galaxy in the host halo or is dispersed into the intracluster light.

Given this general understanding of the relationship between galaxies and dark matter, it is possible to predict the spatial distribution of galaxies from an N-body simulation of dark matter only.  The baryonic matter of the galaxies is a small fraction of all matter, and its effects on the formation of dark matter halos are subdominant, with observable impacts only on small scales \citep{Kra2004, Spr2005, TKP2011}.  However, populating a dark matter simulation with galaxies requires a detailed model to connect the dark matter with the galaxies.  Precise models of this galaxy--halo connection and its evolution are important for constraining galaxy formation models.  They are also of increasing importance in the era of precision cosmology.  In particular, the detailed relationship between the dark matter distribution --- directly related to cosmological parameters --- and the galaxies that trace it is likely to be a dominant systematic in studies of cosmic acceleration with galaxy surveys using a range of probes (e.g., \citealt{Cac2009, More2009, TWC2011, Nuza2012} and references therein).

The most direct approach to understanding the relationship between galaxies and halos is to run a full, hydrodynamic simulation, which may explicitly include the effects of star formation and feedback (e.g., \citealt{BN1998, SH2003b, VSK2012} and references therein).  Unfortunately, this approach remains computationally expensive, and therefore cannot currently be applied to large volumes.  Additionally, the results are complicated by differences in numerical techniques and the treatment of important physics below the resolution limit of the simulation.  An alternative is to use a semi-analytical model of galaxy formation (see, e.g., \citealt{Som2012, Lu2012, Hen2012, Ben2012} for recent examples).  This has the advantage of including many different processes that act on the galaxies in question, such as relations between star formation and feedback.  However, these models tend to be complex, having many parameters and requiring careful tuning, complicating efforts to understand the underlying physics.  A simpler option is to use a Halo Occupancy Distribution (HOD), which is based on knowing the number of galaxies of some type that may be assigned to each halo \citep[e.g.][and references therein]{YMB2008,YMB2009, Zeh2011, LTB2012}.  This approach still has the difficulty of using many parameters, and therefore requires multiple measurements of the galaxy distribution as inputs to constrain the model.  

An alternative to these is a semi-empirical approach known as subhalo abundance matching \citep{Kra2004, VO2004}.  Rather than input galaxy formation processes directly, abundance matching models make the simple assumption that some halo property is monotonically related to some galaxy property, typically galaxy luminosity or stellar mass.  That is, each halo (or subhalo) contains one galaxy at its center, whose luminosity or stellar mass is determined by some property of its host.  This property is often related to host halo mass, but there are many different possibilities.  Additional choices must be made to specify the specific model, such as whether to include nonzero scatter between the given halo property and the galaxy stellar mass.  Nonetheless, abundance models have the advantage of requiring few (or no) parameters, and using the full predictions of numerical simulations to model the dark matter distribution into the fully non-linear regime.

In general, for a given input luminosity or stellar mass function, abundance matching can produce a galaxy population that accurately reproduces measured galaxy statistics and provide insight into galaxy formation \citep{CWK2006, VO2006, Mos2010, BCW2010}.  Previous studies have demonstrated that abundance matching models are generally sufficient to statistically reproduce the observable properties of galaxies, including the two-point clustering, the galaxy bias, and the Tully-Fisher relation \citep{VO2004,CWK2006,TKP2011}.  Recent improvements in numerical dark matter simulations present the opportunity to test this model on a simulation large enough to have excellent statistics for L* galaxies while resolving halos small enough to host galaxies as dim as the Magellanic Clouds.  Bolshoi is one such simulation, which also uses cosmological parameters consistent with WMAP5 and other measurements \citep{KTP2011}.  \cite{TKP2011} showed that an abundance matching model applied to halos in this simulation could provide a good match to clustering statistics and the Tully--Fisher relation.
  
Testing any model requires statistics of the galaxy distribution.  The Sloan Digital Sky Survey \citep{SDSS2009} has provided a quantitative advance in measuring galaxy statistics in the local Universe, yielding increasingly precise measurements of the clustering of galaxies \citep[e.g.][]{Zeh2011} and large numbers of groups or clusters \citep[e.g.][]{Koe2007, YMB2007}.  Because measurements of cosmological parameters depend heavily on galaxies as tracers, systematics of such measures may be reduced by an improved understanding of how galaxies are associated with dark matter \cite[e.g.][]{Rozo2010, Tin2012, More2012}.  

Our intent is two-fold: (1) to examine the ability of different abundance matching models to simultaneously reproduce the correlation function and conditional stellar mass function measured from the Sloan Digital Sky Survey (SDSS), and (2) to systematically test the underlying assumptions in the abundance matching ansatz.  To do so, we also make new measurements of the clustering and conditional stellar mass function from the Sloan Digital Sky Survey.

We first describe the data used in our study (\S~\ref{sec:dr7}).  This is followed by a description of the Bolshoi simulation
and the models considered (\S~\ref{sec:mocks}).  \S~\ref{sec:meas} describes our measurements of the correlation function and the conditional stellar mass function, and additional statistics of the galaxies in groups.  An evaluation of how these vary as the model parameters are varied is presented in \S~\ref{sec:param}.  The principle results of this work are the constraints on the model parameter space derived from these measurements (\S~\ref{sec:constr}).  We then consider the impact of using different stellar mass functions and a comparison with another measurement of the conditional stellar mass function (\S~\ref{sec:altmeas}).  A summary of our results and conclusions may be found in \S~\ref{sec:conclude}.  We find that our best-fit model provides an excellent fit to the data.  We also find that the parameters in the model are well constrained, and that models that abundance match to many commonly used halo properties are ruled out by current data.

Throughout this work, we assume the same cosmology as the Bolshoi simulation, using \LCDM with $\om$=0.27, $\ov=1-\om$, $\Omega_b = 0.042$, $\sigate$=0.82, and $n=0.9$. Absolute magnitudes and stellar masses are quoted with $h=1$.  Except where otherwise specified, stellar masses are those given by the \textsc{Kcorrect} algorithm of \citet{BlRo2007}.  We use $\log$ for the base-10 logarithm, and $\ln$ for the natural logarithm.  Halo masses are given in terms of the virial mass, here defined as the mass within a radius such that the average enclosed density is $\Delta_{\rm{vir}} \rho_{\rm{crit}} \Omega_m$ for $\Delta_{\rm{vir}}=360$ at z=0 as given by \citet{BN1998} unless stated otherwise.

When referring to dark matter halos, the terms "halo" or "host halo" are used to refer to distinct halos only, which do not lie within the virial radius of a more massive dark matter halo.  In contrast, "subhalo" is used to refer to dark matter halos whose centers lie within the virial radius of a more massive halo.  A galaxy group is a set of galaxies that all lie within the virial radius of the same (distinct) halo, which may range in size from only one galaxy up to galaxy clusters.  A central galaxy (or "central") is the galaxy which resides at the center of a halo.  Satellite galaxies (or just "satellites") are those which reside in subhalos inside a more massive dark matter halo.

\section{SDSS DR7 data}\label{sec:dr7}

Our study uses the New York University Value Added Galaxy Catalog (NYU-VAGC) \citep{Bla2005}, based on Data Release 7 of the Sloan Digital Sky Survey (SDSS) \citep{Pad2008,SDSS2009}.  We focus primarily on two measurements: the projected two-point correlation function and the conditional stellar mass function (CSMF).  To measure the clustering, we use a set of volume-limited samples corresponding to a series of cuts in stellar mass.  For the group statistics such as the CSMF, we focus on one volume-limited sample, with a cut in absolute $r$-band luminosity of $M_r -5 \log h< -19$.  The area of the sample we use is $7235 ~\deg^2$, with a median redshift of $z=0.05$.  The $M_r -5\log h<-19$ sample contains a total of 74,987 galaxies with a maximum redshift of $z=0.064$, covering a volume of roughly $4.8 \times 10^6~(\hMpc)^3$.
We focus on the distribution of galaxies in terms of their stellar mass.  Throughout, we quote stellar masses in $\Msun h^{-2}$.  The cut of $\log(M_*)>9.8$ leaves a complete sample of 54,119 galaxies in the same range in redshift.

The details of the group finder are described in the appendix of \citet{TWC2011}, which is based on the algorithm of \citet{YMB2005}.  Galaxy groups are found by initially doing "inverse" abundance matching.  The highest host halo mass expected in the observed volume is assigned to the most massive galaxy.  The next most massive galaxy that is not within the virial radius of the most massive halo, is assigned the second most massive host halo, and so on.  This matching is done with zero scatter, using the mass function of \citet{Tin2008}.  Galaxies within the virial radii of the assigned host halos are treated as satellites.  This initial assignment is used to calculate an initial group stellar mass for each group.  Groups are then reassigned host halo masses using the total stellar mass within virial radius of the initially assigned halos.  This procedure is iterated until group assignments remain unchanged.  These results are distinct from the results of \citet{TWC2011} in that we use $\Delta_{\rm{vir}}=360$, rather than 200, for consistency with the mock catalogs, and in how the initial halo-to-galaxy assignment is done.  This results in a total of $\sim43,000$ groups, of which 17,178 are assigned a host halo mass greater than $10^{12}~\Msun$.  We impose this limit because below a mass of $\sim10^{12}~\Msun$ essentially all "groups" have only one galaxy above the $\log(M_*)>9.8$ threshold.  Therefore, the group assignment is not very informative below this mass.

The group finder introduces two major sources of bias.  First, groups with low total stellar mass may consist of only one or two galaxies.  Because host masses are assigned based on total group stellar mass, the assigned host halo mass relates directly to the stellar mass of the dominant galaxy.  This artificially reduces the scatter between the central galaxy stellar mass and the host halo mass for low-mass host halos.  Second, the assumption that galaxy with the most stellar mass is the central is not always true \citep[e.g.][]{Ski2011} and can bias results based on the central galaxies.  To take these changes into account, we create a galaxy distribution by populating halos in the simulation, and this galaxy distribution is passed through the group finder before making comparisons to the groups found in the volume-limited catalog.  The effects of group finding on our measurements are discussed in more detail in \S~\ref{subsec:obscsmf} and Appendix \ref{app:gf}.

%True for luminosity:  This results in a total of $\sim55,000$ groups, of which 17,178 are assigned a host halo mass greater than $10^{12}~\Msun$.

The NYU-VAGC is based on the SDSS spectroscopic sample.  This allows precision measurements of redshifts, which are required for measuring the projected two-point correlation function and to making group assignments.  However, the spectroscopy was obtained by assigning targets to spectroscopic plates connected to a fiber-fed spectrograph.  The size of the fibers prevents any two targets separated by 55" or less from being observed at the same time on the same plate.  Though overlapping plates partially alleviates this problem, a significant fraction of galaxies in the sample lack redshifts for this reason.  These galaxies are "fiber-collided; " this occurs for $\sim5\%$ of the galaxies in our sample.   A detailed explanation of the SDSS survey and hardware can be found in \citet{Sto2002}.  The tiling algorithm for the spectroscopic plates is described in \cite{BLL2003}.

Our clustering measurements were made on the same volume-limited sample as the groups.  Clustering measurements are presented in \S~\ref{sec:param}, with the error estimation discussed in \S~\ref{sec:meas}.

To use the fiber-collided galaxies, the simplest correction is to assign the galaxy the redshift of the galaxy with which it is fiber-collided.  As demonstrated by \citet{Zeh2005}, this correction is adequate for the correlation function down to scales of $\sim 0.1~\mpc/h$.  However, it has a significant impact on the conditional stellar mass function, since a fiber-collided galaxy is likely to be assigned to the same group as the galaxy it is fiber-collided with.  Our volume-limited sample has a median redshift of $z=0.05$.  At this redshift, the 55" angle corresponds to $\sim40~\rm{kpc/h}$ (comoving).

\section{Simulated galaxy catalogs}\label{sec:mocks}
\subsection{Simulations}\label{subsec:simulations}
The Bolshoi simulation is a recently completed cosmological dark matter simulation, described in \citet{KTP2011}.  The simulation uses $2048^3$ particles and has a volume of $(250~\mpc/h)^3$, roughly three times bigger than the SDSS Mr < -19 volume-limited sample.  The large volume is combined with the capability to resolve subhalos, dark matter halos that lie within the virial radius of larger host halos, down to a circular velocity of $\sim 55$ km s$^{-1}$.  This permits a precise study of subhalos and the satellite galaxies that inhabit them.

Because our models rely on abundance matching, we require knowledge of the dark matter halo distribution.  Therefore, halo finding is necessary to locate the potential wells where galaxies form.  There are several different algorithms used for this purpose, and they may produce different results even when working on the same test halos (see \citealt{KKM2011}, \citealt{Oni2012} and references therein). For our work, we use the \textsc{Rockstar} halo finder \citep{BWW2013}, which has the advantage of using velocity as well as position information to locate substructure.  This halo finder produces results that are comparable to other modern halo finders (e.g. BDM and AHF) on small scales; the use of phase space information allows it to track subhalos better in the inner regions of their hosts \citep{KKM2011, Oni2012, BWW2013}.  The halo (and subhalo) masses and maximum circular velocities ($\vmax$) are calculated using only bound particles, but including substructures.  We also use the merger trees produced by the algorithm described in \citet{BWW2013b}.  The merger trees allow us to use the past history of the halos and subhalos when assigning galaxy properties.  This combination of codes provide better tracking of subhalos over time \citep{BWW2013b}.

\subsection{Abundance matching}\label{subsec:abm}
Abundance matching is a simple and effective method for associating dark matter halos with galaxies (see, e.g., \citealt{Kra2004, VO2004, CWK2006, BCW2010, Mos2010}).  A simple example is that given halo mass and stellar mass functions, halos are assigned galaxies so that the most massive halo hosts the most massive galaxy, the second most massive halo hosts the second most massive galaxy, and so on.  More generally, this approach is complicated by scatter in the halo mass-stellar mass relation \citep[e.g.][]{Tas2004, BCW2010}, and the question of which halo property
is more closely correlated with galaxy stellar mass \citep{CWK2006}.  We consider both the effect of various nonzero values of scatter and the use of different halo properties on observable galaxy properties.

\begin{figure}[t]
\includegraphics[angle=90, width=0.5\textwidth,clip,trim=20mm 0mm 0mm 0mm]{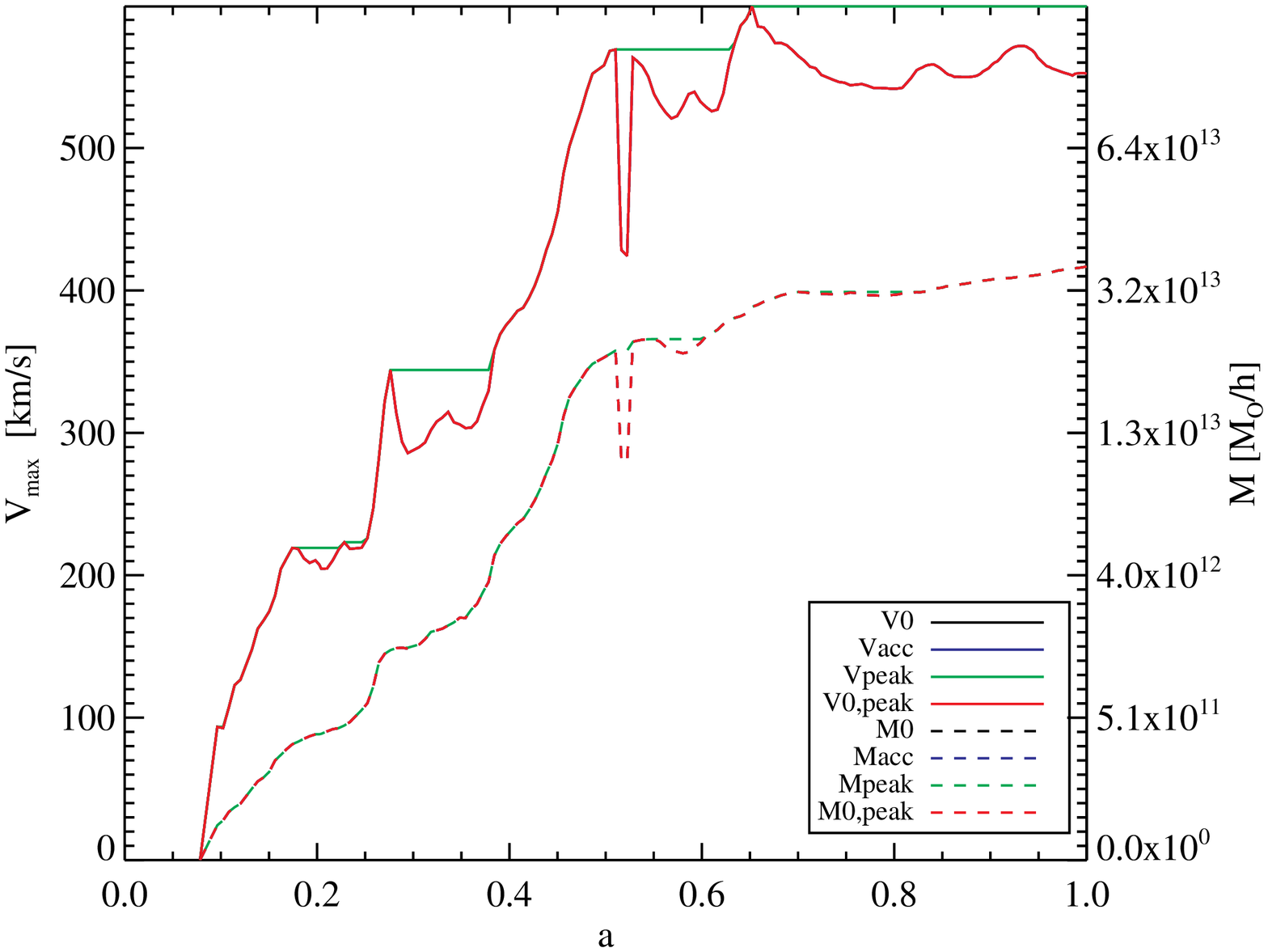}\vspace{-4.2mm}
\includegraphics[angle=90, width=0.5\textwidth]{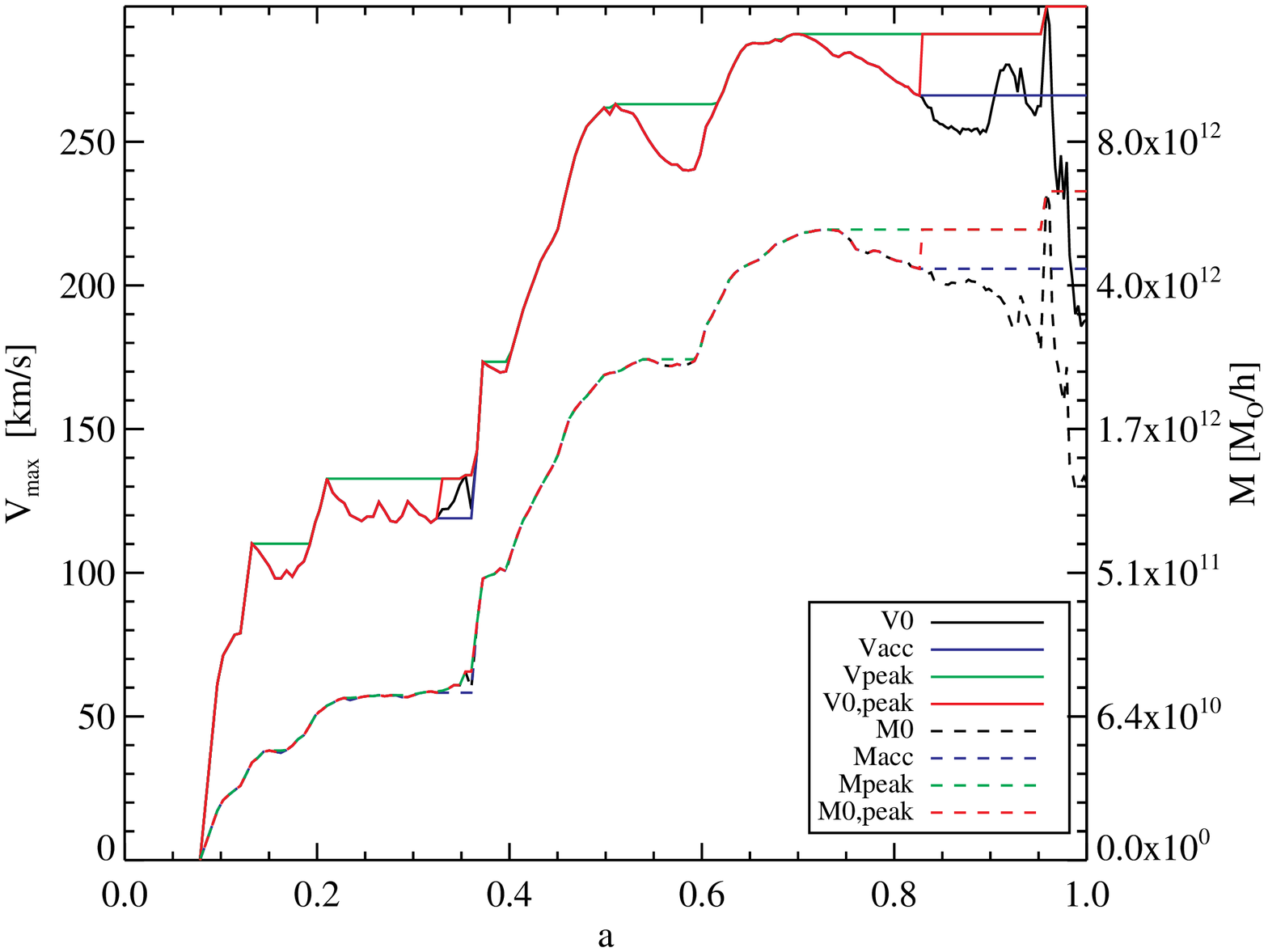}
\caption{\emph{Top:}  Evolution of various halo properties with scalefactor $a$, for 
for a single central galaxy, whose host halo has a mass of $3.7 \times 10^{13}$ at $z=0$.
Note that the distinct halo has no mass loss, so $\mnow=\macc=\mpeak =\mnpeak$.  Further, $\vmax=\vacc=\vnpeak$ by definition.  Only when $\vmax$ drops significantly following a merger (due to the drop in concentration) does $\vpeak$ deviate from $\vmax$.  \emph{Bottom:}  The same plot, but for a galaxy which is a satellite at $z=0$, with a present mass of $1.2\times10^{12}$ in a host of mass $3.1\times10^{13}$.  The satellite is accreted at around $a=0.85$.  Prior to this time, it is a central halo with the same general properties as in the top plot.  After accretion, however, $\vacc$ is fixed, and $\vnpeak=\vpeak$. 
 Because the halo starts being stripped here as well, $\mnow$ is no longer the same as the other mass measures; the rest, however, remain identical.  The jumps at $a=0.95$ are associated with a merger event between this particular subhalo and another subhalo.}
\label{fig:vtype-ev}
\end{figure}

The most natural theoretical expectation may be that galaxy properties are strongly correlated with the depth of their
potential wells.  If this is the case, the property $\vmax$ is likely to be the most relevant for galaxy properties.  
Dark matter halos can be significantly stripped after they are influenced by larger halos (before or after they enter the virial radius), in a way that galaxies are not.  Because of this, is reasonable to expect that galaxy properties 
should be most strongly correlated with their mass before this stripping occurs (see, e.g. discussion in  \citealt{CWK2006}).  At present, there is still a wide range of halo properties used in the literature.  For completeness, 
we consider a range of possible choices for the halo properties, and evaluate their consistency with data:

\begin{itemize}
\item $\mnow$: This is the simplest form of abundance matching, using only the masses of halos (or subhalos) at the present time.  Note that the mass of a subhalo is not measured out to the subhalo's virial radius; the subhalos identified by {\sc Rockstar} include all particles that are bound to the subhalo (see \citealt{BWW2013} for further details).  Because the subhalos' dark matter is more readily stripped than the galaxies hosted at their centers, the $\mnow$ approach generally underestimates satellite stellar masses (or luminosities).BWC2012b
	\item $\macc$:  The mass of halos at accretion, or infall.  For (distinct) halos, this is the mass at the present time, the same as $\mnow$.  For subhalos, this is the mass of the halo when it crosses the virial radius of its host, and is generally greater than $\mnow$.  This boosts the stellar mass of satellites relative to centrals of the same $\mnow$.
	\item $\mpeak$:  The maximum mass that the halo (or subhalo) has ever had in its merger history. This mass is nearly the same as $\mnow$ for isolated halos, but may be significantly greater for subhalos than either their present mass or their mass at infall, as some fraction of halos will be stripped prior to accretion.  \cite{BWC2012b} have found that most subhalos start being stripped at $\sim$ 3 $\Rvir$, regardless of host mass. 
	\item $\mnpeak$:  For isolated halos, this is equal to  $\mnow$; for subhalos, it is equal to $\mpeak$.  
	\item $\vmax$:  Similar to $\mnow$, $\vmax$ is the maximum circular velocity of a halo (or subhalo) at the present time.  This model generally suffers from the same difficulties as $\mnow$, having too few satellite galaxies with a given stellar mass.
	\item $\vacc$:  As with to $\macc$, $\vacc$ is the maximum circular velocity of a halo at the present time (equivalent to $\vmax$ for isolated galaxies), or at the time of infall.  As with $\mnow$, this boosts the stellar mass of satellites over that when using $\vmax$, increasing the satellite fraction at a given stellar mass.
	\item $\vpeak$:  Similar to $\mpeak$, $\vpeak$ is the highest circular velocity a halo has had over its entire merger history.  This is generally slightly greater than $\vmax$ or $\vacc$ for isolated halos and significantly greater than either $\vmax$ or $\vacc$ for subhalos.
	\item $\vnpeak$:  Similar to $\mnpeak$, $\vnpeak$ assigns the halos their present maximum circular velocity, and the subhalos their peak circular velocity.  Because $\vnpeak$ has the largest difference between (distinct) halos and subhalos, this is the model with the most massive satellite galaxies, and consequently the highest satellite fractions.
\end{itemize}  A comparison of how the properties we discuss here change for a single halo can be seen in Fig.~\ref{fig:vtype-ev}.

\begin{figure}
\includegraphics[angle=90, width=0.5\textwidth]{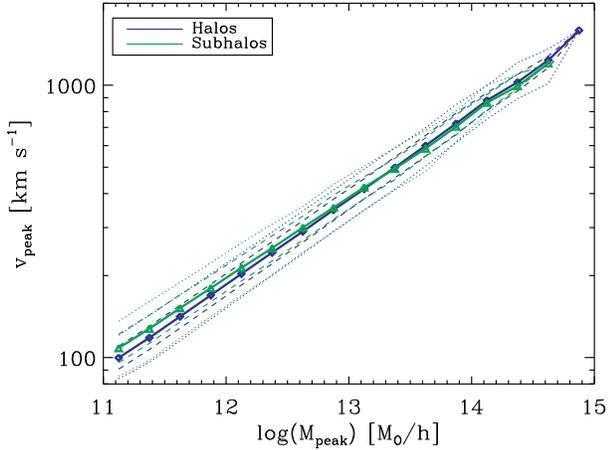}
\caption{Relationship between $\vpeak$ and $\mpeak$ for satellites and central galaxies.  The solid blue line indicates the median $\vpeak$ at fixed $\mpeak$ for distinct halos.  The dashed and dotted lines indicate the 68\% and 95\% bounds, respectively.  The green lines are the corresponding results for subhalos.  Note that subhalos tend to have larger $\vpeak$ and a wider dispersion, particularly at low masses, where the difference in the medians is $\sim 10\%$.}
\label{fig:vpmp}
\end{figure}

Additionally, there is a significant difference between the $\vmax$- and $\mnow$-based matching.  In particular, a direct comparison between $\vpeak$ and $\mpeak$ shows that at fixed $\mpeak$, subhalos tend to have slightly higher peak $\vmax$ (by as much as $\sim7\%$; see Fig. \ref{fig:vpmp}).  This may be due to a combination of two factors.  One is that less concentrated subhalos may be more easily disrupted, and less likely to survive to be included in the sample.  An alternative is halo assembly bias \citep[e.g.][]{Wec2001, GW2007, WZB2006}.  In this case, smaller halos that formed earlier and in lower-density regions, prior to accretion, tend to have higher concentrations.  This alternative is plausible, as it has been demonstrated in \citet{Guo2011} and \cite{RDA2012} that satellite galaxies tend to have slightly more stellar mass than central galaxies with the same (sub)halo mass.  This difference is most significant in less massive host halos.  A test using a lower-resolution simulation (the Consuelo simulation discussed in appendix \ref{app:res}) recovers the same difference in $\vpeak$ between halos and subhalos, suggesting that this difference is not likely due to resolution issues.

The impact of changing the abundance matching parameter is discussed in \S \ref{subsec:vary-abm}.  \citet{CWK2006} considered the use of $\vmax$ and $\vacc$, concluding that $\vacc$ was able to reproduce the two-point correlation function, but $\vmax$ was not.  Most related studies have used one of these two properties.

To perform abundance matching, we use the stellar mass function of the relevant galaxy sample as input.  Because the conditional mass and luminosity functions are sensitive to this input, for consistency with the group catalog, we use the exact stellar mass function of galaxies in the corresponding volume-limited sample to perform the abundance matching instead of using the global relations in the literature \citep[e.g.][]{LiWh2009,YMB2009, Bal2012}.

\begin{figure}
\includegraphics[angle=90, width=0.5\textwidth]{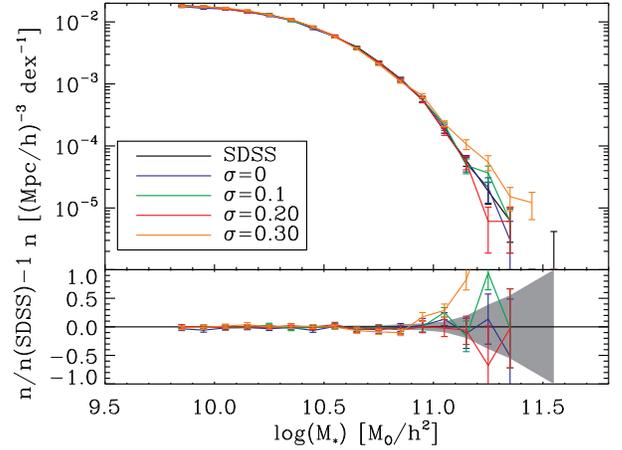}
\caption{Stellar mass function (SMF) from the SDSS sample (black), used as input to the abundance matching, compared against the output results of abundance matching and observational systematics (colored lines; blue, green, red, orange correspond to 0, 0.1, 0.2, and 0.3 dex of scatter).  Note that high values of scatter force the bright end of the stellar mass function high, because this steep region cannot be produced by convolution with a too-broad Gaussian.  Because there is no dependence of the scatter on the matching parameter used or $\mucut$, there is little change in the SMF between models at fixed scatter.  Error bars are derived from jackknife resampling.}
\label{fig:lfcomp}
\end{figure}

Scatter is introduced using the deconvolution method described in \citet{BCW2010}.  In brief, first abundance matching with zero scatter ($\sigma=0$) is performed using the observed stellar mass function.  A log-normal scatter is added to the stellar masses of the galaxies.  The "intrinsic" stellar mass function (SMF), that is, the SMF to which scatter is added in order to produce the observed SMF, is estimated based on the difference between the observed and scattered SMFs.  This new "intrinsic" SMF is then used in abundance matching.  This procedure is repeated until the output of the step where scatter is added is sufficiently close to the observed SMF.  While generally accurate, this approach is incapable of adding extremely high scatter and maintaining the steepness of the SMF above the characteristic stellar mass $M_{*,s}$ (see Fig.~\ref{fig:lfcomp}).  This is not a significant problem, as such large scatter (above $\sim0.3$ dex at fixed stellar mass) appears to be excluded by data at least for galaxies more massive than $M_{*,s}$.  This has been shown by previous authors \citep{More2009, LTB2012}, and is shown to be excluded by our later analysis.  An alternative method of introducing scatter, presented in \cite{TKP2011}, avoids this problem by selecting stellar masses from a predetermined list, guaranteeing that the SMF is exactly reproduced.  This method does not assume constant log-normal scatter in stellar mass, and therefore yields a somewhat skewed distribution of galaxy stellar masses in large dark matter halos compared to a log-normal.  It is not yet clear whether these alternatives can be distinguished by existing data.

In applying this model, we do not include any impact from statistical errors in the stellar mass measurements.  Therefore, the scatter we measure will be a combination of scatter in observed stellar masses and in stellar mass at fixed host halo mass.

In addition to the scatter, we consider the possibility that satellites galaxies are disrupted before their halos are destroyed in the simulation.  To investigate this possibility, we introduce a cutoff on the mass of subhalos.  Once a subhalo falls below some fraction of its maximum past mass $\mpeak$, we consider its galaxy to have been disrupted, similar to the cutoff examined in \cite{WW2010}.  These disrupted subhalos are excluded from abundance matching.  Effectively, we assign disrupted subhalos galaxies with zero stellar mass.  We use the parameter $\mucut$ to define the cutoff fraction of $\mpeak$, ignoring all (sub)halos for which $\mnow <\mucut \mpeak$.  We consider a range of $\mucut$ from zero (all subhalos are assigned a galaxy) to 0.15.  For reference, a value of $\mucut$=0.1 removes $\sim4\%$ of subhalos that would have been included in the sample with $\mucut$=0.

Once the abundance matching has been performed, we convert the Bolshoi snapshot into a lightcone by taking the origin as the point of observation.  This allows us to produce an octant on the sky, including redshifts, to a depth of $z=0.083$.  We use the snapshot at the mean redshift of the data, $z=0.05$, and ignore evolution in the dark matter distribution over this narrow range.  To introduce the same systematics present in the group catalog, we first add fiber collisions (as described below), then use the group finder to find galaxy groups and determine whether galaxies are centrals or satellites.

\subsection{Simulated Fiber Collisions}\label{subsec:simfc}
Once the mock catalog has been converted into a lightcone, it is necessary to consider the effect of fiber collisions.  
%The simplest approach, which would be to find all galaxies in the volume-limited sample within 55" of each other, does not fully emulate the set of possible fiber-collisions.  A galaxy may be fiber-collided with another galaxy that is either too dim or too distant to be in the sample of interest.  Therefore, two samples must be included when creating fiber collisions.  The first is the volume-limited sample of interest.  The second is a flux-limited sample of all galaxies not within that volume-limited sample.  Fiber collisions are then determined using galaxies from both sets, and must be applied before using the group finder.
Fiber collisions must be determined prior to using the group finder.

We use the Bolshoi simulation to provide the volume-limited sample.  The sample of interest extends to a redshift of 0.063.  We use the remaining volume of Bolshoi, to a redshift of 0.083, to provide a background of galaxies that may be collided.  
Following this procedure, we find $\sim4\%$ of galaxies are fiber-collided for the volume-limited sample with $\log(M_*)>9.8$, compared to $\sim5\%$ of galaxies in our sample.
The algorithm that is applied to the SDSS for determining the locations of spectroscopic fibers is discussed in \citet{BLL2003}.  We use a related algorithm applied to the mock lightcones.  We initially include galaxies above the stellar mass limit at any given redshift.  Galaxies that have neighbors within 55" are then placed into "collision groups" of nearby galaxies.  Of these galaxies, one is chosen to be the galaxy for which a true redshift is known.  Some of the other galaxies may also have "measured" redshifts, partly at random and partly depending on the geometry of the collision group.  The remainder are considered fiber-collided with the nearest galaxy on the sky, and assigned its redshift.

After the mock catalogs are completed, we then apply the same group finder as used on the SDSS groups to the mock catalogs.  This allows us to select galaxy groups consistently.

\section{Measurements}\label{sec:meas}

We use multiple measurements on both the SDSS DR7 catalog and the synthetic galaxy catalogs constructed by populating simulations with abundance-matched galaxies.  In particular, we focus on the projected two-point correlation function and the conditional stellar mass function, and use these in constraining our models.  We also consider other measurements, such as the group stellar mass function and the satellite fraction, to provide additional tests and to better understand the underlying galaxy distribution.

\subsection{Projected Correlation Function}
In its most basic form, the two-point correlation function counts pairs of galaxies at different separations, relative to the number of such pairs one would expect from a random distribution \citep[see, e.g.][]{Dav1985,Zeh2005}.
A clustered distribution, such as occurs in dark matter halos and thus, in galaxies, results in a larger value for the correlation function.  Smaller scales ($<\sim1~{\rm Mpc}/h$) generally correspond to clustering in a single host halo, between the central galaxies and its satellites and between pairs of satellites, while larger scales relate to clustering between isolated host halos.

We use the projected two-point correlation function, $w_p(r_p)$ because it does not suffer from peculiar velocities in the radial positions of galaxies.  We present new measurements of the stellar-mass clustering in DR7 based on our volume-limited catalogs, using the Landy--Szalay estimator \citep{LaSz1993}.  We use thresholds in stellar mass of $\log(M_*) > [10.6, 10.2, 9.8]$.  The covariances are drawn from spatial jackknife sampling.

Measurement of $w_p(r_p)$ in the mock catalogs was done using the set of abundance matching models described in section \ref{subsec:abm} applied to Bolshoi, with varying values of scatter and $\mucut$.  Because the simulation volume is similar to the volume of some of the volume-limited catalogs, it is important to understand the errors in the theoretical clustering measurements.  The covariance matrices were estimated by finding the correlation function for each of a set of 300 PM simulations of the same volume as Bolshoi, but with the dark matter down-sampled to the same number density as the observed sample.  These covariances were then scaled to the correlations measured on Bolshoi, according to:
\be
	C_{B,ij} = C_{ij}\frac{w_{B,i} \times w_{B,j}}{\bar{w}_{i} \times \bar{w}_{j}},
\ee
where $C_B$ is the covariance matrix we use, and $C$ that estimated from the multiple simulations.  The $w_B$ are the Bolshoi correlations, while $\bar{w}$ is the mean from the simulations.  The indices [i,j] denote the bin.  We use this procedure for each stellar mass threshold.  We do not include any contributions from stellar mass errors in our errors on the correlation function.

\subsection{Conditional Stellar Mass Function}

The conditional stellar mass function (CSMF) is the expected number of galaxies $\Phi(M_*|M_h)$ in a dark matter halo of mass $M_h$ with a stellar mass of $M_*$.  An equivalent measure, the conditional luminosity function, carries similar information.  The CSMF (or CLF) is a useful measurement for understanding both galaxy properties and cosmology \citep{YMB2003, YMB2009, Cac2009, HSW2009}.  A group catalog may be used to obtain the CSMF directly, by determining the mass of each group, then counting the galaxies in bins of stellar mass for each group mass.  This allows direct counting of the number of galaxies in halos, independent of the clustering described above.

The CSMF may be split into two parts:
\be
	\Phi(M_*|M_h) = \Phi_c(M_*|M_h) + \Phi_s(M_*|M_h).
\ee
Here, $\Phi_c$ is the CSMF of central galaxies only, which are the individual galaxies at the center of each dark matter halo.  $\Phi_c$ is a log-normal function.  $\Phi_s$ is the CSMF of the satellite galaxies, and well approximated by a Schechter function.  In the CLF, $M_*$ may be replaced by L, the luminosity of the galaxies in the groups.

The same procedure is used on both the DR7 volume-limited catalog and the Bolshoi-based mock when measuring the CSMF.  Comparisons are made in observational space, including the impacts of group finding.  Errors are estimated in both cases by using bootstrap resampling of groups, with 100 samples.

\subsection{Properties of satellites and centrals}

We also investigate summary statistics of the CSMF.  This includes the observed scatter in central galaxy stellar masses, as a function of group stellar mass.  We also consider the satellite fraction in our models.  We take this as the fraction of galaxies in our sample that are found to be satellites by the group finder, as a function of stellar mass.

\subsection{Group Stellar Mass Function}

The group stellar mass is the sum of the stellar masses of all galaxies in a group above some threshold in stellar mass, for each group.  The least massive groups correspond to individual galaxies near the stellar mass threshold of $\log(M_*)>9.8$, while the most massive correspond to clusters.  The distribution of group stellar masses is the group stellar mass function (GSMF).  The group luminosity function is the equivalent procedure, using luminosity rather than stellar mass.

\section{Understanding the Parameters}\label{sec:param}

Before discussing explicit constraints on the parameters of the abundance matching models, it is helpful to consider the effect of varying each of them individually on the several measurements that we use.  In \S \ref{subsec:vary-abm}, we consider varying the halo parameter used for abundance matching (Fig.~\ref{fig:comp-vtype}).  In \S \ref{subsec:vary-scat}, we consider varying the scatter in stellar mass at a given halo property (Fig. \ref{fig:comp-sc}).  In \S~\ref{subsec:vary-mucut}, we consider varying a the maximum amount halos can be stripped before galaxies are no longer identified (Fig. \ref{fig:comp-mu}). 

\subsection{Varying the Abundance Matching Parameter}
\label{subsec:vary-abm}

\begin{figure*}[p]
 \centering
\includegraphics[angle=90, width=0.8\textwidth]{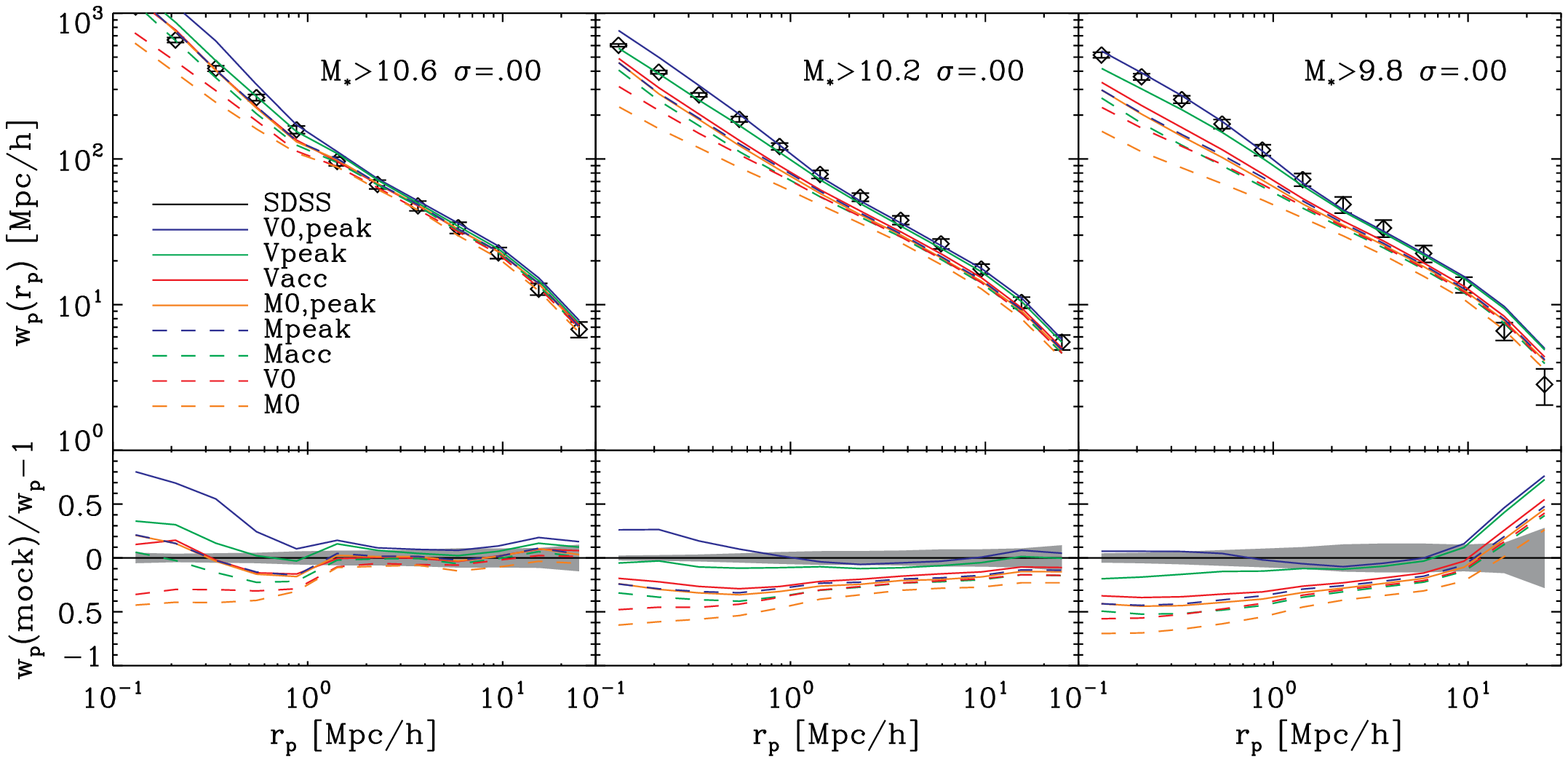}
\includegraphics[angle=90, width=0.8\textwidth]{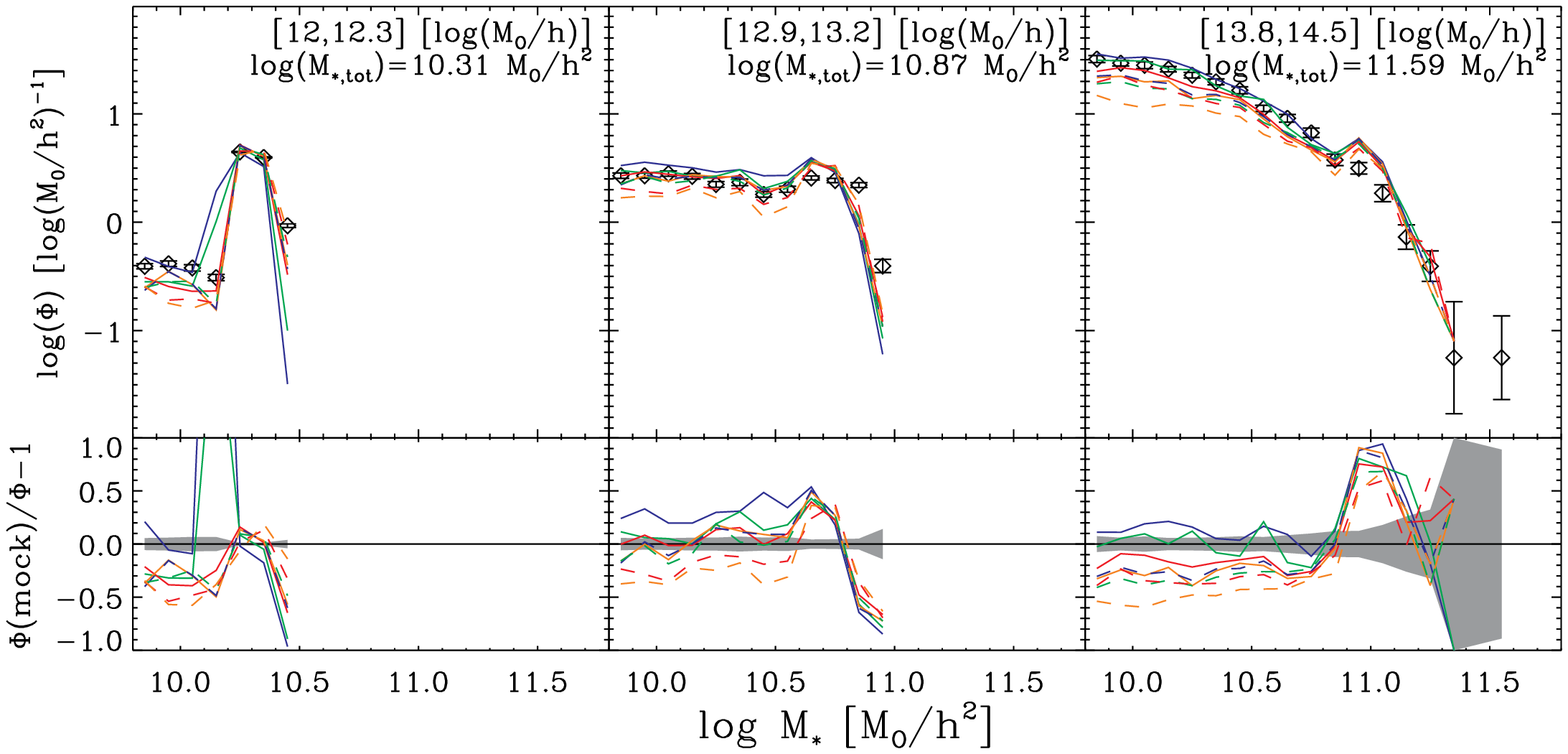}
\includegraphics[angle=90, width=0.8\textwidth]{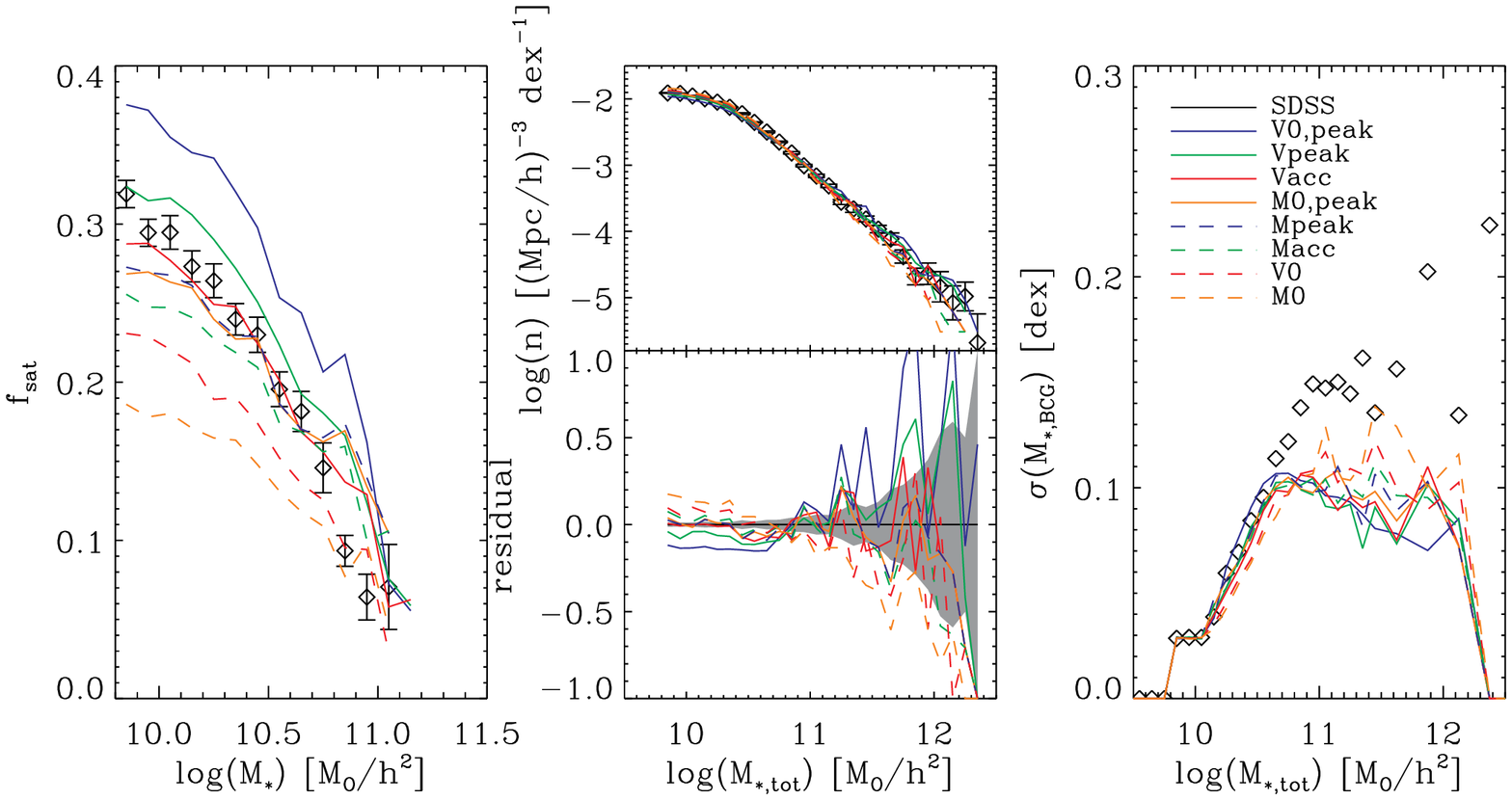}
\caption{Statistical properties of galaxies as measured from simulated galaxy catalogs and galaxy group catalogs, 
constructed using different halo properties for abundance matching.  All shown here have zero scatter and $\mucut=0$.  \emph{Top:} Projected two-point correlation function.  Labels denote the stellar mass threshold, given in $\log(\Msun/h^2)$.  Because increases in scatter or $\mucut$ can only decrease the clustering, it follows that any model which falls significantly below the measured clustering (black) must be excluded.  \emph{Center:}  Conditional stellar mass function (CSMF).  Labels indicate the range in $\log(\mvir)$ for each plot, as well as the median total stellar mass in each bin ($M_{*,tot}$).  Non-zero scatter broadens this part of the distribution.  \emph{Bottom left:}  Satellite fraction as a function of stellar mass.  As should be expected, models with higher satellite fraction also have stronger one-halo clustering and more satellites in the CSMF.  \emph{Bottom center:}  Group stellar mass function and residuals.  \emph{Bottom right:}  Standard deviation (scatter) in stellar mass of central as a function of total group stellar mass.  The models are most readily distinguished by the small-scale clustering and changes in the satellite fraction.
Error bars on the model points have been omitted for clarity.}
\label{fig:comp-vtype}
\end{figure*}

The impact of varying the abundance matching parameter is shown in Fig.~\ref{fig:comp-vtype}.  This figure shows
the two-point correlation functions for three cuts in stellar mass and  the conditional stellar mass function in three bins
of total stellar mass, which are later used to directly constrain the models.  The satellite fraction, the scatter in the 
stellar mass of the central galaxy identified by the group finder, and the group stellar mass function, are also shown.

The impact of changing the abundance matching  parameter on many of the results is best understood in the context of a halo occupation model.  Correlations on small-scales, below $\sim1~\mpc/h$,  are determined by the distribution of galaxies in the same (host) halo, the one-halo term. Larger scales are associated with the two-halo term, from the correlation between galaxies in different halos.  For fixed values of scatter and $\mucut$, the most significant effect of changing the parameter used in the abundance matching assignment is the change in the one-halo term.  Changing the halo parameter
used for abundance matching changes the relative circular velocities of halos and subhalos that are used to assign central and satellite galaxies, respectively.  For example, the difference in the correlation function between $\vmax$ and $\vacc$ is due primarily to the fact that subhalos are stripped after accretion.  This difference can be seen in Fig.~\ref{fig:vtype-ev} at $a=1$:  $\vacc>\vmax$ for the example subhalo shown, but $\vacc=\vmax$ for the distinct halo.  Thus, when abundance matching to $\vacc$, this increases the fraction of galaxies that are satellites (hosted by subhalos) at a fixed number density (and therefore above a fixed threshold in stellar mass) relative to the same procedure applied to $\vmax$.  This increase in number of satellites enhances the one-halo term due to additional satellites in clusters, but has little effect on the two-halo term.

The same pattern can be seen among all four different abundance matching methods using $\vmax$.  The parameter $\vnpeak$ results in the highest satellite fraction and the most small-scale clustering.  This is followed by $\vpeak$ and $\vacc$; $\vmax$ the least clustered.  A similar trend can be seen among the models using mass, though the differences tend to be smaller due to the smaller relative differences between mass definitions, as discussed in \S \ref{subsec:abm} and as can been seen for a pair of example halos in Fig.~\ref{fig:vtype-ev}.  The mass-based matching is also less clustered than the equivalent $\vmax$ method; for example, $\vpeak$ is more clustered than $M_{\rm{peak}}$.  This is because, as shown in Fig.~\ref{fig:vpmp}, satellites tend to have higher $\vpeak$ than centrals at fixed $M_{\rm{peak}}$.  The results of all eight models with no scatter and $\mucut$=0 are shown in Fig.~\ref{fig:comp-vtype}.

As is shown in the following two sections, using nonzero values of either scatter or $\mucut$\ can only reduce the clustering, not increase it.  Therefore, any model shown here that falls significantly below the measured projected correlation function cannot reproduce the clustering by any variation of these values, and is excluded from further consideration.  This leaves only $\vpeak$ and $\vnpeak$ as viable models.  Because these are the models with the highest values of the matching property for subhalos relative to distinct halos, this implies that stripping of the subhalo begins prior to the time of accretion, but that the stripping of the satellite galaxy it hosts does not begin until significantly later.

Our exclusion of all matching parameters other than $\vpeak$ and $\vnpeak$ is dependent on our particular abundance matching method.  For example, we do not consider including "orphan" galaxies which may be present despite the disruption of their subhalos.  We cannot rule out such models.

\subsection{Varying Scatter}
\label{subsec:vary-scat}
\begin{figure*}[p]
\centering
\includegraphics[angle=90, width=0.85\textwidth]{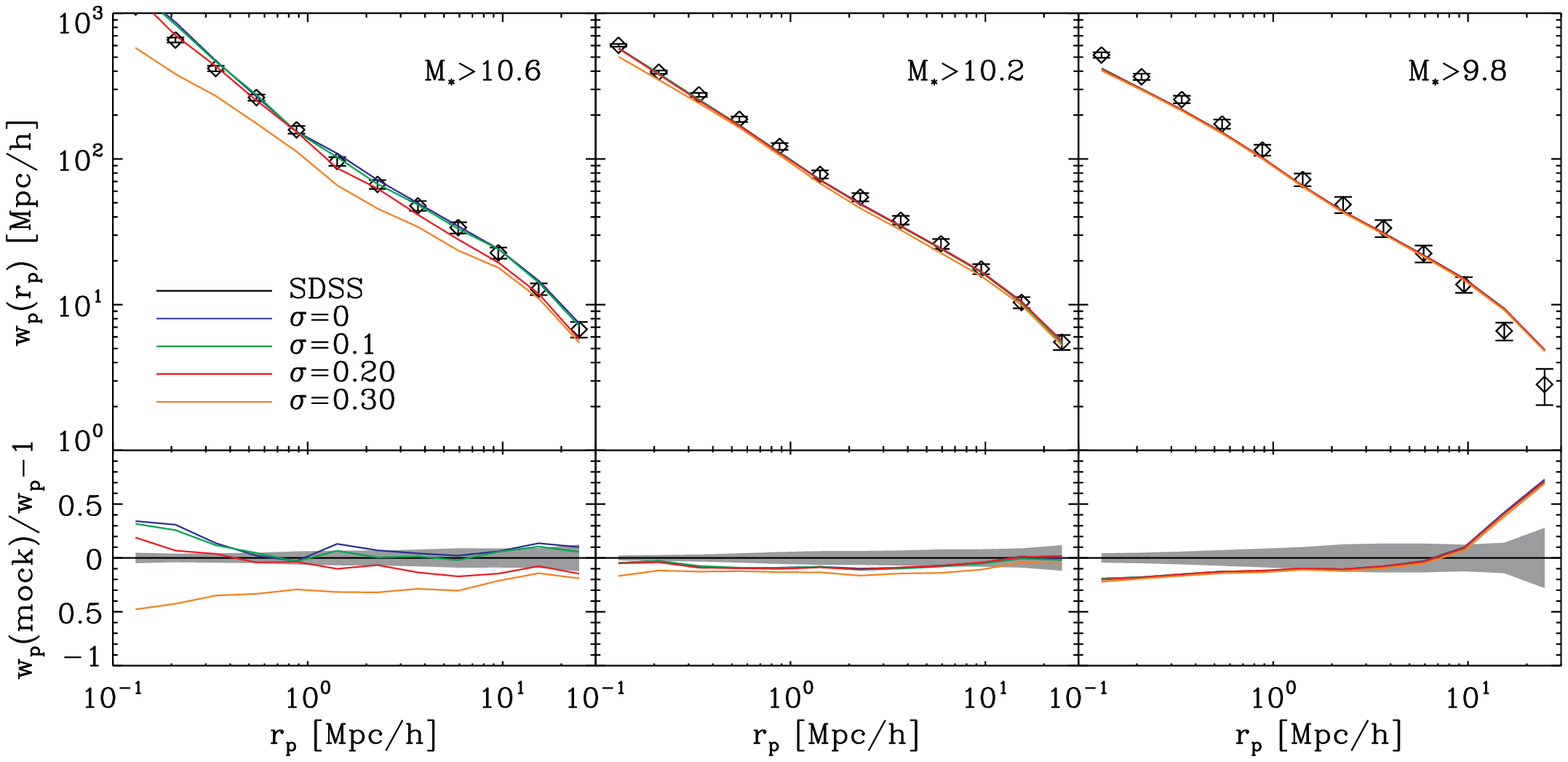}
\includegraphics[angle=90, width=0.85\textwidth]{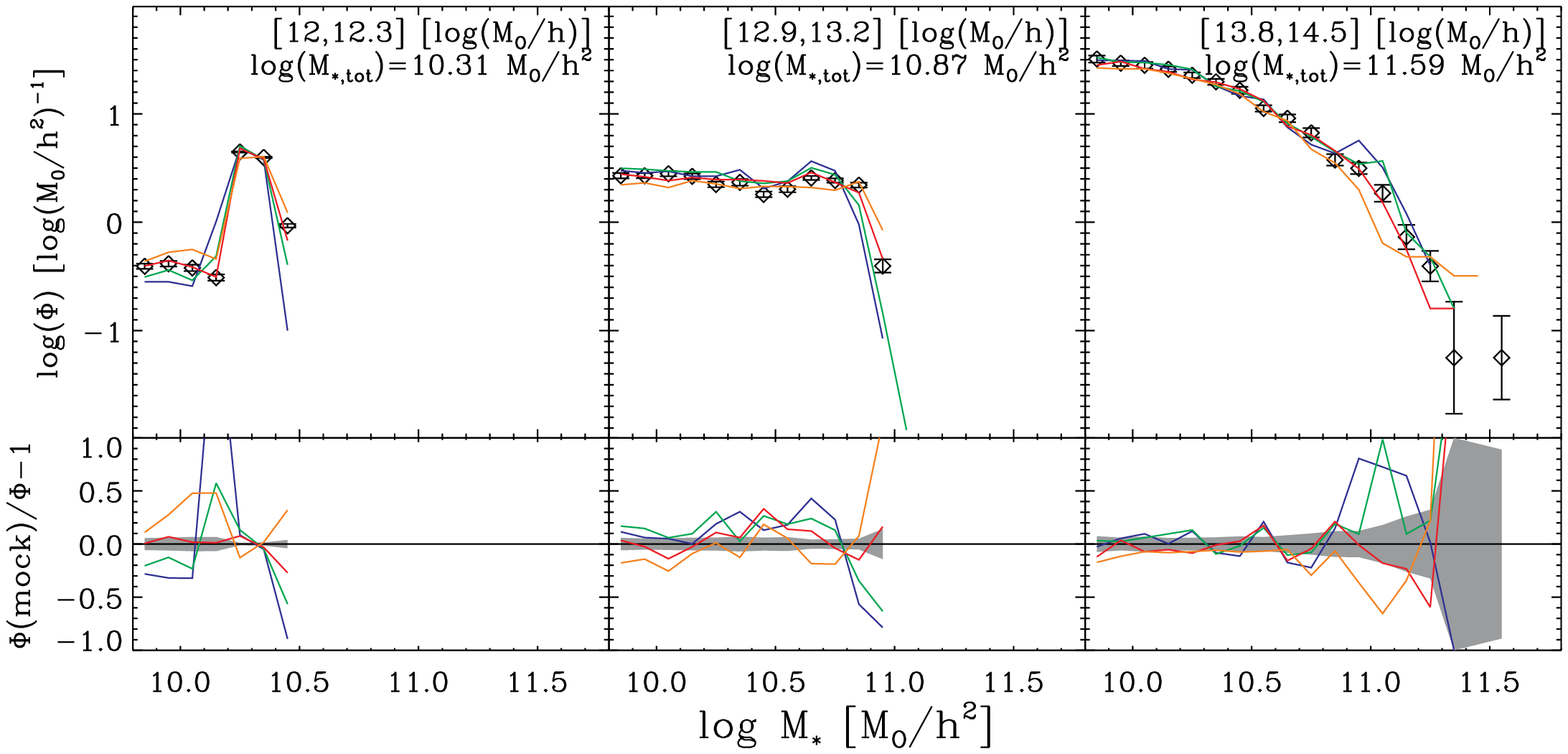}
\includegraphics[angle=90, width=0.95\textwidth]{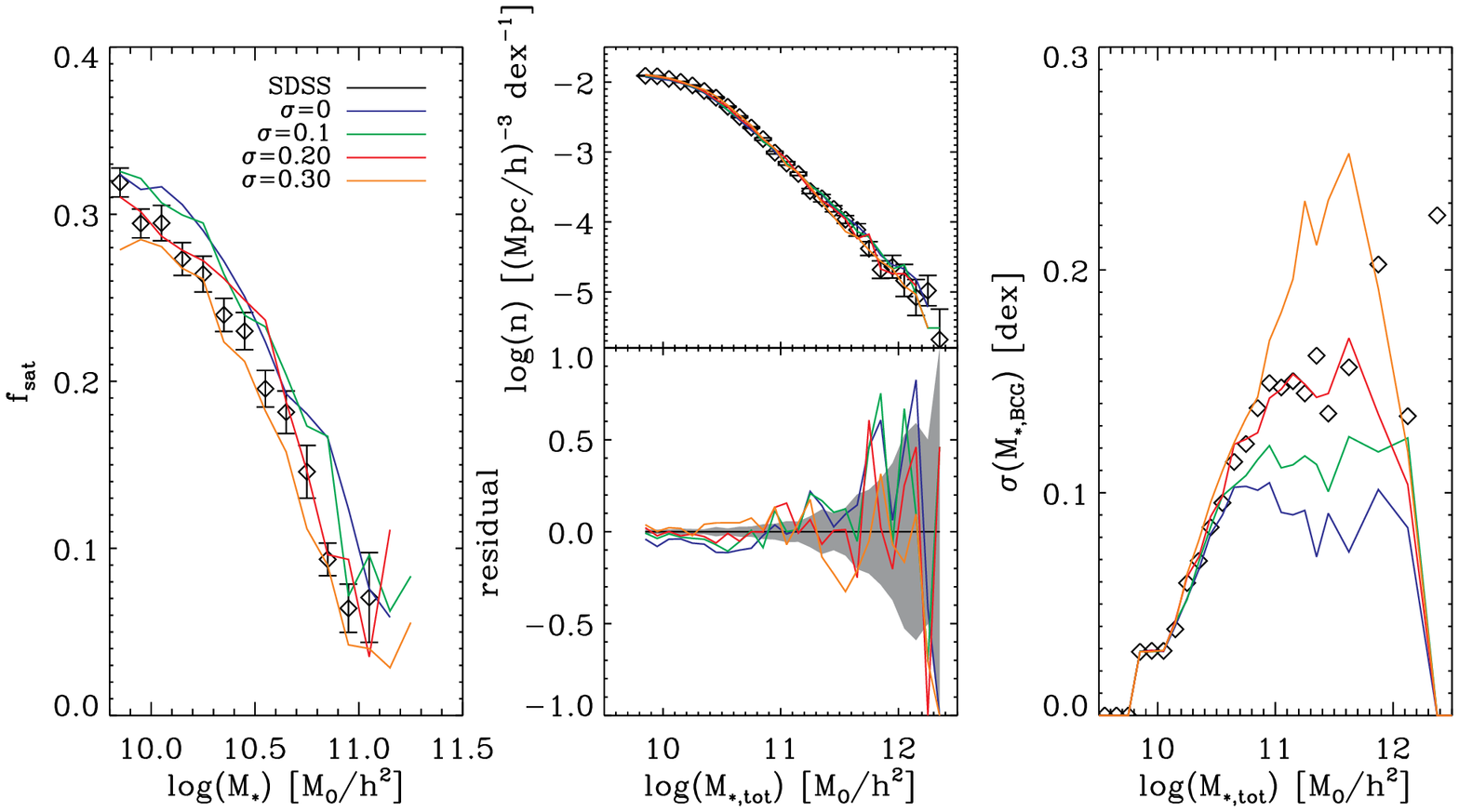} \hspace{-0.5 cm}
\caption{
Impact of scatter in galaxy stellar mass at a given $\vpeak$ on observed statistics of the galaxy distribution. 
The models shown abundance match to $\vpeak$ with fixed $\mucut$=0, with varying values of scatter.  Increasing scatter reduces the clustering, but does not strongly affect clustering for thresholds below the characteristic stellar mass of the volume-limited sample.  Individual plots are the same as described in Fig. \ref{fig:comp-vtype}.}
\label{fig:comp-sc}
\end{figure*}

We evaluate the impact of scatter on galaxy statistics in Fig.~\ref{fig:comp-sc}.  For a fixed method of abundance matching, and fixed $\mucut$, the effect of adding scatter is to reduce the clustering amplitude; this effect is most noticeable for the brightest, and most strongly-biased, samples. This is due to the steepness of the stellar mass function above the characteristic mass scale, where the falloff becomes exponential.  It is more likely that less massive galaxies will be scattered to higher stellar mass than the reverse, decreasing the bias of galaxies above a fixed stellar mass threshold.  However, this effect is reduced significantly for stellar mass thresholds less massive than this scale, since in this range the bias is only weakly mass-dependent, and the stellar mass function flattens.

Similarly, increasing the scatter directly broadens the central peak of the CSMF.  In general, this scatter should increase the width of the stellar mass distribution of central galaxies in host halos of any mass.  However, the assumption that the brightest galaxy is the central galaxy, combined with the use of the group finder, reduces this scatter dramatically in poorer groups.  This effect is most striking in the smallest halos, where there may be one or no satellite galaxies, and the stellar mass of the central galaxy becomes directly related to the host halo mass determined by the group finder.

The scatter has some impact on the satellite portion of the CSMFs, tending to slightly reduce the number of satellites in clusters, and increase the number in small halos.  This may be most easily understood by first considering the satellite fraction, which also tends to decrease at low stellar masses with increasing scatter.

More massive galaxies are more likely to be centrals, because the fraction of halos of a given $\vmax$ which are subhalos generally decreases with $\vmax$ (or mass) \citep{Kra2004, CWK2006}.  As scatter increases, this relationship weakens and the likelihood that a central galaxy is not the most massive galaxy -- and therefore determined to be a satellite by the group finder -- should increase.  That is, there is a significant likelihood that a satellite is more massive than the central in a particular host halo.  The intrinsic satellite fraction of less massive galaxies should change only weakly with scatter, since most such low mass galaxies are centrals with no satellites of sufficiently high stellar mass to scatter to a higher mass than the central.  On the other hand, particularly in richer groups, some satellite galaxies will be scattered to higher stellar mass, possibly more massive than the true central.  This suggests that the satellite fraction of low mass galaxies should remain roughly constant with increasing scatter, and should increase at high stellar mass with increasing scatter.  If this is surprising, consider the case of infinite scatter, where galaxy stellar mass is completely unrelated to the (sub)halo mass.  In that case, the satellite fraction will be constant with stellar mass, because satellites are as likely to be the most massive as centrals.

However, in the data, we do not know whether a galaxy is a central or satellite a priori.  As a consequence, when the group finder assumes that the most massive galaxy is the central, it artificially reduces the satellite fraction of massive galaxies.  Furthermore, this assignment changes the center of the measured halo away from the true center, which means that some galaxies that should be assigned as satellites are now outside the inferred virial radius.  This tends to reduce the satellite fraction of low mass galaxies.  This same effect reduces the number of galaxies in massive clusters, as can be seen in the CSMFs.

The opposite effect is seen in the least-massive groups that we consider, where the number of satellites increases with scatter.  This is due to our method of host mass assignment, where group stellar mass is used as the host mass proxy.  When a small group, with one or no satellites, gains a new satellite above the stellar mass threshold due to scatter, the group will be pushed up in group stellar mass and added to the host mass selection.  This effect is negligible on halos which host many satellites, which are dominated by the miscentering issue.  (For more details, see Appendix \ref{app:gf}.)

The impact of scatter on the group stellar mass function is also similar to that of $\mucut$.  That is, it increases the number of low-stellar mass groups, and reduce the number of large clusters, steepening the group stellar mass function.

In sum, increased scatter reduces the overall clustering amplitude, more strongly for higher stellar mass thresholds.  It also broadens the central part of the observed CSMF in massive groups, and alters the shape of the observed satellite CSMF in a way that depends on the size of the group.  The clustering prohibits high scatter, while the CSMF requires some moderate, nonzero scatter.  The two parts of the CSMF provide the strongest constraint in this regard.

\subsection{Varying $\mucut$}
\label{subsec:vary-mucut}

%As discussed in \S \ref{subsec:abm}, the $\mucut$ parameter defines a cutoff in subhalo mass (see Fig. \ref{fig:comp-mu}).  This allows inclusion of satellite galaxy disruption prior to the disruption of the simulated subhalo (see, e.g., \citealt{WW2010}).  Those subhalos whose mass at the present time falls below $\mucut \rm{M}_{peak}$ are assumed to have been destroyed, where $\mpeak$ is the largest mass the (sub)halo ever had in its history.  The effect of this parameter is to reduce the overall number of satellites at fixed stellar mass.  This reduces the number of small-scale pairs and depresses the one-halo term in the correlation function.  Because this removes satellites, the satellite fraction drops, especially at lower stellar masses, and the satellite part of the CSMF is depressed.  While the number of groups overall is unchanged by increasing $\mucut$, the groups with satellites tend to lose satellites, reducing their total group stellar mass.  This tends to make the group stellar mass function steeper, pushing more groups to lower total stellar masses.

\begin{figure*}[p] \centering 
\includegraphics[angle=90, width=0.85\textwidth]{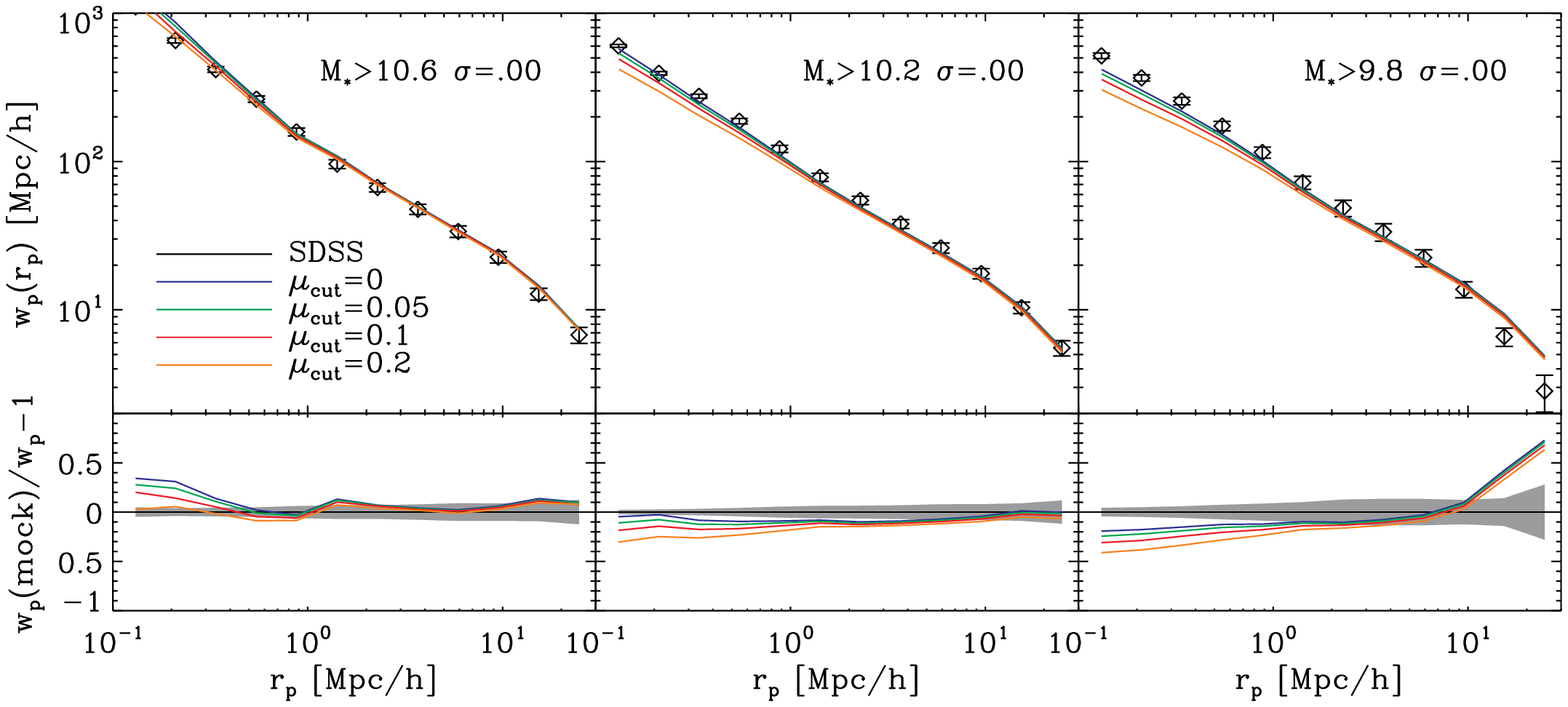}
\includegraphics[angle=90, width=0.85\textwidth]{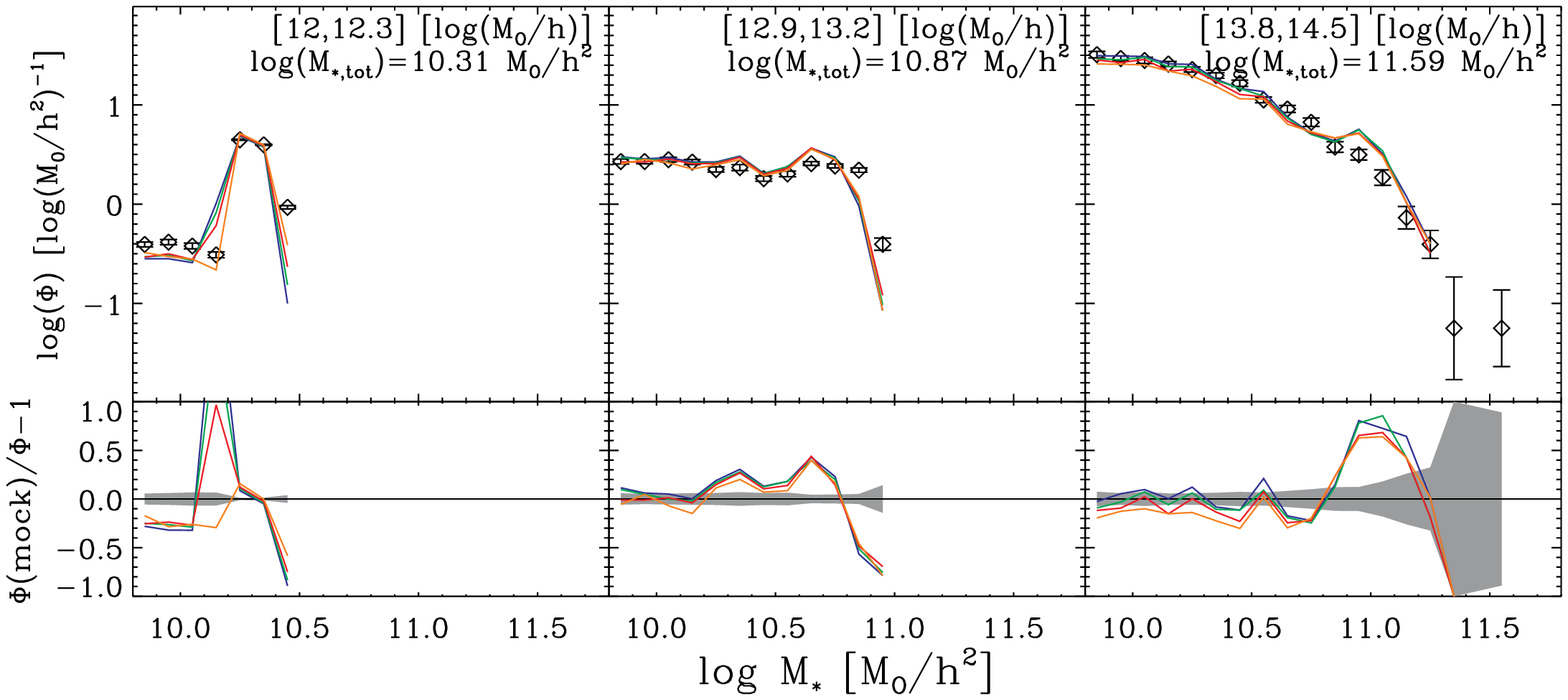}
\includegraphics[angle=90, width=0.95\textwidth]{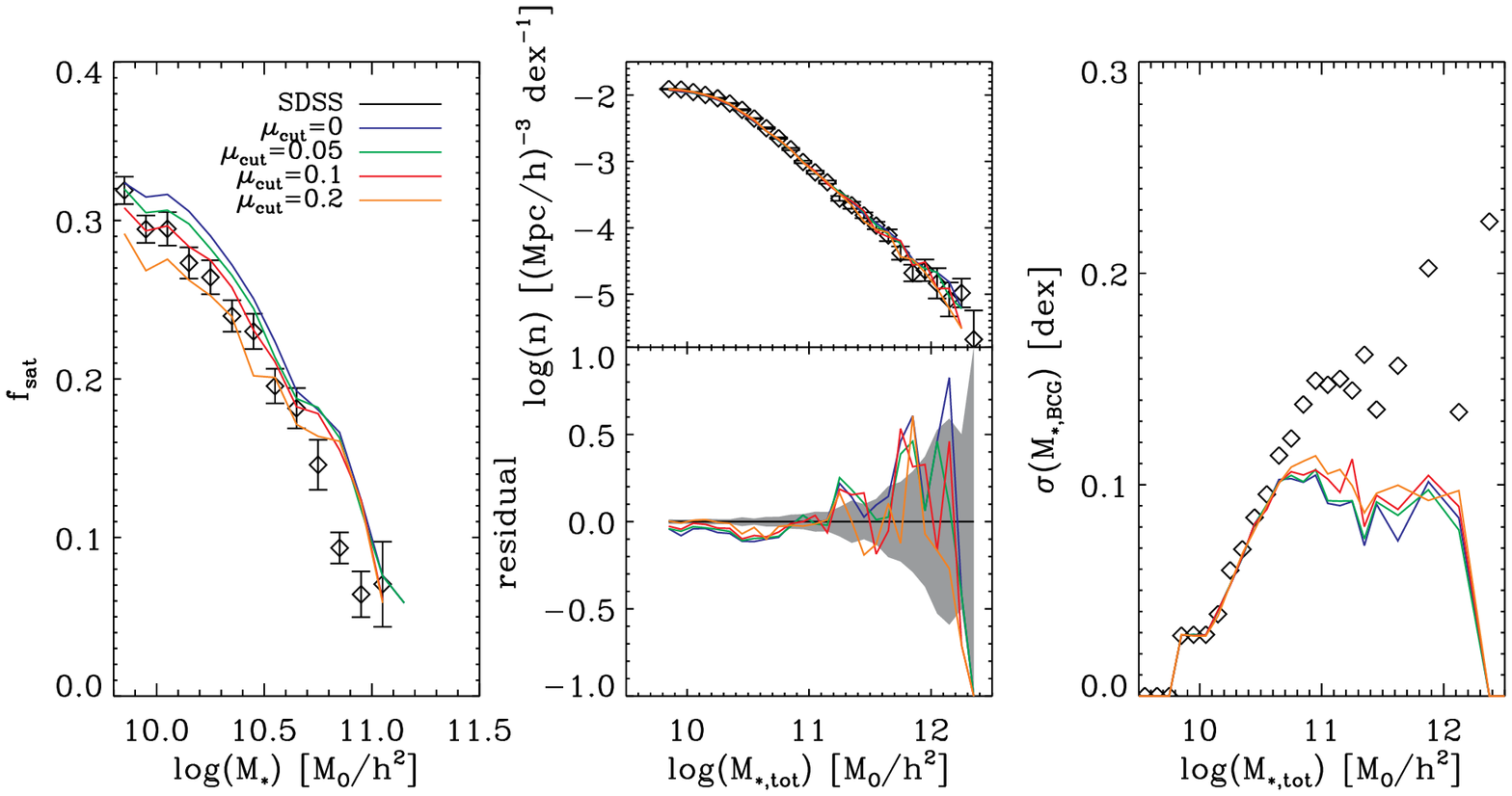}
\caption{ Impact of the $\mucut$ parameter, related to galaxy stripping, on observed statistics of the galaxy distribution.  The models shown abundance match to $\vpeak$ with zero scatter in stellar mass, with varying values of $\mucut$.  Increasing $\mucut$ pushes down the clustering on small scales only, and decreases the satellite fraction.  Individual plots are the same as described in Fig. \ref{fig:comp-vtype}.}
\label{fig:comp-mu}
\end{figure*}

Because $\mucut$ effectively removes satellites, and therefore most strongly affects small scales, it cannot be too large.  Details of how $\mucut$ acts, however, depend somewhat on other details of the model in question.  Fig.~\ref{fig:comp-mu} demonstrates the impact of $\mucut$ on our measurements.

To summarize the implications of these initial tests:

\begin{enumerate}
	\item Any model, to reproduce the clustering, must have at least as many satellite galaxies as a model using $\vpeak$ as the abundance matching property.  Of the set of properties we consider, only $\vpeak$ and $\vnpeak$ pass this criterion.
	\item The $\mucut$ parameter most strongly affects small scales and the number of satellite galaxies, removing those whose subhalos were most stripped.  To have enough satellite galaxies to reproduce the clustering and CSMF, $\mucut$ cannot be too large.
	\item Increasing scatter reduces the clustering for the high stellar mass thresholds, widens the central CSMF distribution, and alters the shape of the satellite CSMF.  It also reduces the satellite fraction.  Scatter is most strongly constrained by the two parts, satellite and central, of the CSMF.  Large scatter is also excluded by the two-point clustering measurements (zero scatter is only weakly disfavored by the clustering statistics alone).
\end{enumerate}

\section{Constraints on the Local Galaxy--Halo Connection}\label{sec:constr}
\subsection{Parameter Constraints}
We now investigate the two candidate models which plausibly have enough substructure to match the data, abundance matching stellar mass to $\vpeak$ and $\vnpeak$.  We systematically vary the parameters in these models to determine which are allowed by the data.  For each model, we consider a large grid of models in the scatter and $\mucut$ parameters described above, and evaluate which range in these parameters provides an acceptable fit to the correlation function and the conditional stellar mass function measured in the SDSS data.

\begin{figure*}
\centering
\includegraphics[angle=90, width=\textwidth]{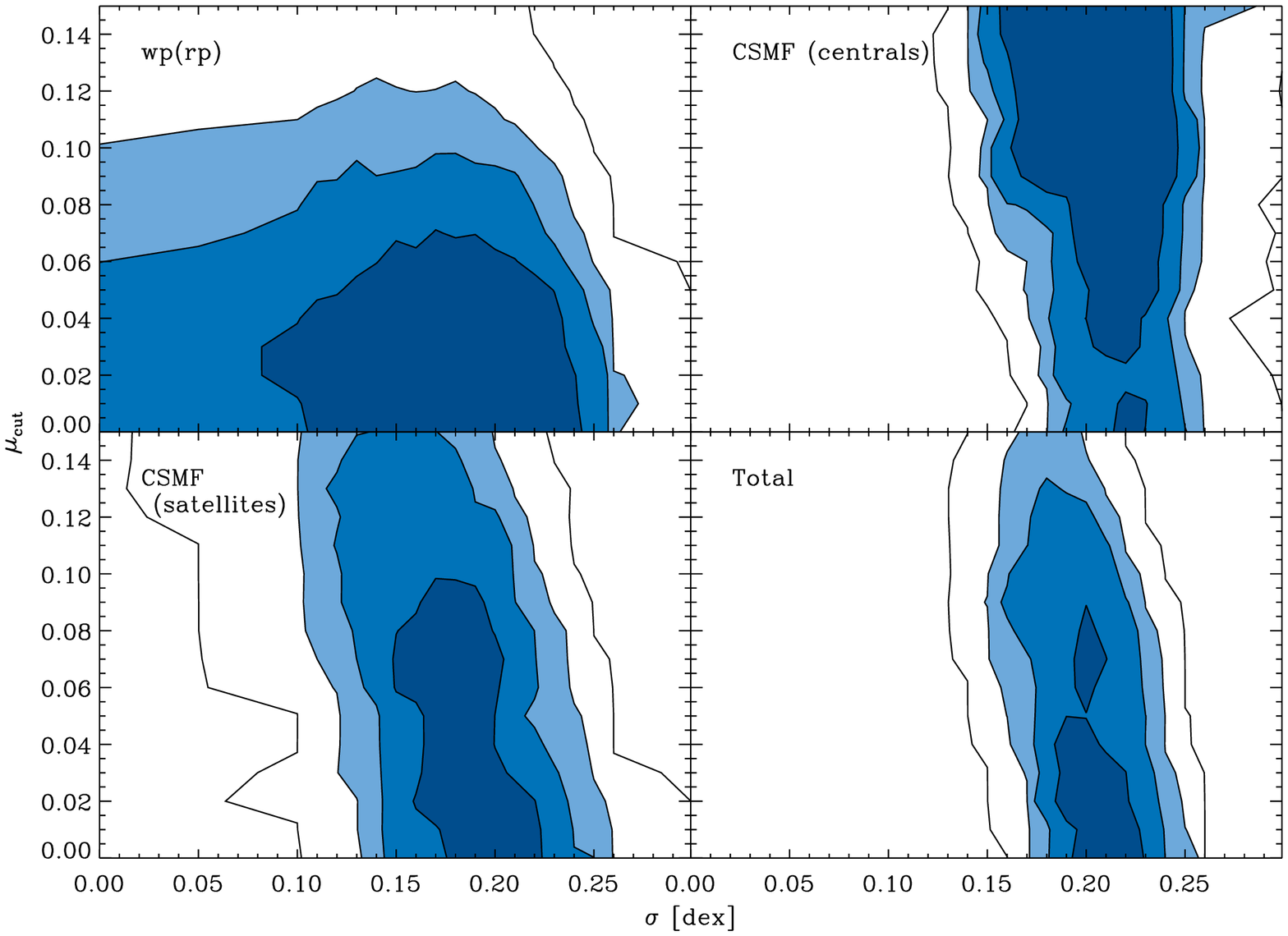}
\vspace{-1.5 cm}

\caption{Constraints for the scatter and $\mucut$ parameters, for abundance matching models which assign galaxies to $\vpeak$ of both halos and subhalos.  Clustering constraints use data for galaxies with $\log(M_*) > 10.2$.  Levels give $P(>\chi^2)$, corresponding to 1, 2, 3, and 5-$\sigma$ contours.  \emph{Upper left:} Constraint from clustering only.  \emph{Upper right:} Constraint from central part of CSMF only.  \emph{Lower left:} Constraint from satellite part of CSMF only.  \emph{Lower right:} Parameter constraints using the total $\chi^2$ from all three measurements.}
\label{fig:constr}
\end{figure*}

\begin{figure*}[p]
\centering
\includegraphics[angle=90, width=0.9\textwidth]{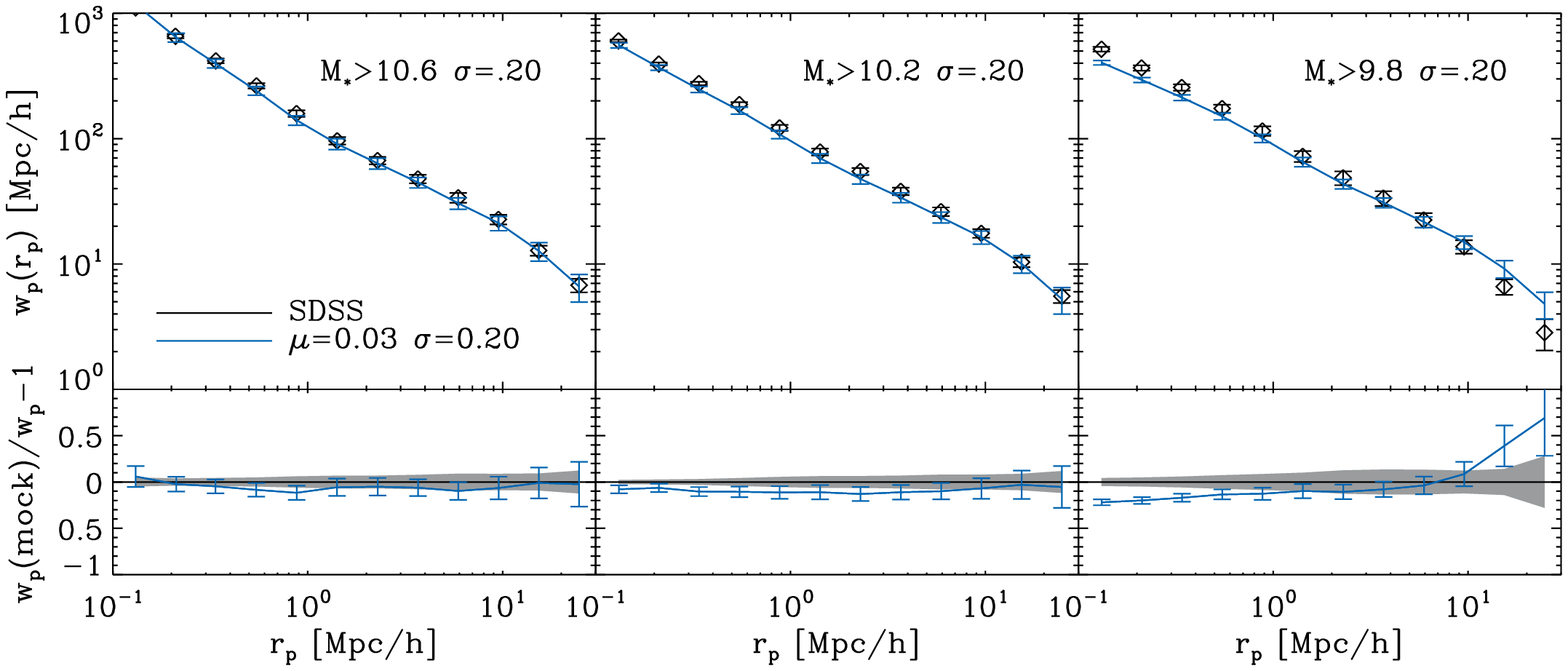}
\includegraphics[angle=90, width=0.9\textwidth]{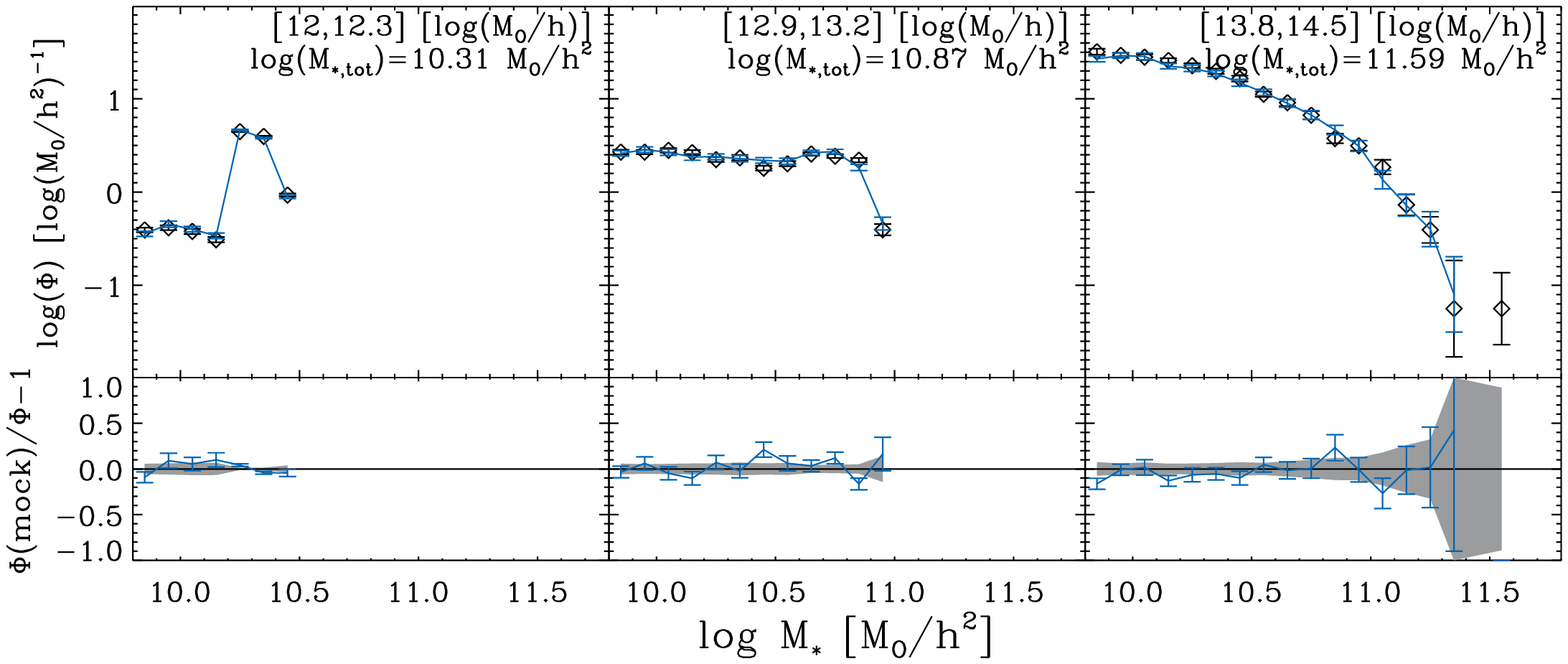}
\includegraphics[angle=90, width=0.9\textwidth]{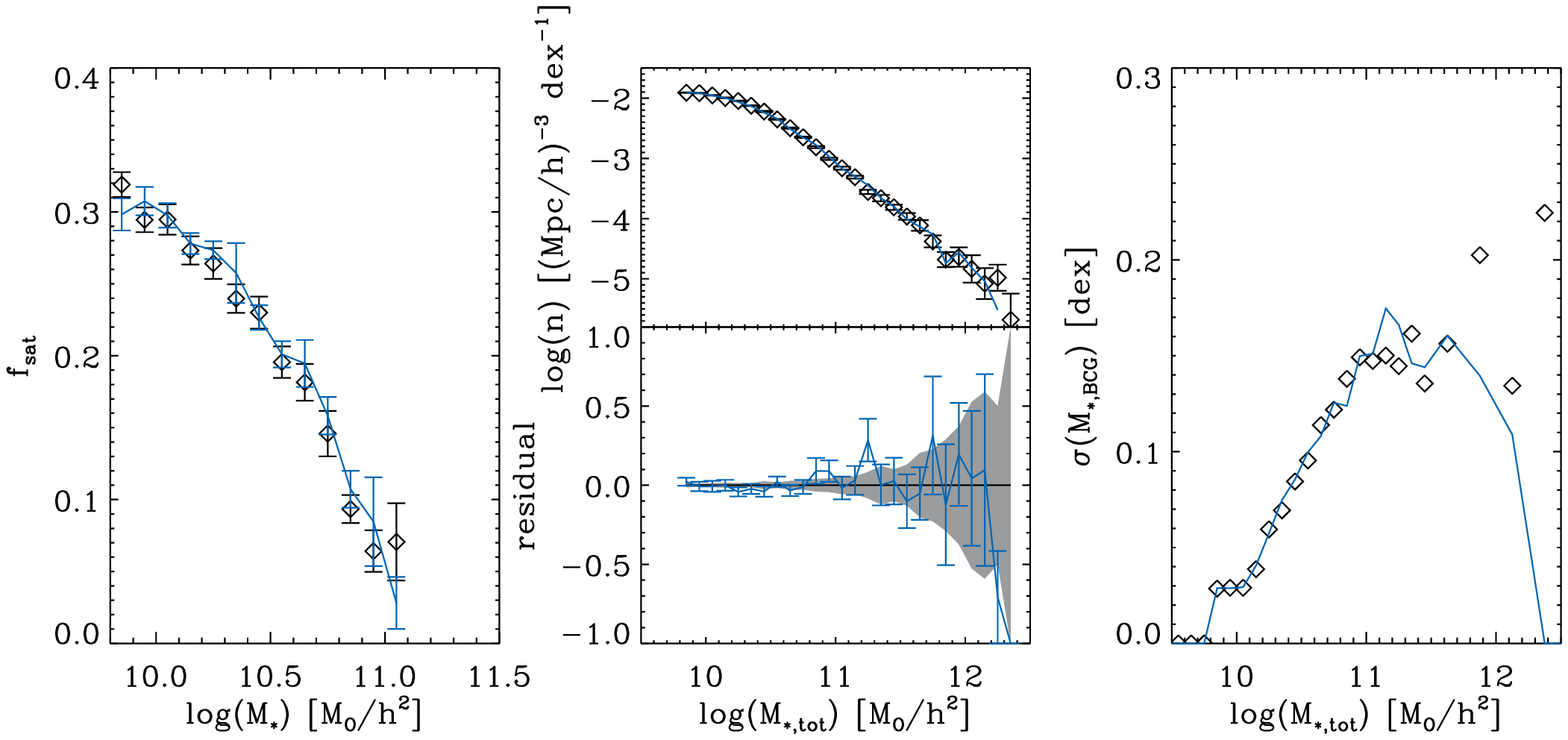}
\caption{Comparison of observed galaxy statistics between SDSS DR7 and our best-fit model, which uses $\vpeak$, $\mucut$=0.03 and scatter=0.20 dex.  Note that only the CSMF and correlation functions with $\log(M_*)>10.2$ are used for fitting.  Plots are the same as described in Fig.~\ref{fig:comp-vtype}.}
\label{fig:vp-bestfit}
\end{figure*}

At every point in parameter space, we measure the CSMF after passing the mock catalog through the group finding procedure and add fiber collisions, as discussed in \S~\ref{sec:mocks}.  This ensures that we accurately mimic the systematic effect these have on the galaxy groups.  Additionally, we add a systematic error to account for shot noise in the galaxy assignment, which is due to using a finite number of halos.  For a fixed set of model parameters, we produced 25 mock catalogs.  Though these have the same input parameters and stellar mass function, the stochasticity of the algorithm produces a certain amount of variation between individual implementations.  We estimated the point-by-point variation between these models for all the measures we use to constrain the fit, and add this estimated variance to the diagonals of the covariance matrices.  Table~\ref{tab:fitq} lists the overall fit results for $\vpeak$ and $\vnpeak$, including this systematic error.  (Unless otherwise noted, error bars shown in plots are statistical only.)  Systematic errors are of roughly the same magnitude as the statistical errors.  There is no large change in our conclusions when we do not include these systematic errors.

To fully accommodate the variation between individual implementations of any given model, we take the mean of each data point and all of its neighbors in parameters space, and the mean variances.  For instance, for a point at $\mucut$=0.02 and $\sigma=0.20$, we take the mean CSMF and two-point clustering of the nine data points within $\mucut$=$0.02\pm0.01$ and $\sigma=0.20\pm0.01$.  This is a reasonable procedure as nearby points in parameters space have relatively small changes in output observables and it smooths fluctuations in the likelihood due to occasional individual outlier points in the CSMF.

We find that only the model based on $\vpeak$ can produce an adequate fit to both the CSMF and the clustering combined.  This model provides an excellent fit to the CSMF and clustering above $\log(M_*) \sim 10$.  However, in general, even the best-fit versions have slightly low clustering on small scales for the $\log(M_*)>9.8$ samples.  Because we cannot cleanly determine whether this is due to a systematic issue with the simulation or a problem with the model, we exclude this lowest threshold from the total $\chi^2$ calculated for the combined measures.  The $M_h=[12.6, 12.9]$ host mass bin from the CSMF estimated $\chi^2$, has significant fluctuations in neighboring bins in stellar mass, which suggest some problematic behavior in the SDSS measurement in that bin, and we omit this bin from our combined fits.

Parameter constraints for this model are shown in Fig.~\ref{fig:constr}.  Here we show the constraints from clustering alone, from the central and satellite parts of the CSMF separately, and from all of these statistics together.  Notably, all three data sets require scatter of $<0.25$ dex.  Marginalizing over scatter to obtain $\mucut$ provides only upper limits: $\mucut$$<0.07$ (68\%) and $\mucut$$<0.11$ (95\%).  Marginalizing over $\mucut$ and interpolating between points in parameter space, the resulting constraints on scatter using the $\vpeak$ model are $\sigma=0.200\pm0.02$ dex (68\%) or $\sigma=0.200\pm0.03$ dex (95\%).  The scatter is most strongly constrained by the two components of the CSMF, while $\mucut$ is determined largely by the clustering.

The measured statistics of the best-fit model are shown in Fig~\ref{fig:vp-bestfit}.  For the best-fit case, we use scatter of 0.20 dex, and $\mucut$=0.03, both well inside the constraints.  This is the best-fit model in the absence of the local averaging procedure described above for estimating the constraints.  We show the clustering and stellar mass functions used to constrain the model, which are in excellent agreement except for the dimmest galaxies.  We also compare the total group stellar mass function, the satellite fraction, and the scatter in central galaxy properties.  All statistics are in excellent agreement with the data
for galaxies with stellar masses greater than $\log(M_*) \sim 10$; there is slightly less clustering and a smaller substructure fraction in the lowest bin of stellar mass.

\begin{figure}
\includegraphics[angle=90, width=0.5\textwidth]{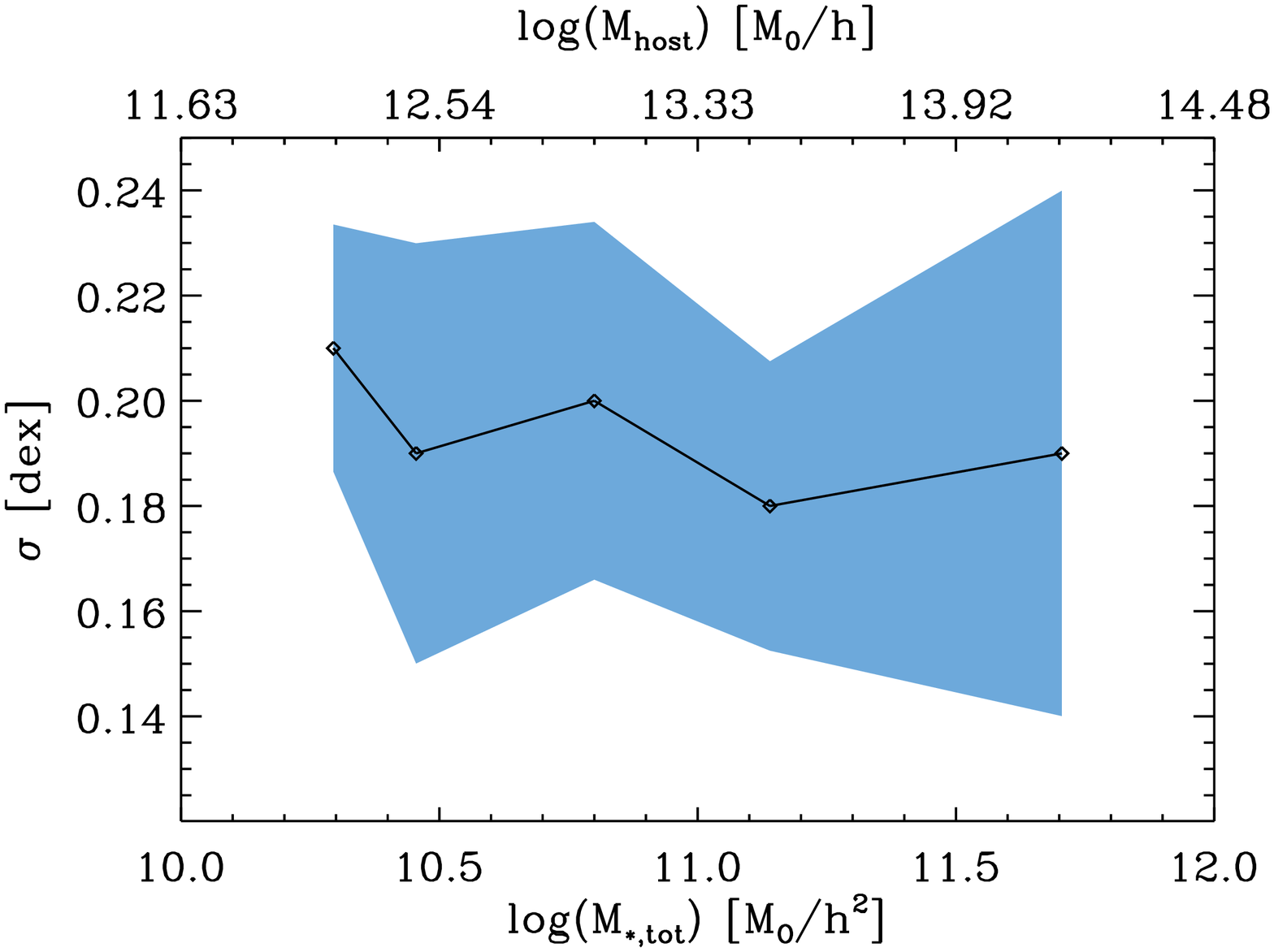}
\caption{Maximum likelihood (black points) value of the scatter in each bin in inferred host halo mass, marginalized over $\mucut$, using constraints from the conditional stellar mass function alone.  Gray bands show the 68\% bounds.  The scatter value is consistent with our overall best-fit scatter of 0.20 dex in the full mass range from $10^{12}$--$10^{14}$.}
\label{fig:sigma-mass}
\end{figure}

As shown in Fig.~\ref{fig:constr}, both the central and satellite parts of the CSMF constrain the scatter in stellar mass at fixed (sub)halo mass in our model.  To check our assumption that scatter is constant with respect to (sub)halo $\vpeak$, we can obtain the best fit in each bin in inferred host halo mass, or total group stellar mass, which is strongly correlated with $\vpeak$.  This result is shown in Fig.~\ref{fig:sigma-mass}.  Here we are using the CSMF only (and not the clustering), and use the results from the mass bins independently, thus the constraints at a given mass are weaker than the full model constraint.  However, it is clear that a scatter of 0.20 dex is in excellent agreement with the result in each individual mass bin, within the 68\% bounds, after marginalizing over $\mucut$.  A very mild trend in the scatter parameter with mass would still be consistent with these constraints.

\begin{figure}
\includegraphics[angle=90, width=0.5\textwidth]{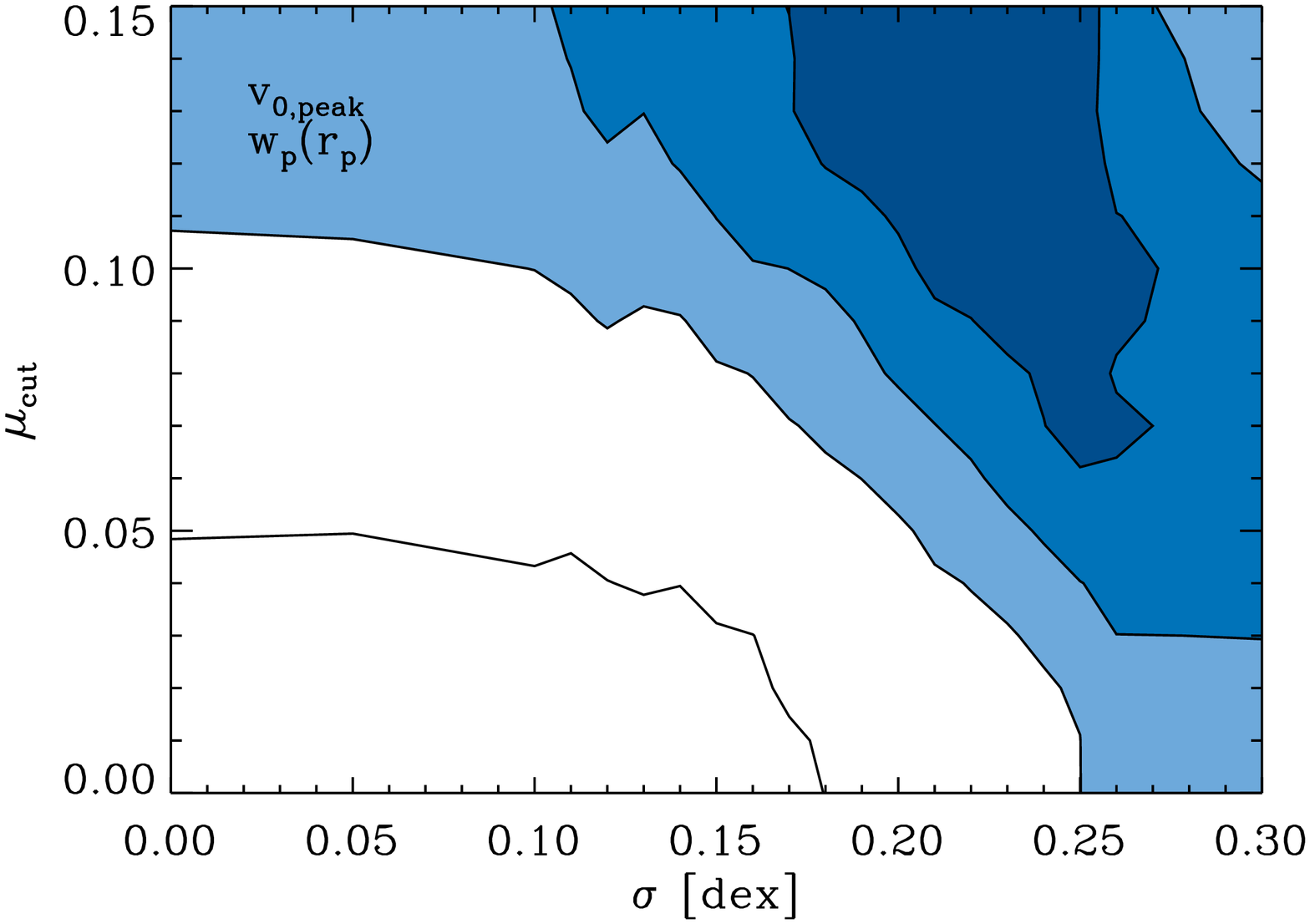}
\caption{Same as Fig.~\ref{fig:constr}, but using $\vnpeak$, and using data for galaxies with $\log(M_*) > 10.2$.  Levels give $P(>\chi^2)$, corresponding to 1, 2, 3, and 5-$\sigma$ contours.  The only constraint plot shown is that for the two-point correlation function.  The CSMFs have such high $\chi^2$ values that they are all completely excluded over this parameter space at the $5-\sigma$ level.}
\label{fig:vnpconstr}
\end{figure}

\begin{table}
	\center
	\caption{Quality of Fit}
	\begin{tabular}{l c c c c r}
	\hline \hline
	Model type					&	$\mucut$	&	$\sigma$ (dex)	&	$\chi^2$	&	N	&	$P(>\chi^2)$ \\
	\hline
	%$\vpeak$ (no systematics)		&	0.02		&	0.20		&	113		&	116	&	0.57	\\
	$\vpeak$					&	0.02		&	0.20		&	107		&	116	&	0.70	\\
	%$\vnpeak$ (no systematics)		&	0.14		&	0.24		&	279		&	116	&	$<10^{-4}$	\\
	$\vnpeak$					&	0.15		&	0.24		&	260		&	116	&	$<10^{-4}$	\\	
	\hline
	\end{tabular}
	\label{tab:fitq}
\end{table}

The low clustering for the dimmest sample considered implies that the model catalogs are missing dim satellites in general; a deficit of satellites in groups and clusters will reduce the small-scale clustering.  A hint of this is also visible in the satellite fraction, which is slightly low in the lowest stellar mass bin.  Further hints are seen in the radial profiles of galaxies, which show a slight deficit in the density of galaxies in the innermost regions (see Appendix~\ref{app:rp}).  It is possible that this is due to a lack of resolution in the N-body simulation on the smallest scales, which could artificially destroy subhalos that correspond to these galaxies.  Equivalently, this may imply support for the inclusion of "orphan" galaxies, which still exist yet whose dark matter halos have already been significantly disrupted (see, e.g., \citet{Guo2011} and references therein for a discussion of orphans).  Adding a small number of orphan galaxies may be able to correct the correlation function without significantly increasing the number of satellites.  Alternatively, it is possible that some form of assembly bias becomes important at low stellar masses, or that the $\mucut$ parameter varies with stellar mass.  A model similar to the last suggestion was considered by \cite{WBZ2012} and found to provide a good match.  However, these possibilities are degenerate and we postpone a full consideration of these degeneracies to future work.  We note that for the Bolshoi simulation considered here, there is no indication that orphans, assembly bias, or non-constant parameters are required for galaxies with $\log(M_*) > 10$.

We find that the $\vnpeak$ model is not able to provide an acceptable fit to the data for any region in parameter space.  With respect to the correlation function alone, $\vnpeak$ is capable of matching or exceeding the correlation function in all bins, as shown in Fig.~\ref{fig:comp-vtype} and with the $w_p(r_p)$ constraint shown in Fig.~\ref{fig:vnpconstr}.  In fact, only the $\vnpeak$ model can produce a good fit to all three stellar mass thresholds simultaneously.  However, it is not able to match either the central or satellite portions of the CSMF.  The central portion of the CSMF is offset somewhat low in stellar mass, due to the increased number of bright satellites.  The high scatter and $\mucut$ needed to match the width of the central CSMF and the high-stellar mass $w_p$ also reduces the number of satellites too much for both the central and satellite parts of the CSMF to be fit simultaneously.  Although this model is ruled out by the data, the values with the best fit for the $\vnpeak$ matching parameter are $\mucut$ = $\sim0.14$ and scatter of $\sim 0.24$ dex.

\begin{figure*}
\centering
\includegraphics[angle=90, width=0.45\textwidth]{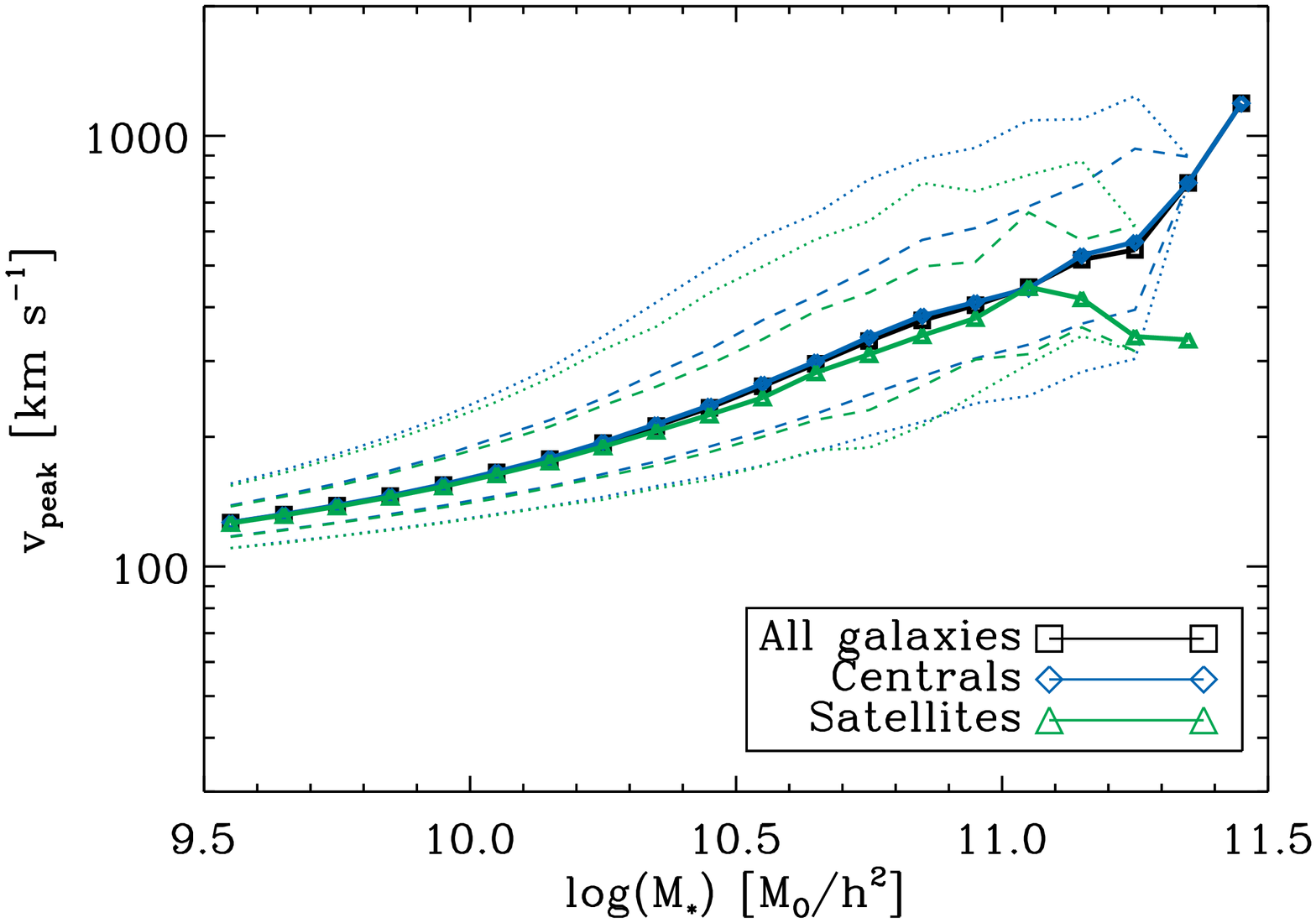}
\includegraphics[angle=90, width=0.45\textwidth]{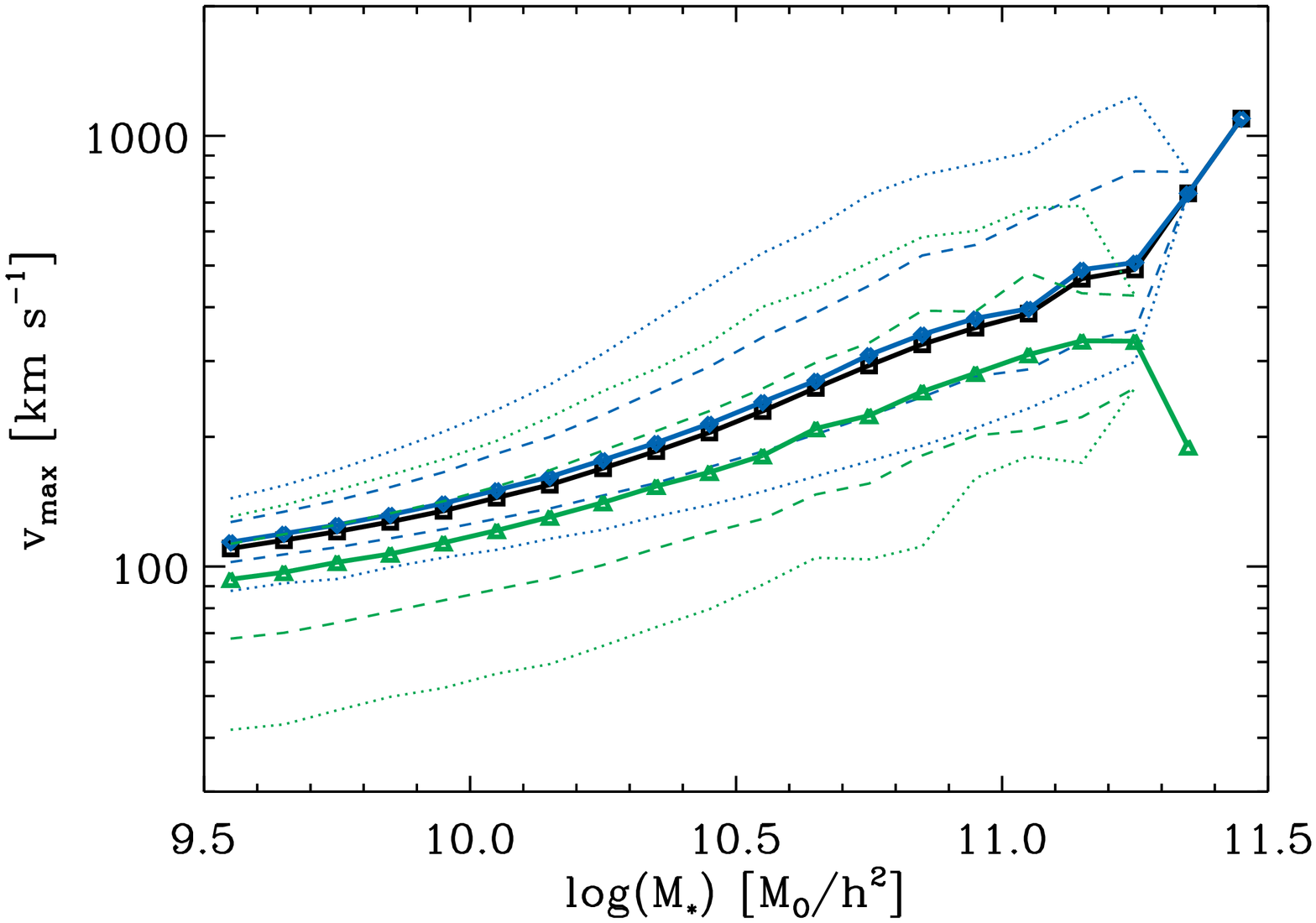}
\includegraphics[angle=90, width=0.45\textwidth]{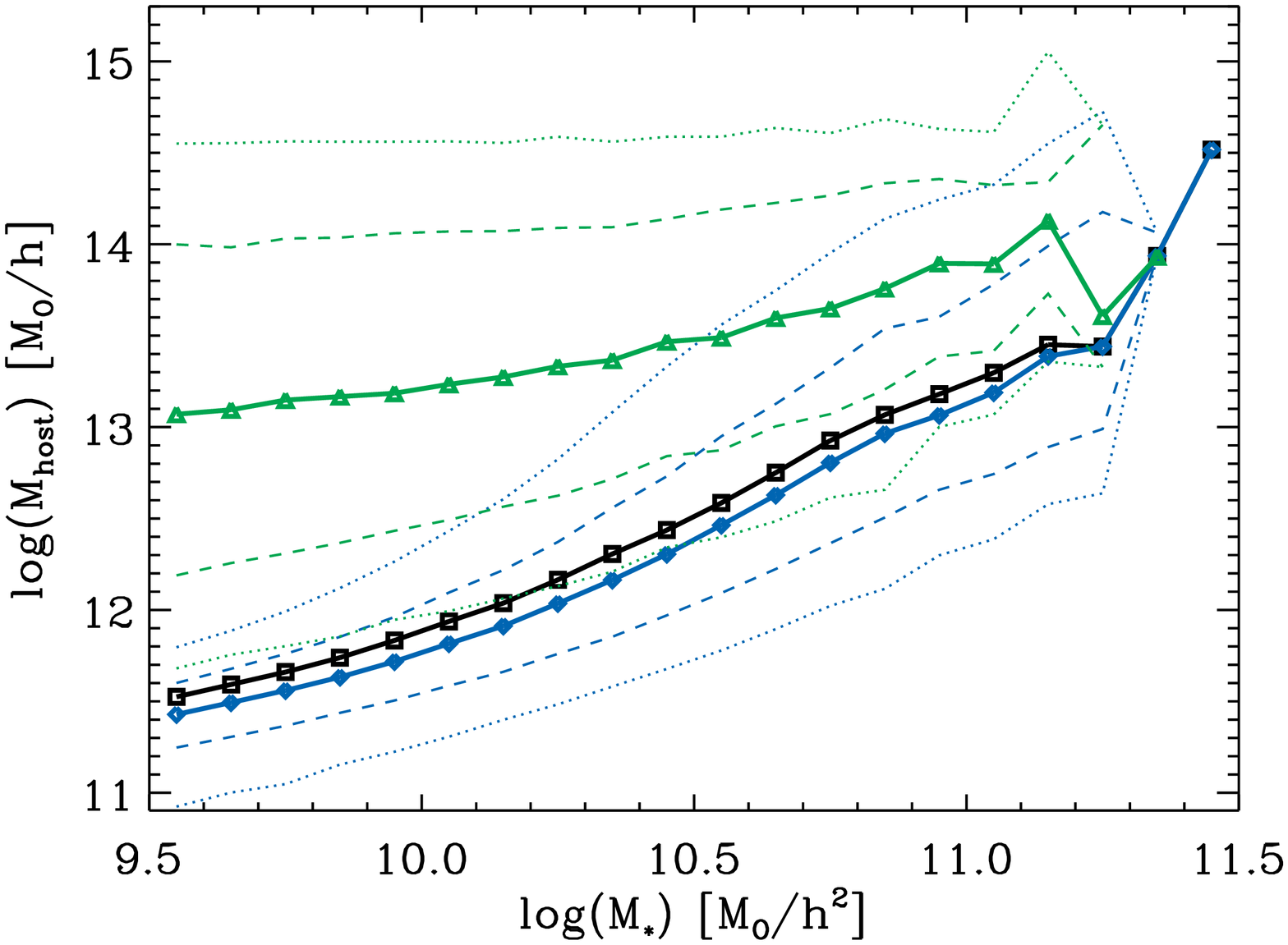}
\includegraphics[angle=90, width=0.45\textwidth]{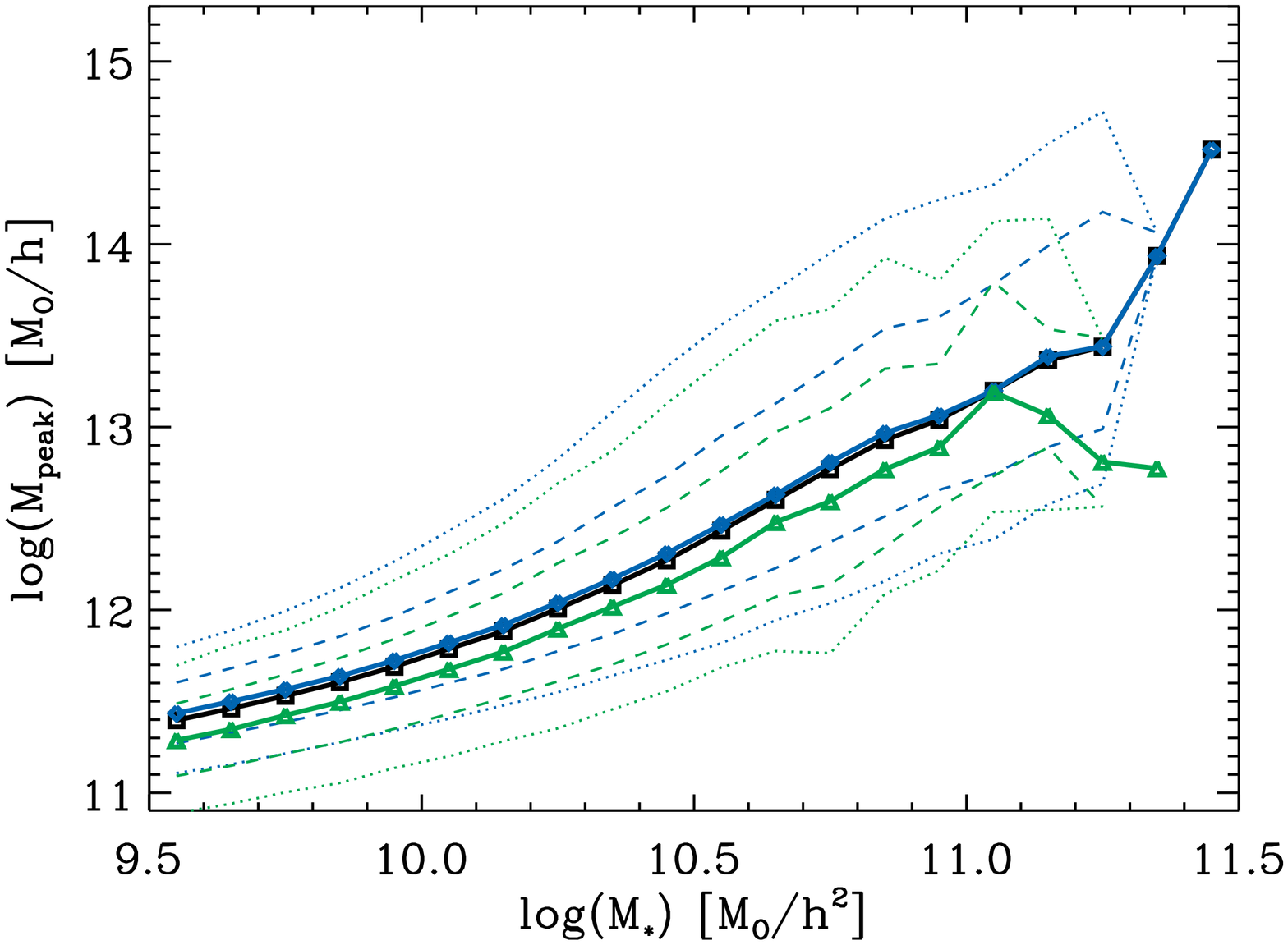}
\caption{ $M_*$ relationship with $\vpeak$ (top left), $\vmax$ (top right), host halo mass (bottom left) and peak (sub)halo mass (bottom right) for the best-fit model, with matching based on $\vpeak$, with 0.20 dex scatter and $\mucut$=0.03.  Blue indicates centrals, green, satellites.  Solid black lines are the median of the total (satellites plus centrals).  Solid lines are the median values of $\vmax$ or $\vpeak$ for bins in $M_*$.  Dashed and dotted lines contain given the 68\% and 95\% bounds on galaxies in each bin, centered at the median.  
Although the central and satellite distributions are similar in $\vpeak$ due to how the catalog is constructed, satellites typically have lower $\vmax$ and larger dispersion due to stripping after accretion.  (All units are given with $h=1$.)}
\label{fig:vp-mst-reln}
\end{figure*}

\subsection{Halo properties for Satellite and Central Galaxies in the Best-Fit Model}

The results shown in the previous section were all in observed space.  We now consider the properties of the underlying model in our best-fit case.  For the best-fit case, we use scatter of 0.20 dex, and $\mucut$=0.03, both well inside the constraints.  This is the best-fit model in the absence of the local averaging procedure described above for estimating the constraints.

A series of general relationships between halo (or subhalo) properties and galaxy stellar mass for our best-fit model are shown in Fig. \ref{fig:vp-mst-reln}.  This shows the median values of various halo properties in bins of stellar mass, split between satellite and central galaxies.  The relationship between $\vpeak$ and stellar mass is nearly the same for both satellites and centrals.  This is as expected, since when abundance matching stellar mass to halos sorted by $\vpeak$ we make no distinction between satellites and centrals.

On the other hand, the satellite galaxies have significantly lower $\vmax$ at the present time.  This is sensible, as (sub)halos with the same $\vpeak$ host galaxies with comparable stellar mass, but satellite galaxies at that same stellar mass are in subhalos with lower $\vmax$ due to stripping following accretion.  As a result, central galaxies with $\log(M_*)<10.5$ are in halos with roughly 25\% higher $\vmax$ than subhalos hosting satellite galaxies with the same stellar mass.  This difference increases to as much as $\sim35\%$ at higher stellar mass.  This result may be in tension with a recent study of the variation of the Tully-Fisher relation on environment using SDSS galaxies \citep{Mocz2012}, which finds no dependence on environment.  However, a direct comparison is complicated by differences in the environment definition from our designation of central and satellite galaxies, as well as differences in sample selection, so we leave a precise comparison to future work.

It is also noteworthy that for (sub)halos hosting lower stellar mass galaxies, the subhalos have a much larger variation in $\vmax$ than do the distinct halos.  This is due to the wide variety in $\vmax$ that may be associated with the same past $\vpeak$, depending on how much the individual subhalo has been stripped since it was accreted.  

The distribution of galaxies in host halo mass at a fixed stellar mass is an interesting complement to the CSMF.  As one might expect,
satellite galaxies (and their subhalos) tend to be hosted by significantly more massive distinct halos than central galaxies of the same stellar mass.  The variation in satellites' host masses is also much larger at lower stellar mass, since a relatively small subhalo may reside in a low mass halo, as well as a very massive dark matter halo.  At higher stellar mass, this relationship narrows, since only sufficiently massive dark matter halos can host massive subhalos, and, hence, very massive satellite galaxies.  We refer to this host mass, of the distinct halo containing a central or both a satellite and its subhalo, as $\mhost$.

The variation in $\vpeak$, $\vmax$ or $\mhost$ at fixed central stellar mass is reduced as stellar mass decreases.  This is most likely due to the fact that at high stellar mass, the stellar mass function, as well as the halo mass function and the circular velocity function, is much steeper.  Thus, at high stellar masses, a bin of fixed width yields a wider range of values in the circular velocities or host halo mass.

\subsection{Best-Fit Conditional Stellar Mass Function}

\begin{table*}
	\center
	\caption{Intrinsic CSMF Fit Parameters for Best-Fit Model}
	\begin{tabular}{l c c c c c r}
	\hline
	\hline
	$\mhost$	&	$\log(M_{*,c})$ &	$\sigma_c$  	&	$\phi_*$  	&	$\alpha$	&	$\log(M_{*,Sch})$	&	No. of hosts \\
  	$[\log(\Msun/h)]$	&	$[\log(\Msun/h^2)]$	&	$[\log(\Msun/h^2)]$	&	$[\log(\Msun/h^2)^{-1}]$	& &	$[\log(\Msun/h^2)]$	&\\
	\hline
	12.0-12.3	&	$10.232\pm0.001$	&	$0.218\pm0.001$		&	$0.652\pm0.059$	&	$-0.98\pm0.16$	&	$9.92\pm0.04$		&	27948	\\
	12.3-12.6	&	$10.383\pm0.002$	&	$0.212\pm0.001$		&	$1.56\pm0.08$		&	$-0.76\pm0.10$	&	$10.01\pm0.02$	& 	14983	\\
	12.6-12.9	&	$10.500\pm0.002$	&	$0.205\pm0.001$		&	$3.40\pm0.09$		&	$-0.41\pm0.08$	&	$10.04\pm0.02$	&	7814		\\
	12.9-13.2 &	$10.591\pm0.003$	&	$0.209\pm0.002$		&	$6.07\pm0.22$		&	$-0.62\pm0.06$	&	$10.17\pm0.02$	&	4000		\\
	13.2-13.8 &	$10.656\pm0.004$	&	$0.206\pm0.002$		&	$13.5\pm0.5$		&	$-0.74\pm0.04$	&	$10.27\pm0.01$	&	2896		\\
	13.8-14.5 &	$10.748\pm0.009$	&	$0.213\pm0.004$		&	$42.5\pm2.3$		&	$-0.95\pm0.05$	&	$10.38\pm0.02$	&	595		\\
	\hline
	\end{tabular}
	\label{tab:clf}
	\vspace{10 pt}
	\caption{Intrinsic HOD Fit Parameters for Best-Fit Model}
	\begin{tabular}{l c c c c c c r}
	\hline \hline
	$M_*$ threshold	&	$(M_r - 5\log(h))$	&	$\log(M_{\rm{min}})$		&	$\sigma_m$	&	$\log(M_1)$	&	$\log(M_{\rm{cut}})$	&	$\alpha_{\rm HOD}$	&	No. of galaxies \\
	$\log(\Msun/h)$	&	&	[$\log(\Msun/h)$]	&	$[\ln(\Msun/h)]$	&	$[\log(\Msun/h)]$	&	$[\log(\Msun/h])$	&	&	\\
	\hline
	10.76	&	-21.5		&	$13.71\pm0.03$		&	$2.30\pm0.06$			& 	$14.31\pm0.13$	&	$13.1\pm0.5$	&	$0.97\pm0.30$		&	4437		\\
	10.54	&	-21.0		&	$12.924\pm0.006$		&	$1.75\pm0.01$			& 	$13.74\pm0.15$	&	$12.8\pm0.3$	&	$0.94\pm0.21$		&	18062	\\
	10.31	&	-20.5		&	$12.318\pm0.002$		&	$1.161\pm0.002$		& 	$13.30\pm0.17$	&	$12.6\pm0.2$	&	$0.93\pm0.17$		&	49715	\\
	10.07	&	-20.0		&	$11.950\pm0.001$		&	$0.9000\pm0.0007$		& 	$12.98\pm0.18$	&	$12.4\pm0.2$	&	$0.94\pm0.15$		&	103904	\\
	9.82		&	-19.5		&	$11.6336\pm0.0001$	&	$0.6248\pm0.0001$		& 	$12.76\pm0.17$	&	$12.2\pm0.2$	&	$0.95\pm0.13$		&	174932	\\
	9.54		&	-19.0		&	$11.4588\pm0.0002$	&	$0.6047\pm0.0001$		& 	$12.59\pm0.16$	&	$12.0\pm0.2$	&	$0.96\pm0.11$		&	261915	\\
	\hline
	\end{tabular}
	\label{tab:hod}
\end{table*}

Following \citet{YMB2009} and \citet{Cac2009}, we fit the central galaxies with a log-normal function.  We find that a Schechter function is sufficient for the satellite galaxies (as has been found previously, see e.g. \citealt{Mos2010}).
When we perform fits to the CSMF, we adopt the following parameterization of these quantities, using in all cases the differential $d\log(M_*)$:
\be\label{eq:csmf-cen}
	\Phi_c(M_*|\mhost) = \frac{1}{\sqrt{2 \pi \sigma_c^2} } \exp\left({-\frac{(\log{M_*}-\log{M_{*,c}})^2}{2 \sigma_c^2}}\right)
\ee

\be\label{eq:csmf-sat}
	\Phi_s(M_*|\mhost) =\phi_* \left(\frac{M_*}{M_{*,s}}\right)^{\alpha+1} \exp\left(-\frac{M_*}{M_{*,s}}\right)
\ee

Thus, the central galaxies are characterized by two parameters:  $M_{*,c}$, which is the geometric mean of the central stellar mass, and $\sigma_c$, which is the width of the log-normal distribution in dex.  Both are closely related to the scatter in the model, as described below.  The satellite galaxies are described by the usual three parameters of a Schechter function.  Here, $M_{*,s}$ is the cutoff luminosity, $\alpha$ the faint-end slope, and $\phi_*$ the overall normalization.  Unlike in \citet{YMB2008, YMB2009}, we choose not to fix the relationship between $M_{*,c}$ and $M_{*,s}$ explicitly.
These results are combared to others in the literature in \S 7.

\begin{figure*}
\centering
\includegraphics[angle=90, width=0.9\textwidth]{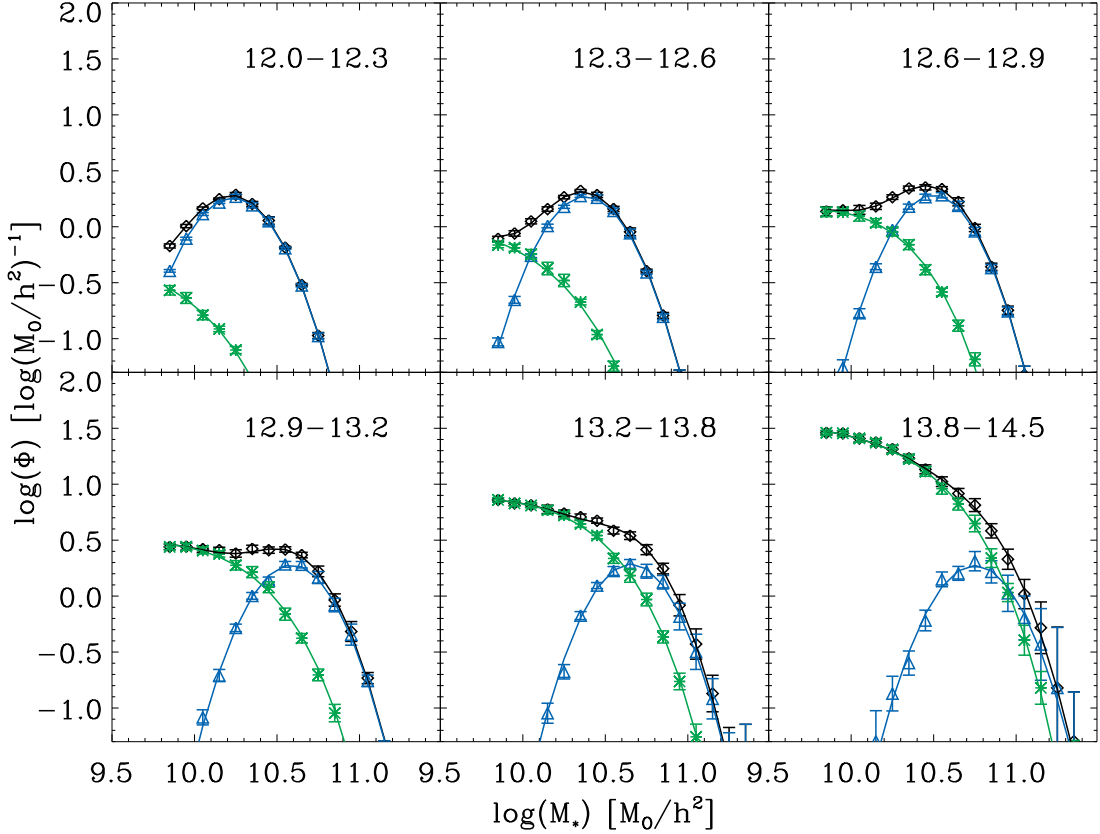}
\caption{ CSMF fits for the best model.  Black is the overall CSMF; blue, central galaxies only; green, satellite galaxies only.  Solid lines are the respective fits.  Labels give the host mass range in $\log(\Msun/h)$.  Eq.~\ref{eq:csmf-cen} and \ref{eq:csmf-sat} describe the fit, while Table~\ref{tab:clf} lists the parameters.  Error bars include estimated systematic errors.
}
\label{fig:fit-clf}
\end{figure*}

The results of fitting to the intrinsic CSMF can be seen in Fig. \ref{fig:fit-clf}.  This is the CSMF in the Bolshoi simulation, using our best-fit model, and without observational complications (e.g., group-finding).  Here, a galaxy is a satellite if its halo is a subhalo.  This is the same model as shown in Fig. \ref{fig:vp-bestfit};  the main difference between the two is that the intrinsic CSMF does not require that the central galaxy has the most stellar mass, a necessary assumption of the group-finding algorithm.
%The impact is strongest for the least massive halos, or groups with the least total stellar mass.  In particular, if a "group" has only one or two galaxies, the stellar mass is dominated by the most massive one.  That most massive galaxy is assumed to be the central galaxy.  Because our earlier analysis used the group stellar mass to assign host halo mass, at low host halo masses, we  obtain a nearly zero-scatter correspondence between central stellar mass and host halo mass.
This produces the sharp central peak that can be seen in Fig.~\ref{fig:vp-bestfit} and the other comparison figures.  However, as can be seen in Fig. \ref{fig:fit-clf}, the underlying distribution is much broader.  This is primarily due to the 0.20 dex scatter in this model, with a small contribution from the finite size of the mass bin.

\begin{figure}
\includegraphics[angle=90, width=0.5\textwidth]{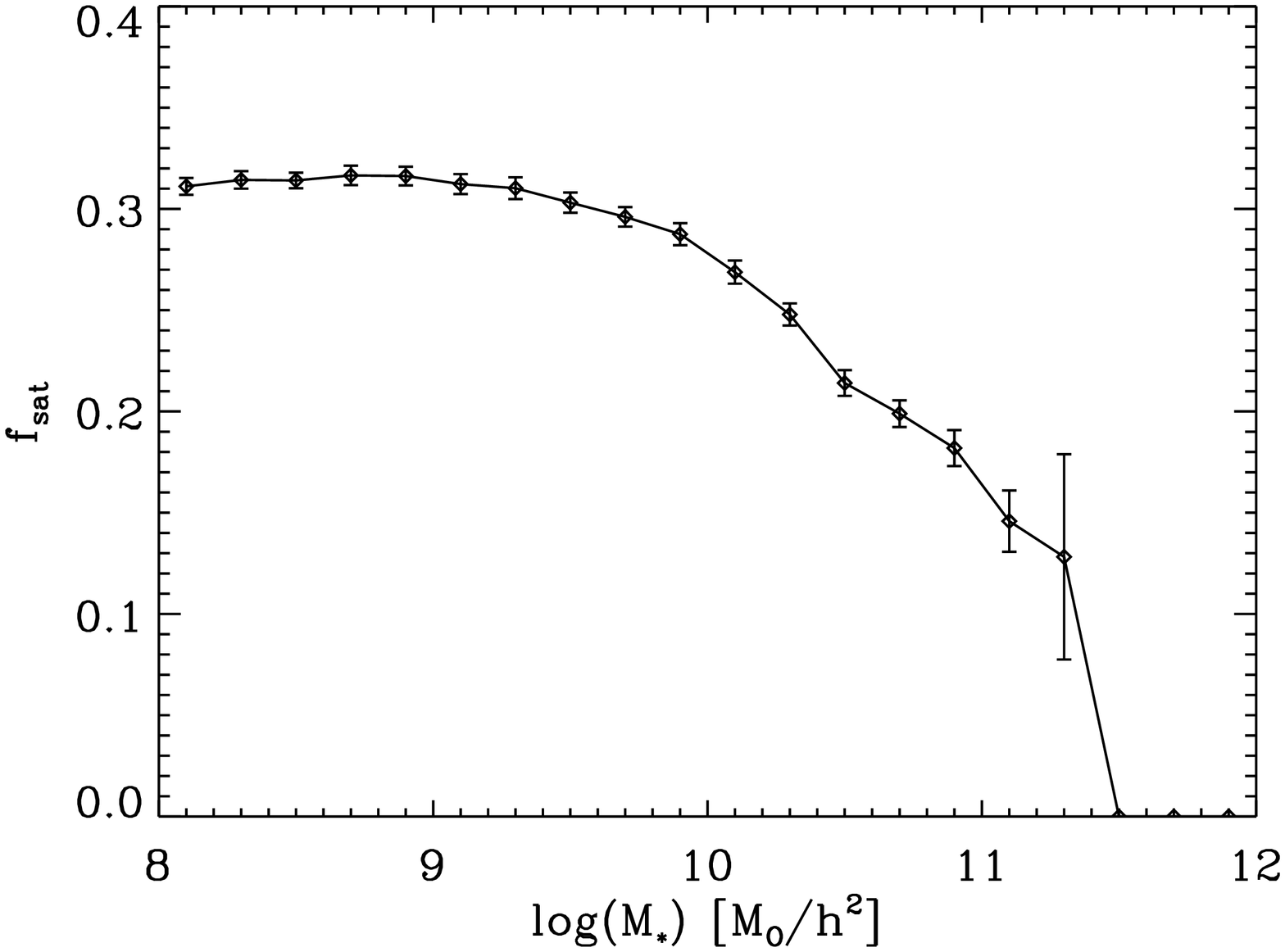}
\includegraphics[angle=90, width=0.5\textwidth]{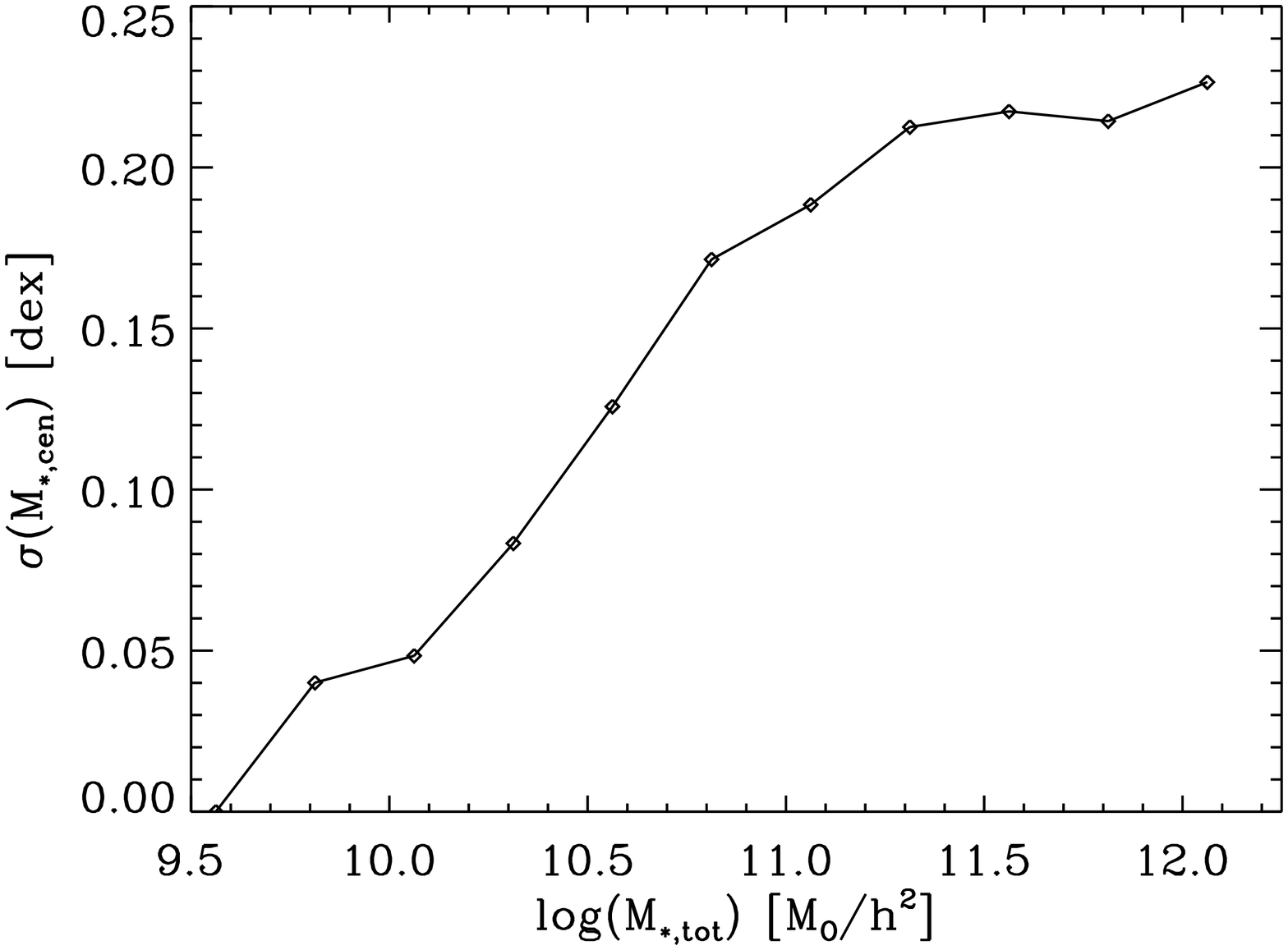}
\caption{Additional measures of the intrinsic distribution of galaxies in our best-fit model.  {\em Top:} Intrinsic satellite fraction as a function of stellar mass.  Because the input SMF only extends down to $\log(M_*)=9.8$, stellar masses below this cutoff are drawn from a power-law extrapolation to the input SMF.  {\em Bottom:}  Scatter in central galaxy stellar mass as a function of total group mass.  Note the difference between the intrinsic scatter shown here and the smaller ``observed'' scatter after group finding shown in Fig. \ref{fig:vp-bestfit}.  In both cases, this scatter
becomes poorly defined for groups with no galaxies above the stellar mass cutoff.  
}
\label{fig:smtruth-other}
\end{figure}

\begin{figure}
\centering
\includegraphics[width=0.5\textwidth]{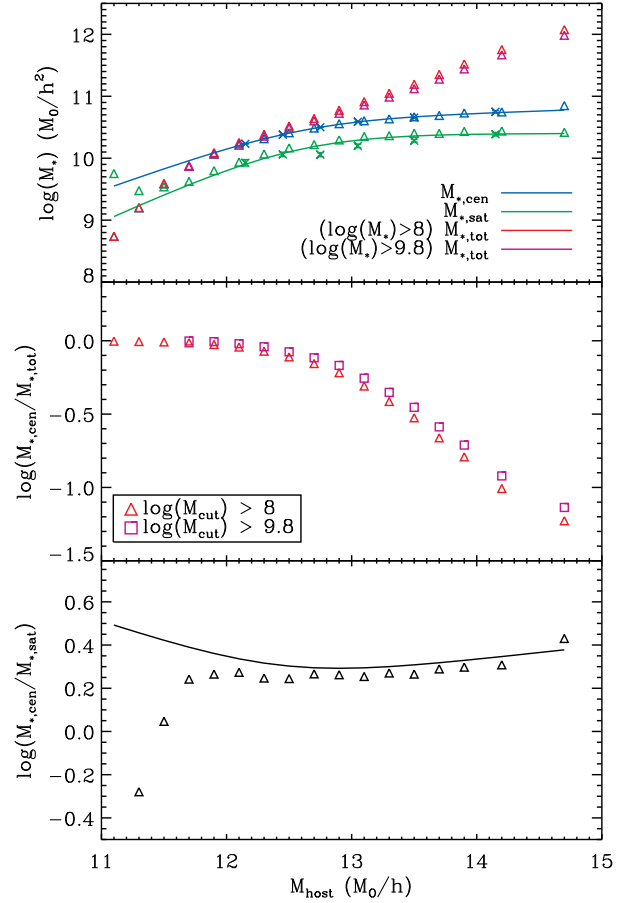}
\caption{Measures of the intrinsic distribution of galaxies in our best-fit  model.    {\em Top:}  Median central mass ($M_{*,c}$), median total group stellar mass ($M_{*,tot}$) for two different stellar mass thresholds, and the fitted $M_{*,s}$ to a Schechter function in narrow mass bins (triangular points).  Solid lines are the fitted values of $M_{*,c}$ and $M_{*,s}$ as discussed in \S~\ref{subsec:csmf-mhost}.  The x's with error bars indicate the $M_{*,c}$ and $M_{*,s}$ fitted values in the individual mass bins used for observational comparisons.  {\em Center:}  Ratio of the median central stellar mass to the median total group stellar mass, as a function of host halo mass.  This becomes less meaningful as the central comes to dominate the group's stellar mass.  {\em Bottom:}  Ratio of characteristic satellite stellar mass $M_{*,s}$ to the median central stellar mass.  Note that this is fairly constant at $\log(\frac{M_{*,c}}{M_{*,s}})\sim 0.28$.  Solid line indicates the difference in the host mass dependent fits for $M_{*,c}$ and $M_{*,s}$.  In all cases, cuts in stellar mass are given in $\log(\Msun/h^2)$.}
\label{fig:smtruth}
\end{figure}

A few additional intrinsic measurements are shown in Figs.~\ref{fig:smtruth-other} and \ref{fig:smtruth}.  For all of these plots, we extrapolate our stellar mass function down to stellar masses of $10^8~\Msun/h^2$.  Fig.~\ref{fig:smtruth-other} shows the intrinsic satellite fraction and scatter, which may be contrasted with the mock observed values in Fig.~\ref{fig:vp-bestfit}.  Notably, in the intrinsic case, the satellite fraction flattens below the cutoff stellar mass of $\log(M_* h^2/\Msun)=9.8$ in our volume-limited sample.  The scatter in central stellar mass at fixed group total stellar mass shows the same trend as in the observed case, with low scatter at low stellar masses due to the fact that the central contributes nearly all of the stellar mass.  However, because no group finding is involved to artificially reduce the scatter for groups with many galaxies, it reaches $\sim0.2$ dex at the massive end.

We also show the more finely binned trends in characteristic group stellar mass, central galaxy stellar mass, and satellite galaxy stellar mass in Fig~\ref{fig:smtruth}.  At low host masses, there are few satellite galaxies with even $10^8~\Msun/h^2$ solar masses, and so the measured $M_{*,s}$ is not reliable below $\log{\mhost}\sim11.5$.  The central stellar mass and satellite stellar mass $M_{*,s}$ are only slowly changing for host halo masses above $\sim10^{13}\Msun/h$, and then fall off at lower host halo masses.   Note that the ratio between central galaxy stellar mass and satellite stellar mass $M_{*,s}$ is roughly constant over a broad range in host halo mass, which is in general agreement with results from \cite{YMB2009}.  This figure includes some of the results of a fit parameterized to host halo mass, which works well for $\mhost>10^{12}\Msun/h$ and is discussed in the next section.

\subsection{Conditional Stellar Mass as a Function of Halo Mass}\label{subsec:csmf-mhost}

\begin{figure*}\centering
\includegraphics[angle=90, width=0.85\textwidth]{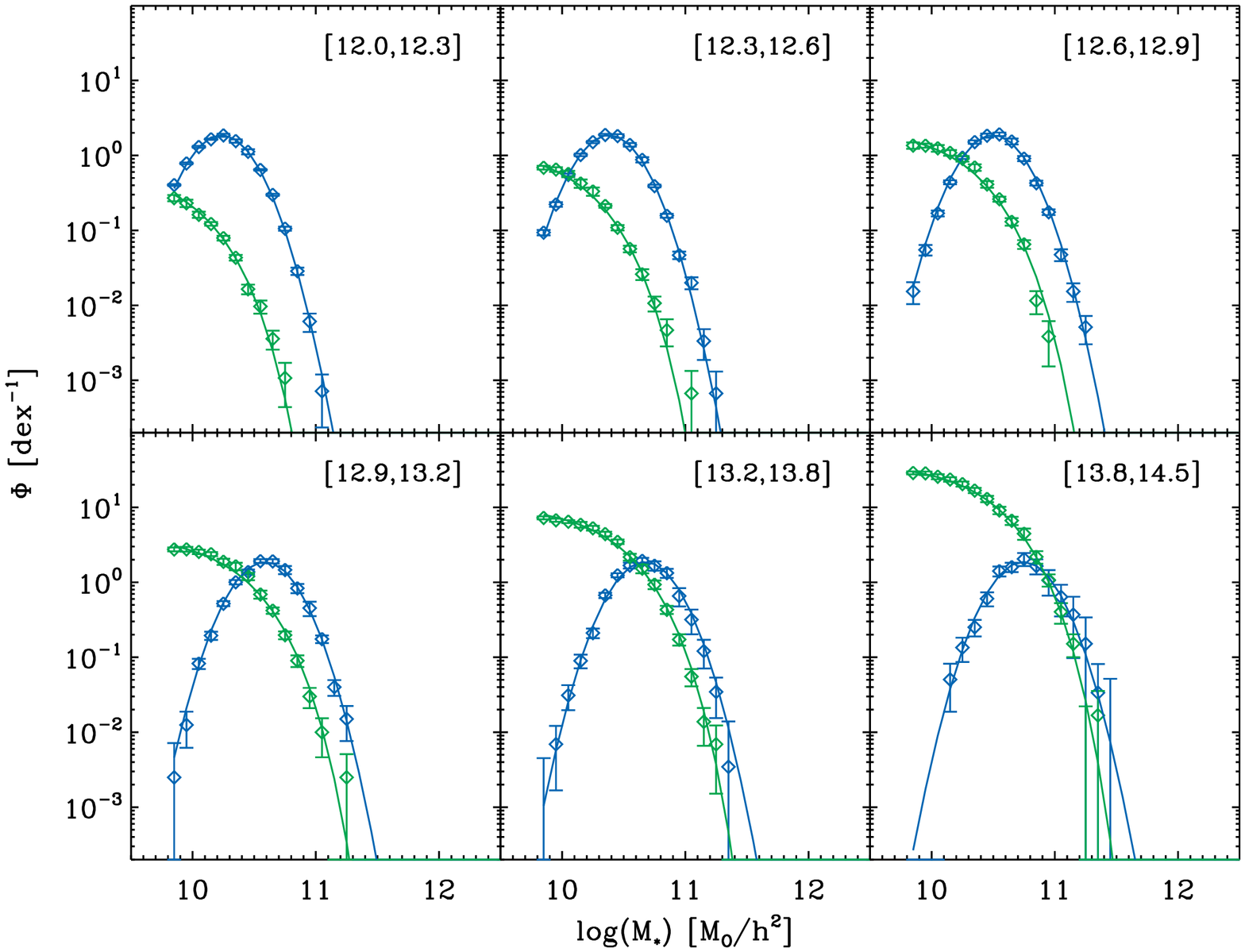}
\caption{ Comparison of the best-fit model with the DR7 SMF (points) against the full fit using host halo-mass dependent parameters (lines).  Host halo mass ranges are given in $\log(\Msun/h)$.  Error bars include estimated systematic errors.}
\label{fig:fullfit-dr7}
\includegraphics[angle=90, width=0.85\textwidth]{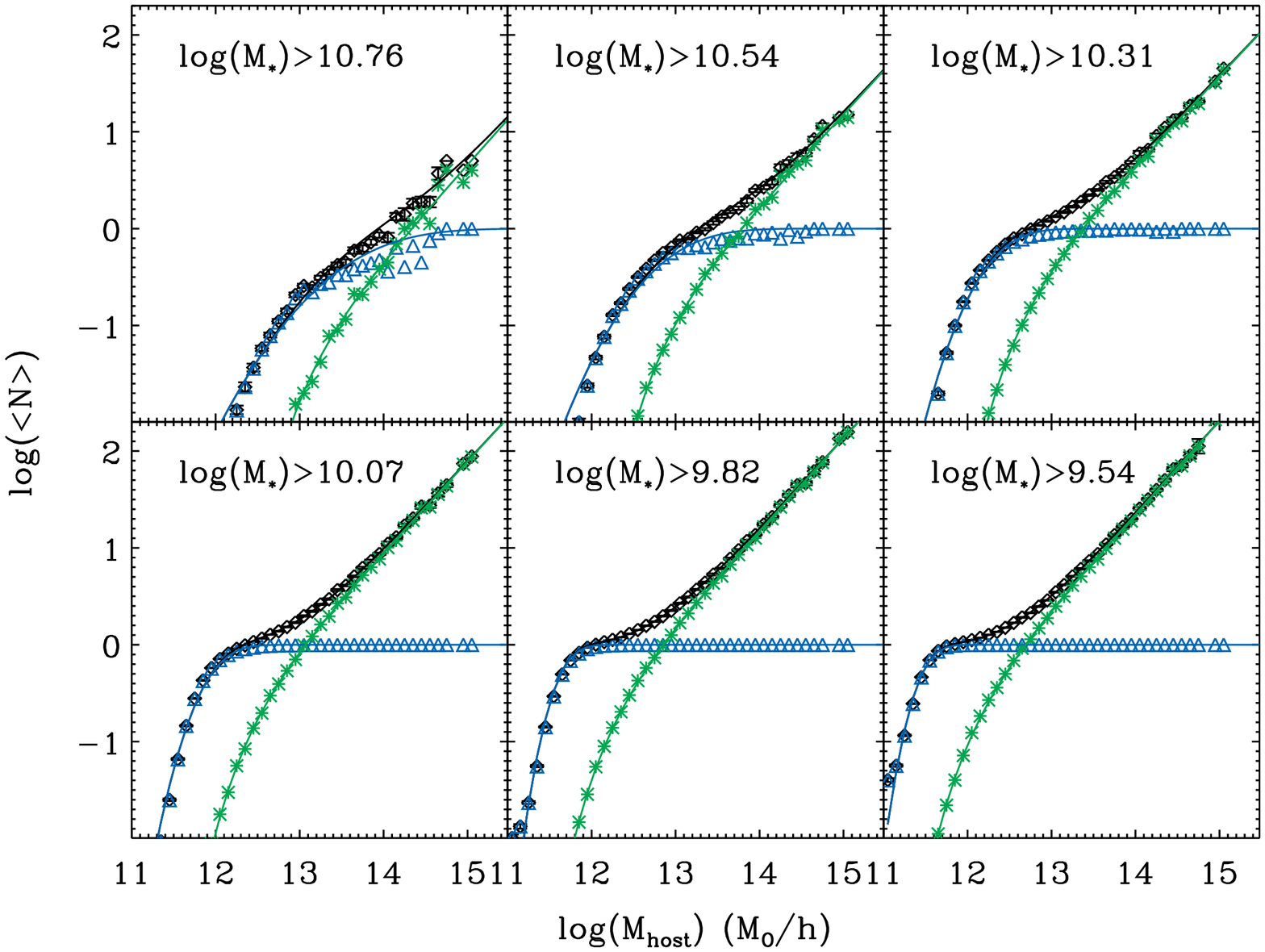}
\caption{ HOD fits for the best model.  Black is the overall HOD; blue, central galaxies only; green, satellite galaxies only.  Solid lines are the respective fits.  Cuts in stellar mass are given in  $\log(\Msun/h^2)$.  Error bars have been omitted from the centrals and satellites for clarity.  The HOD fit is presented in Eq.~\ref{eq:hod-cen} and \ref{eq:hod-sat}, with parameters listed in Table \ref{tab:hod}.
}
\label{fig:fit-hod}
\end{figure*}

To more generally describe the CSMF, we take the parameters from equations \ref{eq:csmf-cen} and \ref{eq:csmf-sat} to be functions of host halo mass.

For the central CSMF, the mean stellar mass is defined by:

\be
	\log(M_{*,c}) =  \log(M_0) + g_1\log\left(\frac{\mhost}{M_1}\right) + (g_2-g_1)\log\left(1+\frac{\mhost}{M_1}\right)
\ee

where $M_0$ is a characteristic stellar mass, $M_1$ is a characteristic host halo mass, and $g_1$ and $g_2$ are power-law slopes.  $M_h$ is the host halo mass.  The width $\sigma_c$ of the log-normal function is assumed to be constant as a function of host halo mass.

The satellite CSMF is determined by the three Schechter function parameters, $\phi_*$, $\alpha$, and $M_{*,s}$.

\be
	\phi_* = \left(\frac{\mhost}{M_\phi}\right)^{a} %* \left(1+\frac{M_h}{M_\phi}\right)^{a_2}
\ee
\be
	\log(M_{*,s}) = \log(M_{*,0}) + b \log\left(\frac{\mhost}{M_{*,1}}\right) - b \log\left(1+\frac{\mhost}{M_{*,1}}\right)
\ee

The slope $\alpha$ is assumed to be constant as a function of halo mass.  Based on Fig.~\ref{fig:fit-clf} and the individual fit results in Table~\ref{tab:clf}, it is evident that $\alpha$ varies significantly from one fit to another without a commensurate variation in the shape of the satellite CSMF.  This is due to the fact that when limiting the fit to stellar masses $\log(M_*)>9.8$ we lose constraining power on the low-mass slope, and it becomes degenerate with the other satellite parameters.  When we consider the extrapolation to lower stellar mass, we find that the slope at all host masses converges to $\alpha\sim-1$.  There, we hold $\alpha=-1$ fixed.

We then fit this functional form to the binned CSMF data.  The parameters for the resulting fit are in Tables~\ref{tab:yang-comp} and \ref{tab:yang-comp-sat} for the DR7 input stellar mass function.  The overall result of this fit is shown in Fig. \ref{fig:fullfit-dr7}, which clearly reproduces the data well.  Some comparisons of the parameters as a function of halo mass are shown in Fig.~\ref{fig:smtruth} as discussed in the previous section.

\subsection{Best-Fit Halo Occupancy Distribution}

The halo occupancy distribution (HOD) may be used, for instance, to predict or fit to galaxy clustering \citep{ZCZ2007, WBZ2011, Zeh2011}.  The HOD is defined in part by $P(N|M_h)$, the probability of finding $N$ galaxies of some type in a halo of mass $M_h.$  The common procedure takes galaxies brighter than some fixed stellar mass $M_{*,min}$ as the type of interest.  In this case, the expectation of the HOD may be obtained directly from the CSMF:

\be\label{eq:hod-int}  \langle N(\mhost) \rangle =\int_{M_{*,min}}^\infty \Phi(M_*|\mhost) ~\mathrm{d} M_* \ee

Similar to the CSMF, the HOD may also be split into central and satellite contributions, with $<N(M)>=<N_c>+<N_s>$.  The central portion may be described by a step function, with a cutoff of some width.  Thus, there is some minimum host mass, $M_{\rm{min}}$, below which the halo is too small to host a central galaxy brighter than $M_{*,min}$.  Above $M_{\rm{min}}$, each halo typically hosts one central galaxy; below $M_{\rm{min}}$, each typically hosts none.  The satellite galaxies are a different matter, generally well-described by a power law, with some cutoff at or above $M_{\rm{min}}$.  Below this cutoff there are very few satellite galaxies.

While the usual approach to determining the HOD is to perform a fit to the clustering and number density data, we instead use the information on group association available in the simulations to measure the HOD directly.  This is done by counting all galaxies above some stellar mass for each (host) halo of a given mass, then averaging over all halos.

We fit the following functional form to the HODs drawn from these catalogs:

\be\label{eq:hod-cen} 
\langle N_c\rangle=\frac{1}{2} \left(1+\mathrm{erf} \left( \frac{\ln \mhost-\ln M_{\rm{min}}}{\sigma_m}\right)\right)
\ee 
\be\label{eq:hod-sat} 
\langle N_s\rangle=\left( \frac{\mhost}{M_1}\right)^{\alpha_{\rm HOD}} \exp\left(-\frac{M_{\rm{cut}}}{\mhost}\right)
\ee

$M_{\rm{min}}$ is, as described above, the cutoff in the central galaxies.  The error function provides a smoothed step function that reproduces the form of the central galaxies, whose width is characterized by the parameter $\sigma_m$.  The satellites are characterized by $M_{\rm{cut}}$, the cutoff below which galaxies of the given type are not expected to have satellites, the scale $M_1$ at which the galaxies typically have one satellite, and $\alpha_{\rm HOD}$, the power-law slope.  All mass scales increase as the stellar mass of the selected sample increases.  These fits are presented in Fig.~\ref{fig:fit-hod}.

Our model may be compared against the \cite{Zeh2011} HODs fitted from clustering.  An exact comparison requires the use of luminosity rather than stellar mass (see Appendices \ref{app:lum} and \ref{app:hod} for the results using $r$-band luminosity).  Our stellar mass results show the same general trends, that is, a satellite slope of $\alpha_{\rm HOD}$ consistent with one for all thresholds, decreases in all three mass scales with decreasing stellar mass, and decreasing $\sigma_m$ with decreasing stellar mass.  However, there are differences in detail.  We find that $\sigma_m$ is significantly larger, and necessarily nonzero, for all thresholds we consider.  We also find a higher value of $M_{\rm{min}}$ at each threshold.  This is likely due in part to the degeneracy between $M_{\rm{min}}$ and $\sigma_m$ when estimating the HOD from clustering.  However, it remains possible that these differences are attributable to the use of stellar mass rather than luminosity.

\section{Comparisons with Other Measurements}\label{sec:altmeas}

\subsection{Stellar Mass Function}

\begin{figure}
\centering
\includegraphics[angle=90, width=0.5\textwidth]{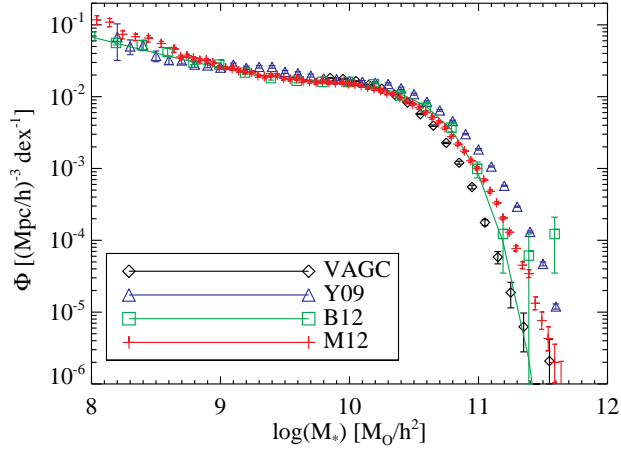}
\caption{Four stellar mass functions from the SDSS local data.  The NYU-VAGC (black) was used to fit our model parameters and tests its validity; we repeat our calculations using the others to understand the sensitivity to this global measurement.  The \citet{YMB2009} stellar mass function (green) is drawn from a sample used in a previous study of the CSMF.  For \citet{Bal2012}, we show both the data (square points) and their fit (line), the latter of which we use in later model tests.  Finally, we also show \citet{Mou2013}, a recent result based on SDSS combined with additional multi-wavelength data and a full Bayesian analysis of SEDS to derive stellar masses.}
\label{fig:smf-all}
\end{figure}

The precise stellar mass function we use has a significant impact on the results and implications of our model.  In other words, abundance matching is systematically dependent on the stellar mass function used.
For comparison, we consider several different stellar mass functions from the literature, with the intent of examining how abundance matching behaves with different input.
The set of stellar mass functions we now consider is shown in Fig.~\ref{fig:smf-all}.

We give significant attention to the previous study of groups from \citet{YMB2009}, of which further related details are available in \citet{YMB2005, YMB2007, YMB2008}.  While they use the mass-to-light ratios and $g-r$ colors based on \citet{Bell2003}, the SMF from DR7 in our volume-limited catalog uses \textsc{Kcorrect} stellar masses from the template method of \citet{BlRo2007}.  This difference in approach introduces in effect an offset and scatter between the two definitions of stellar mass, preventing a straightforward galaxy-by-galaxy comparison.
Additionally, the \citet{Bell2003} stellar masses effectively assume a \citet{Kro2001} initial mass function (IMF), while we assume \citet{Cha2003}.  The change in IMF produces an offset in stellar mass (see Figs.~\ref{fig:smf-all}, \ref{fig:yang-comp-jt}).

We note that the \citet{YMB2009} results treat fiber-collided galaxies differently.  In general, the \citet{YMB2009} group catalog results we consider in the next section exclude fiber-collided galaxies for which redshifts from other surveys are not available.  We do not expect this to produce significant changes, as only $~5\%$ of galaxies are fiber collided in our mocks, and many of those are collided near their true redshifts.
%There is an additional observational systematic which we have not previously considered in detail.  Because some fraction of the galaxies are fiber collided (as discussed in \S \ref{sec:dr7} and \S \ref{subsec:simfc}), their true redshift is unknown.  The correction for fiber collisions assumes that the fiber-collided galaxy is at the same redshift as the galaxy with which it is collided.  This can put a galaxy at the wrong distance, resulting in an incorrect inference of its luminosity and stellar mass.  Appropriately taking this effect into account for our stellar mass mocks would require knowledge of the colors in addition to the stellar mass.  This generally has only a small effect, since only $~5\%$ of galaxies are fiber collided in our mocks, and many of those are collided near their true redshifts.  However, in general, the \citet{YMB2009} group catalog results we consider in the next section exclude fiber-collided galaxies for which redshifts from other surveys are not available. 

In addition to the group catalog and associated stellar mass function of \citet{YMB2009}, we consider two additional recent measurements of the stellar mass function.  The first is that of \citet{Bal2012}, which applies a color-based method of estimating stellar mass which is similar in form to that of \citet{Bell2003}.  The data they use are drawn from the Galaxy and Mass Assembly (GAMA) survey at $z<0.06$.  The second is \citet{Mou2013}, 
which combines SDSS data with additional UV and IR photometry.  
%which uses PRism MUlti-object Survey (PRIMUS) low-resolution spectra, combined with 
From this data, they obtain accurate stellar masses using spectral energy distribution (SED) modeling.  
Their stellar population synthesis assumes a \citet{Cha2003} IMF.

\subsection{Intrinsic Conditional Stellar Mass Function}

\begin{figure*}
\centering
\includegraphics[angle=90, width=0.9\textwidth]{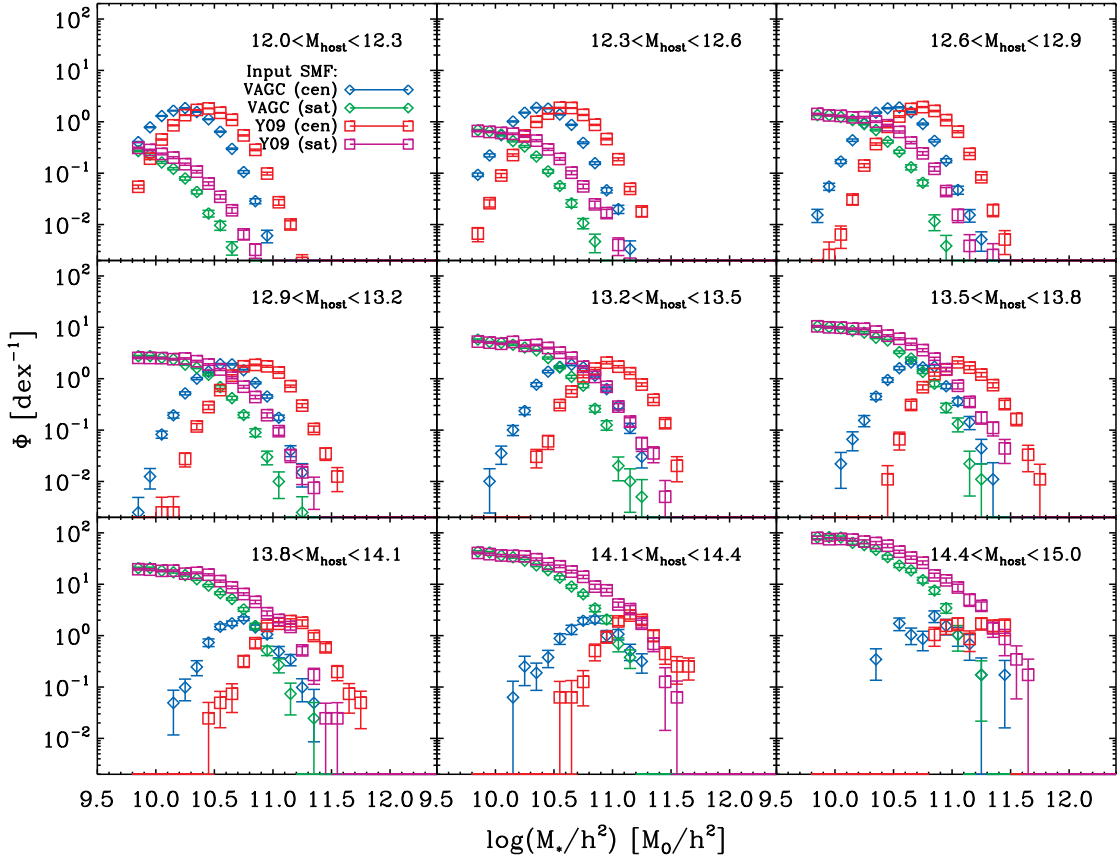}
\caption{Comparison of the results of our best-fit abundance matching model using the SMF drawn from our volume-limited samples (centrals in blue, satellites in green) and using the SMF reported in \citet{YMB2009} (centrals in red, satellites in magenta).  Ranges in host halo mass are given in $\log(\Msun/h)$.  The primary difference between the two cases is the stellar mass definition: while we use the stellar masses from \textsc{Kcorrect} as described in \citet{BlRo2007},  \citet{YMB2009} use stellar masses from \citet{Bell2003}, resulting in an offset.}
\label{fig:yang-comp-jt}
\end{figure*}

Two different intrinsic CSMFs can be seen directly compared in Fig.~\ref{fig:yang-comp-jt}, where the difference is the SMF input.  Here, abundance matching was performed using both our VAGC derived SMF and that of \citet{YMB2009}.  We use the best-fit parameters found in \S \ref{sec:param} in both cases.  It is clear that the \citet{YMB2009} CSMF generally has higher stellar mass, as expected from the change in input SMF seen in Fig.~\ref{fig:smf-all}.  To more precisely quantify this difference, we fit to the intrinsic CSMF found in each of the mock catalogs produced for all four input SMFs.  The fit is done as a function of host halo mass, using the parameters from equations \ref{eq:csmf-cen} and \ref{eq:csmf-sat} as described in \S \ref{subsec:csmf-mhost}.

\begin{figure*}\centering
\includegraphics[angle=90, width=0.85\textwidth]{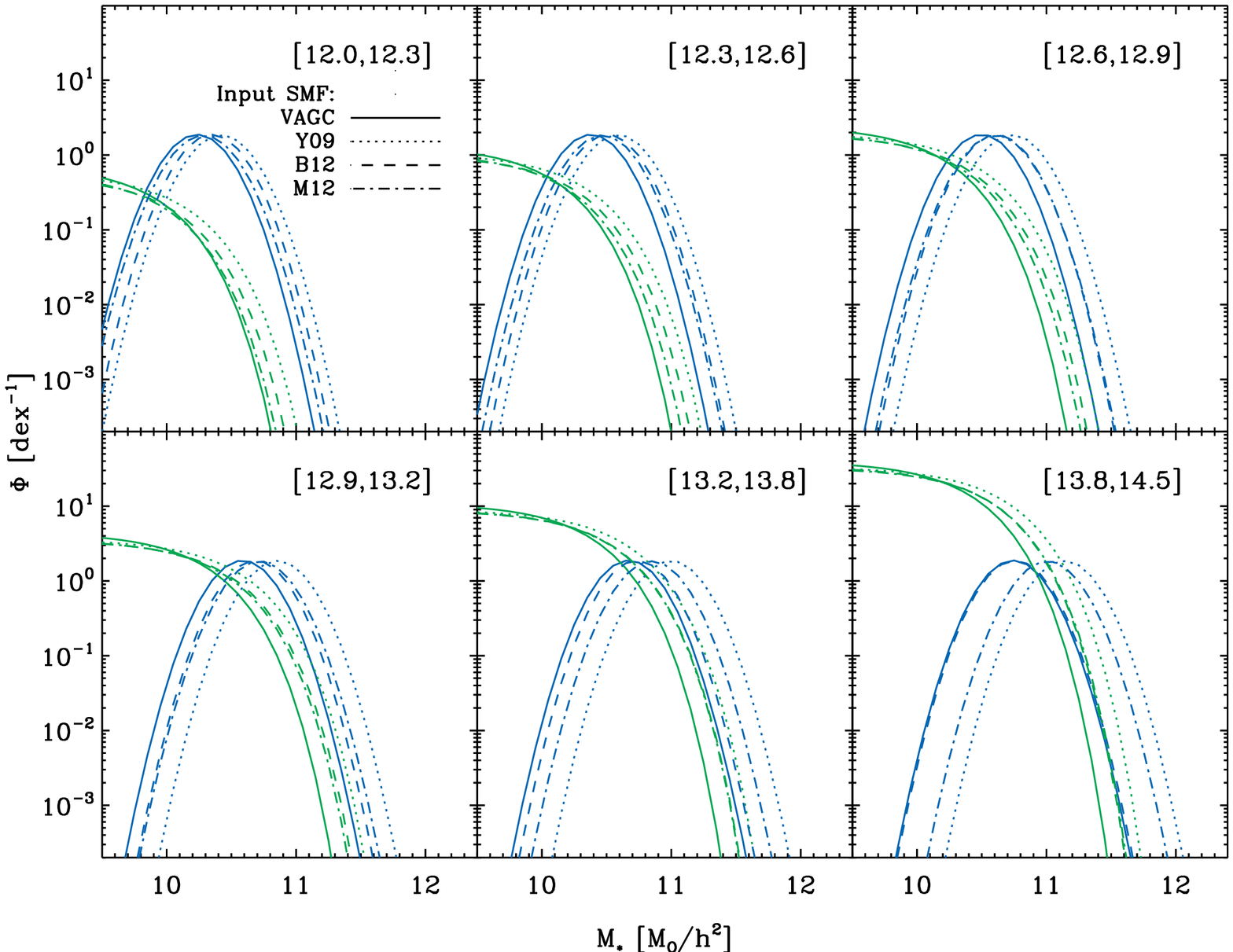}
\caption{Comparison of fits to the intrinsic CSMF for our model using four different stellar mass functions, using the prescription discussed in \S \ref{subsec:csmf-mhost}.  Blue lines indicate the central part of the CSMF, and green, the satellites.  Solid lines show our main results, using the VAGC CSMF, the same as shown in Fig.~\ref{fig:fullfit-dr7}.  Dotted lines show the \citet{YMB2009} SMF.  Dashed lines indicate the fit to our model using \citet{Bal2012}.  Dot-dashed lines show \citet{Mou2013}.  Ranges in host halo mass are given in $\log(\Msun/h)$.   Note how the cutoff of the satellite stellar mass and the mean central stellar mass vary with the massive end of the SMFs shown in Fig.~\ref{fig:smf-all}.}
\label{fig:fullfit-all}
\end{figure*}

Using this overall parameterization allows a comparison between the two different stellar mass function cases, as shown in Tables~\ref{tab:yang-comp} and \ref{tab:yang-comp-sat}, by comparing just these eleven parameters for the two cases.  Fits were done using the midpoint host mass value in each bin.  The VAGC fit is demonstrated if Fig.~\ref{fig:fullfit-dr7}, and the fits to all four intrinsic CSMFs are shown in Fig.~\ref{fig:fullfit-all}.  The parameters in Tables \ref{tab:yang-comp} and \ref{tab:yang-comp-sat} demonstrate primarily the shift in stellar mass that is also visible in the figure.  Note the increase in the central mass scale $M_0$ from our VAGC SMF to the \citet{YMB2009} result.  The host halo mass scale, where the central stellar mass turns over from increasing significantly with host halo mass to a more shallow increase, is also higher in the \citet{YMB2009} case.  This is most likely indicative of the change in the SMF relative to the host halo mass function, particularly since only the high host mass slope changes significantly.  The scatter in the centrals remains about the same, as expected from the fixed input model.  The other two stellar mass functions generally produce intermediate mean central stellar masses, in agreement with the different SMFs presented in Fig.~\ref{fig:smf-all}.

The VAGC version does have lower $M_{*,s}$ in general, as suggested by the slightly lower intercept value.  The slightly steeper change in $M_{*,s}$ with host halo mass, as indicated by the b parameter, also pushes the characteristic stellar mass higher in the \citet{YMB2009} case.  Changes in $\phi_*$ are somewhat more difficult to interpret, though the individual values remain similar in normalization.  This is likely due to the presence of the same subhalos determining how many satellites are in each group.  Most of the variation in the satellite parameters among the different SMFs stems from changes in the $M_{*,s}$ value and how it changes with $\mhost$.  On the other hand, $\phi_*$ has similar variation with group host halo mass, regardless of the SMF used.  

\begin{table*}
	\center
	\caption{CSMF Mass Dependent Fit Parameters -- Centrals}
	\begin{tabular}{l c c c c c}
	\hline \hline
	\label{tab:yang-comp}
	SMF	&	$\log(M_0)$	&	$\log(M_1)$	&	$g_1$	&	$g_2$	&	$\sigma_c$\\
		&	$[\log(\Msun/h^2)]$	&	 [$\log(\Msun/h)$]	&	&	&	  $[\log(\Msun/h^2)]$\\
	\hline	
	VAGC	&	$10.64\pm0.03$	&	$12.59\pm0.10$	&	$0.726\pm0.055$	&	$0.065\pm0.021$	&	$0.212\pm0.001$ \\	
	Y09		&	$10.96\pm0.05$	&	$12.94\pm0.12$	&	$0.644\pm0.028$	&	$0.155\pm0.031$	&	$0.215\pm0.001$ \\
	B12 		& 	$10.77\pm0.01$	&	$12.40\pm0.05$	&	$0.947\pm0.061$	&	$-0.003\pm0.003$	&	$0.213\pm0.001$ \\ 
	M12		&	$10.56\pm0.07$	&	$12.21\pm0.20$	&	$1.19\pm0.26$		&	$0.224\pm0.017$	&	$0.218\pm0.002$\\
	\hline
	\end{tabular}
	\vspace{10 pt}
	\caption{CSMF Mass Dependent Fit Parameters -- Satellites}
	\begin{tabular}{l c c c c c c}
	\hline \hline
	\label{tab:yang-comp-sat}
	SMF & $\log(M_{*,0})$	&	$\log(M_{*,1})$	&	b	&	$\log(M_\phi)$  	&	a \\
		&	  [$\log(\Msun/h^2)$]	&	[$\log(\Msun/h)$]	&	&	[$\log(\Msun/h)$] & \\
	\hline
	VAGC 	&	$10.401\pm0.008$	&	$12.71\pm0.08$	&	$0.753\pm0.063$	&	$12.30\pm0.01$	&	$0.866\pm0.010$	\\
	Y09		&	$10.664\pm0.008$	&	$12.60\pm0.07$	&	$0.948\pm0.083$	&	$12.42\pm0.01$	&	$0.881\pm0.006$	\\
	B12		&	$10.538\pm0.006$	&	$12.35\pm0.09$	&	$1.26\pm0.16$		&	$12.43\pm0.01$	&	$0.951\pm0.007$ 	\\
	M12		&	$10.553\pm0.009$	&	$12.65\pm0.08$	&	$0.986\pm0.092$	&	$12.41\pm0.01$	&	$0.875\pm0.007$	\\
	\hline
	\end{tabular}
\end{table*}

\subsection{Observed Conditional Stellar Mass Function}\label{subsec:obscsmf}

\begin{figure*}
\centering
\includegraphics[angle=90, width=0.9\textwidth]{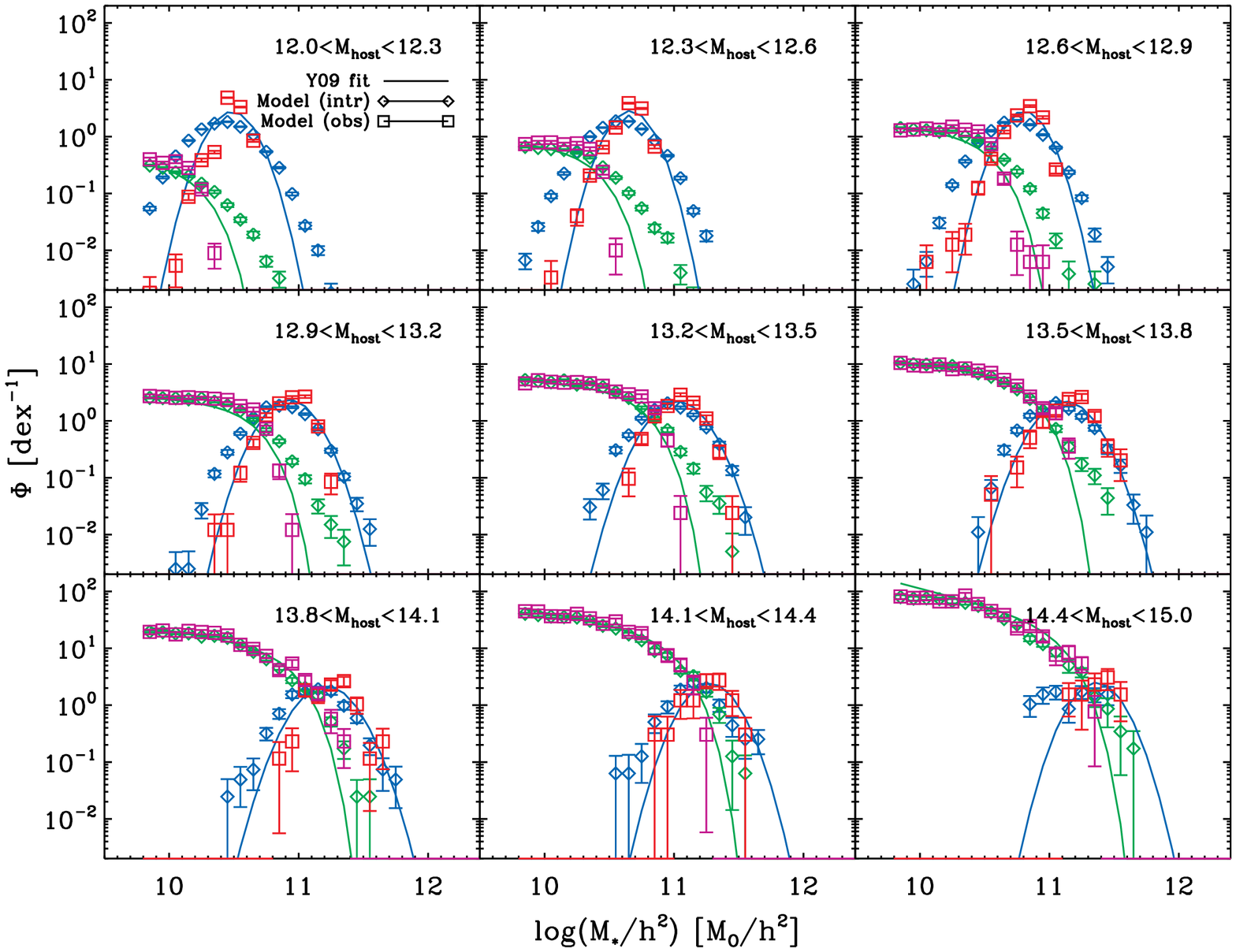}
\caption{ Results of our best-fit model using the SMF of \citet{YMB2009} before (diamonds, centrals in blue, satellites in green) and after (squares, centrals in red, satellites in magenta) the application of observational effects (group finding and fiber collisions), compared to the measurements of \citet{YMB2009} (solid lines, blue for centrals and green for satellites).  Ranges in host halo mass are given in $\log(\Msun/h)$.   The main difference in these two cases lies in the details of the group finding procedure.}
\label{fig:yang-comp-sham-obs}
\end{figure*}

Direct comparisons made of the fitted CSMF results drawn from \citet{YMB2009} to our model CSMF using their stellar mass function are shown in Fig.~\ref{fig:yang-comp-sham-obs}.  Both versions, with and without observational systematics, were done using our best-fit model ($\vpeak$, scatter=0.20 dex, $\mucut$=0.03) applied with the stellar mass function of \citet{YMB2009}.

It is important to note the systematic differences imposed by the slightly different group finding done in these two cases.  The \citet{YMB2009} results use both $r$-band luminosity and stellar mass information.  They define their groups by requiring that at least one galaxy in each group to have $^{0.1}M_r<-19.5$.  They then use either the group total luminosity or stellar mass of all galaxies that pass that luminosity limit to assign host halo masses.  They find limited differences between these using total luminosity or stellar mass.  They also use the same assumption we do that the galaxy with the most stellar mass is the central galaxy.

However, the fact that their limit is a cut in luminosity rather than stellar mass significantly alters the shape of the CSMF at low host halos masses (poor groups).  This effect is most clearly seen in the $12<\log(\mhost)<12.3$ bin of Fig.~\ref{fig:yang-comp-sham-obs}, which compares their results with our model, including the effects of group finding.
%In our model, there are effectively two types of groups in this bin.  Those consisting of only a single galaxy (which then provides all the stellar mass) form the high part of the peak, and are most common.  The rest are groups with two galaxies just above the stellar mass threshold.  In this case, the more massive of the pair makes up the lower part of the central peak while the other provides all the satellites seen in this host mass range.  Therefore, the stellar mass of centrals, as well as the location of the few satellites, is directly determined by the range in total group stellar mass associated with the inferred host halo mass bin.  On the other hand, 
In the \cite{YMB2009} result, their overall cut on galaxies to include is in luminosity, rather than stellar mass.  This means that stellar mass of the central galaxy is not directly determining the host halo mass at low host masses, smoothing out the distribution.  Aside from this difference in the low host mass bins, there is generally good agreement between our "observed" model results and these measurements.

A comparison of the intrinsic model results with these measurements is also shown in Fig.~\ref{fig:yang-comp-sham-obs}.  This demonstrates directly some of the effects of the group finding.  Most obvious is the fact that the group finding reduces the width of the central distribution, as well as introducing the extra feature in low-mass host halos described above.  There is also some offset in the centrals between these two cases, most likely due to the fact that the group finding assumes that the most massive galaxy in a group must be the central, pushing the observed centrals to being more massive in general.  Additional, the cutoff in the satellite distribution is much sharper after group finding.  This is also due to the assignment of the most massive galaxy in the group as the central, since more massive satellites are more likely to be reassigned as the central.  This imposes an extra cut on the satellite distribution.  Therefore, it is likely that the sharp cutoff imposed on the satellite galaxies in the CSMF fits of \citet{YMB2009} is not purely physical, but convolved with the group finding.

\subsection{Comparisons to Previous Work}

There has been significant work in the literature regarding the question of the galaxy-halo connection.  We consider a few recent examples in relation to our study.

The work of \cite{WW2010}, using an abundance matching model based on $\macc$, considered in detail the effect of satellite disruption in a form similar to our $\mucut$ on the clustering and satellite fraction of galaxies.  They examine the disruption of satellites when the fraction $f_{inf}=\macc/\mnow$ of the subhalo falls below some threshold, up to $f_{inf}=0.1$.  They find that values of $f_{inf}=0.1-0.3$ at $z=0.1$ best reproduces observables, which is reassuringly similar to our preferred values for $\mucut$.  Another study was done in \cite{WBZ2012} using a similar abundance matching method.  They specifically addressed the stellar mass loss of satellite galaxies and the transfer of stellar mass into the intra-halo light.  They considered two separate models for stellar mass loss after a subhalo was accreted.  The main property of the model was gradual stellar mass loss at a rate related to the loss of dark matter after the subhalo was accreted.  This is related to our consideration of the $\mucut$ parameter, though our simpler implementation assumes that the galaxy in the subhalo is rapidly destroyed after the subhalo mass falls below a threshold.  They succeed in reproducing the clustering measured in \citet{Zeh2011}, including the low-luminosity thresholds.  This difference may be accounted for by several differences in implementation.  They use a slightly lower scatter (0.15 dex) which increases the overall clustering.  They also use an analytic model for substructure \citep{ZBB2005} rather than an N-body simulation, which permits them to track subhalos at far lower circular velocities.  Nonetheless, their successful implementation is supportive of the general principle of abundance matching.  Because their work shows that the satellite galaxies with the least stellar mass should also be those that are most stripped of stellar mass relative to their dark matter stripping, we suspect that the low clustering in our low stellar mass bin may be due to the loss of a few subhalos in the simulated clusters.

Another related study was done by \cite{Mos2010}.  They assign stellar masses using the peak subhalo mass and the present halo mass.  Their work also relies on the inclusion of orphan galaxies, which may be more necessary in their work as they use a dark matter simulation with lower force resolution than Bolshoi.  Rather than performing strict abundance matching using an input SMF, they assume an analytic form for the relationship between galaxy stellar mass and halo (or subhalo) mass.  They then require that the model SMF is an adequate fit to that measured in SDSS (in this case, SDSS DR3).  Because they use a different stellar mass function and cosmological model, the results are slightly tricky to compare, but we note that overall their central galaxies are brighter with respect to satellites than both our model and the model of \cite{YMB2012}.  They successfully reproduce the two-point clustering, but do not compare with the observed conditional stellar mass function, which we show provides a tighter constraint.   They also note that when they use abundance matching instead of their stellar mass-halo mass relation, that the low halo mass end ($\mhost<10^{12}~\Msun$) of the relationship is significantly different from the power law that they assume, and add another parameter to fit this result.  The general \cite{Mos2010} form may be too restrictive at low stellar masses \citep[see discussion in][]{BWC2012b}, but this halo mass is generally below what we consider.

The simple assumption in our models that scatter is constant may be modified by allowing the scatter to vary with galaxy stellar mass, halo mass, or some other halo property such as $\vmax$.  While the analytical model of \cite{YMB2012} incorporates these effects, it is likely that not all are necessary modifications.  Another related approach was used by \cite{Nei2011a}, who use a shuffle test to determine that abundance matching may require a dependence on the host halo mass, in addition to $\macc$, which is explored further in \cite{Nei2011b}.  However, they consider only the stellar mass function and the correlation function of galaxies in their sample, and they use only the infall mass (and host halo mass) for their abundance matching.  Our analysis considers only a model with no dependence on the host halo mass.  However, a more direct comparison to the results of \cite{Nei2011b} is not immediately possible due to the difference in matching statistics ($\macc$ as opposed to our preferred $\vpeak$).  Regardless, degeneracies between their different models would be broken by including a comparison to the CSMF or similar group statistics.

An alternative abundance matching approach involves dividing subhalos and isolated host halos prior to abundance matching, and applying different matching functions to each.  \citet{RDA2012} investigate this, decomposing the overall stellar mass function into central and satellite components, and matching these separately to the halos and subhalos, respectively.  They find that when matching against the mass of subhalos at accretion or at the present time, the satellites must have more stellar mass than would be inferred from applying the stellar mass-halo mass relation derived for the central galaxies.  This is in general agreement with our findings as well, since the $\mnow$ and $\macc$ direct abundance matching models have a deficit of satellites.  Further, the preferred matching to $\vpeak$ naturally gives the subhalos of satellites higher $\vpeak$ than the halos of central galaxies, and thus, more stellar mass at fixed $\mpeak$, as shown in Fig.~\ref{fig:vpmp}.

In contrast with our comparisons to observations, \cite{Sim2012} make a comparison between abundance matching in a purely dark matter simulation and in a dark matter simulation with the addition of gas hydrodynamics and prescriptions for star formation and feedback.  The two simulations use the same initial conditions.  They generally find good agreement between these cases, but there are indications of incompleteness or premature galaxy disruption at low stellar masses.  However, the resolution of their dark matter simulations is not as good as that of the Bolshoi simulation that we use.  Based on the results of a resolution test presented in App.~\ref{app:res}, we find that these discrepancies are all below the mass at which the simulation used there is able to track the full population.  We thus expect that these discrepancies are primarily due to limited resolution, and not to failures of the abundance matching approach.  Higher resolution hydrodynamical simulations will be required to verify this.

One set of measurements complementary to our own are presented in \cite{More2009}.  Rather than using the total group stellar mass or luminosity to determine the mass of a halo, they instead use satellite kinematics to determine the mass of a halo around a central galaxy.  They obtain a relationship between central galaxy luminosity and host halo mass, with a scatter of of $0.16\pm0.04$ dex at fixed host halo mass.  This is somewhat low relative to our constraints for the luminosity model ($\sigma=0.22^{+0.01}_{-0.02}$, see Appendix~\ref{app:lum} for details), but our result is still within two standard deviations of theirs.

%\section{Discussion and Conclusions}\label{sec:conclude}
\section{Summary}\label{sec:conclude}
We have used an analysis of the Bolshoi cosmological simulation to examine the correlation functions and CSMFs of several different models for the connection between galaxies and halos which are variants of the subhalo abundance matching approach.  We have compared these models against data drawn from SDSS, using new measurements of the two-point correlation function as a function of stellar mass and the conditional stellar mass function in groups.  All CSMF comparisons between models and data are done in ``observed space'', after applying group finding and fiber collisions to our models.  Our study is the first to combine this set of measurements in a fully self-consistent way to test a model which assigns all galaxies to resolved subhalos in a simulation.  From these results, we have reached the following conclusions:

\begin{enumerate}
\item  An examination of the correlation function shows that most of the halo mass properties used as proxies for stellar mass that we considered cannot reproduce the data
regardless of the parameters used.  This includes abundance matching models where
the halo property used is $\mnow$, $\macc$, $\mpeak$, $\mnpeak$, $\vmax$ and $\vacc$.  Each of these models is insufficiently clustered even in cases with no scatter and $\mucut$=0.  
Because non-zero scatter and $\mucut$ only reduce galaxy clustering, we exclude those models.  The only exceptions are $\vpeak$ and $\vnpeak$.  This exclusion applies only to this particular family models, and cannot be applied to models with significantly different methodology, such as those which incorporate orphan galaxies.

\item  Our best-fit model uses $\vpeak$, with $\mucut$=0.03 and scatter of 0.20 dex.
This scatter is effectively the combination of intrinsic scatter in stellar mass and scatter from the measurements because we do not distinguish between them.
This model provides a good fit to the combined constraints of the clustering for galaxies with  $\log(M_*) > 10.2$,
the mean and dispersion of the central galaxies in bins of host mass (in the CSMF),
and the satellite distribution in the CSMF, both for galaxies brighter than  $\log(M_*) > 9.8$.

\item The $\vnpeak$ model provides significantly poorer fits to the data overall that $\vpeak$.  It can marginally fit the clustering data alone, but cannot fit the satellite CSMF and is strongly ruled out by the combined data.  The increased stellar mass of satellites relative to central galaxies forces the mean stellar mass of the central CSMF slightly low.  The high $\mucut$ needed to match the clustering also reduces the satellite fraction at low stellar masses too much to reproduce the satellite distribution.

\item  The scatter is most strongly constrained by the width and mean of the distribution of galaxies in groups, both centrals and satellites.  Thus, the central CSMF provides the sharpest limit.  This strongly excludes zero (or very low) scatter, and scatter above 0.25 dex.  We estimate scatter of $\sigma=0.20\pm0.03$ dex in stellar mass at fixed  $\vpeak$.

\item We explicitly test the mass dependence of the scatter value, using the conditional stellar mass function in bins of total stellar mass, and find that it is consistent with being constant for the galaxies living
in halos from $10^{12}$--$10^{14}~\Msun/h$.  Changes by more than 0.1 dex over this range are ruled out.

\item  The value of $\mucut$ is only weakly constrained for the $\vpeak$ model.  A value of zero is weakly 
disfavored by the CSMF; the correlation function disfavors values above 0.08.
Marginalizing over scatter results in a one-sigma upper limit of $\mucut < 0.07$.

\item The projected correlation function using this $\vpeak$ model is low for the $\log(M_*)>9.8$ threshold at small scales.  This may be due to loss of a few low-stellar mass satellites, suggesting that even the Bolshoi simulation may be inadequate at tracking subhalos at these masses, and that properly reproducing the galaxy distribution may require the inclusion of orphan galaxies.  Another possibility is that our model is too simple; loss of substructures is degenerate with a mass-dependence in the $\mucut$ parameter, which could have similar impact on the satellite fraction.  Alternatively, the discrepancy may be due to inadequately modeling the observational effects on galaxies at these stellar masses when calculating the correlation function.

%\item  The $\vpeak$ model can fit the clustering alone only for $\log(M_*) > 10.2$  For the $\log(M_*) > 9.8$ threshold, it is insufficiently clustered.

\item  The fact that only the $\vpeak$ model is capable of reproducing the data indicates that satellites typically have more stellar mass than central galaxies for a given (sub)halo mass such as $M_{\rm{peak}}$.  This is in general agreement with other recent models, such as those of \cite{Guo2011, Nei2011b, RDA2012}.
\end{enumerate}

The subhalo abundance matching model presented here is capable of reproducing all the trends expected from the measurements we consider, particularly the projected correlation function and the CSMF, when specific assumptions are made about the parameter on which to abundance match, the value of the scatter, and the halo stripping required to remove a galaxy from the sample.  This is true even for the simple assumptions used -- fixed scatter in stellar mass, and no dependence on when $\vmax$ is assigned to satellites.  

Using this model, the data are only reproduced within the very small statistical errors for $\log(M_*) \gsim 10.0$.  Below this stellar mass there appears to be slightly fewer satellites in the model.  Possible explanations include observational systematics, required variation in the mass threshold for destroying satellites, or the need for inclusion of subhalos below the resolution limit of the simulation. In the context of the current approach, we cannot distinguish between these.  We intend to revisit this issue in the future
using a combination of data that is complete to lower stellar masses and higher-resolution simulations.

In this work, we have only tested a single cosmology.  The fact that the CSMF and correlation function can be well reproduced suggests that our chosen cosmology is very close to the correct model.  This is further supported by the fact that we well-reproduce other measures not directly used to constrain the model parameters, in particular, the group total stellar mass function, which depends on the halo mass function (and thus on $\sigma_8$) for a given clustering strength.

We also focus primarily on the results using the \textsc{Rockstar} halo finder.  Using the BDM (Bound Density Maxima) halo finder \citep{Kly1999} does not produce significantly different results.  However, there are slightly fewer galaxies in the model applied to the BDM halos than in the \textsc{Rockstar} case, most likely because \textsc{Rockstar} finds more substructure, particularly near the centers of halos.

This same analysis may be applied to samples based on luminosity, rather than stellar mass.  While the framework remains unchanged, the results may be slightly different, as a galaxy remaining at fixed stellar mass after being accreted will dim in luminosity as its stars age.  This will reduce the luminosity of satellites compared to centrals, unlike stellar mass.
At a given number density of objects, this will mean that the satellite fraction at the specified luminosity should be slightly lower 
than the satellite fraction at the equivalent stellar mass.  A demonstration of this difference may be seen in Appendix \ref{app:lum}.  While the scatter estimated by this method is similar ($\sim 0.20$ dex), it produces a significantly higher value of $\mucut=0.13$ (vs. 0.03 for stellar mass), and a resulting lower satellite fraction.
%  This selection of $\mucut$ is driven almost entirely by the central part of the CLF, which is due in part to suppression of the tails of the central CLF.  The underlying physical reason for this change is currently unclear, and the central CLFs are a significantly poorer fit in general than the equivalent CSMF.  

In the local universe, further improvements may be possible by including additional measurements in a self-consistent approach, including the velocity dispersion of galaxies in groups, galaxy-galaxy lensing, the Tully-Fisher relation \citep[as was done by][]{TKP2011} and the properties of bright galaxies \citep[e.g.][]{Hea2012}.  Additional constraints on the bright sample are also possible using larger volume. Future work may determine how well this model performs at higher redshift.  At present, the study is only possible at this level of detail in the local Universe, but larger spectroscopic samples are becoming available at higher redshift.  An extension of our modeling approach to photometric data will be important to take account of the large amount of information from upcoming imaging surveys.

The detailed understanding of the galaxy--halo connection we have presented here has implications for a wide range of areas in galaxy formation and cosmology.  We expect the constraints provided on the intrinsic conditional luminosity function will be very helpful in constraining semi-analytic galaxy formation models and hydrodynamical simulations.  These constraints can also be used to implement CLF or CSMF-based modeling on larger, lower-resolution simulations.  This will be important for accurately modeling the distribution of dimmer galaxies and forecasting how well future imaging surveys, such as DES and LSST, can constrain cosmological parameters.  Uncertainty in the connection between galaxies and halos is an important systematic in several methods to constrain cosmological parameters.  Examples include the precise determination of galaxy bias required for clustering and lensing constraints, understanding the galaxy content of clusters for cluster cosmology \citep{Rozo2010, Tin2012}, and modeling the mass along the line of sight to strong lensing time delays \citep{Suy11}.  The precise constraints we now provide in the nearby Universe are a step towards minimizing these systematics and achieving the precision required for next generation cosmological measurements.

%\section*{Acknowledgements}

%RW go over Andrew Wetzel and Surhud More's comments before resubmission.

\acknowledgements
RMR is supported by a Stanford Graduate Fellowship.  This work was additionally supported by the U.S. Department of Energy under contract number DE-AC02-76SF00515, and a Terman Fellowship to RW at Stanford University.  This research was also supported in part by the National Science Foundation under Grant No. NSF PHY11-25915, through a grant to KITP during the workshop ``First Galaxies and Faint Dwarfs''.  This work used computational resources at SLAC.

We thank Yu Lu, Michael Busha, Frank van den Bosch, Andrew Zentner, Anatoly Klypin, and Cameron McBride for useful discussions, and Andrew Wetzel, Douglas Watson, Matt George, and Surhud More for comments on a draft.  We thank the anonymous referee for several suggestions that improved the presentation of this work.  We are grateful to Anatoly Klypin and Joel Primack for providing access to the Bolshoi simulation, which was run on the NASA Ames machine Pleiades.  This work also uses data from the LasDamas simulations.  These were run using computational resources at Teragrid/XSEDE and at SLAC.  Further information is available at http://lss.phy.vanderbilt.edu/lasdamas/ .  We are grateful to our LasDamas collaborators, and especially Michael Busha and Cameron McBride, who ran the Consuelo and Esmerelda simulations used here.  We would also like to thank Cameron McBride for providing a copy of the program for making mock fiber collisions and John Moustakas for providing his derived SDSS stellar mass function prior to publication.

Funding for the Sloan Digital Sky Survey (SDSS) has been provided by the Alfred P. Sloan Foundation, the Participating Institutions, the National Aeronautics and Space Administration, the National Science Foundation, the U.S. Department of Energy, the Japanese Monbukagakusho, and the Max Planck Society. The SDSS Web site is http://www.sdss.org/.  The SDSS is managed by the Astrophysical Research Consortium (ARC) for the Participating Institutions. The Participating Institutions are The University of Chicago, Fermilab, the Institute for Advanced Study, the Japan Participation Group, The Johns Hopkins University, Los Alamos National Laboratory, the Max-Planck-Institute for Astronomy (MPIA), the Max-Planck-Institute for Astrophysics (MPA), New Mexico State University, University of Pittsburgh, Princeton University, the United States Naval Observatory, and the University of Washington.

\bibliography{shamclf}

\newpage
\begin{appendices}
\section{Effects of the Group Finder}\label{app:gf}

\begin{figure}
\centering
\includegraphics[angle=90, width=0.48\textwidth]{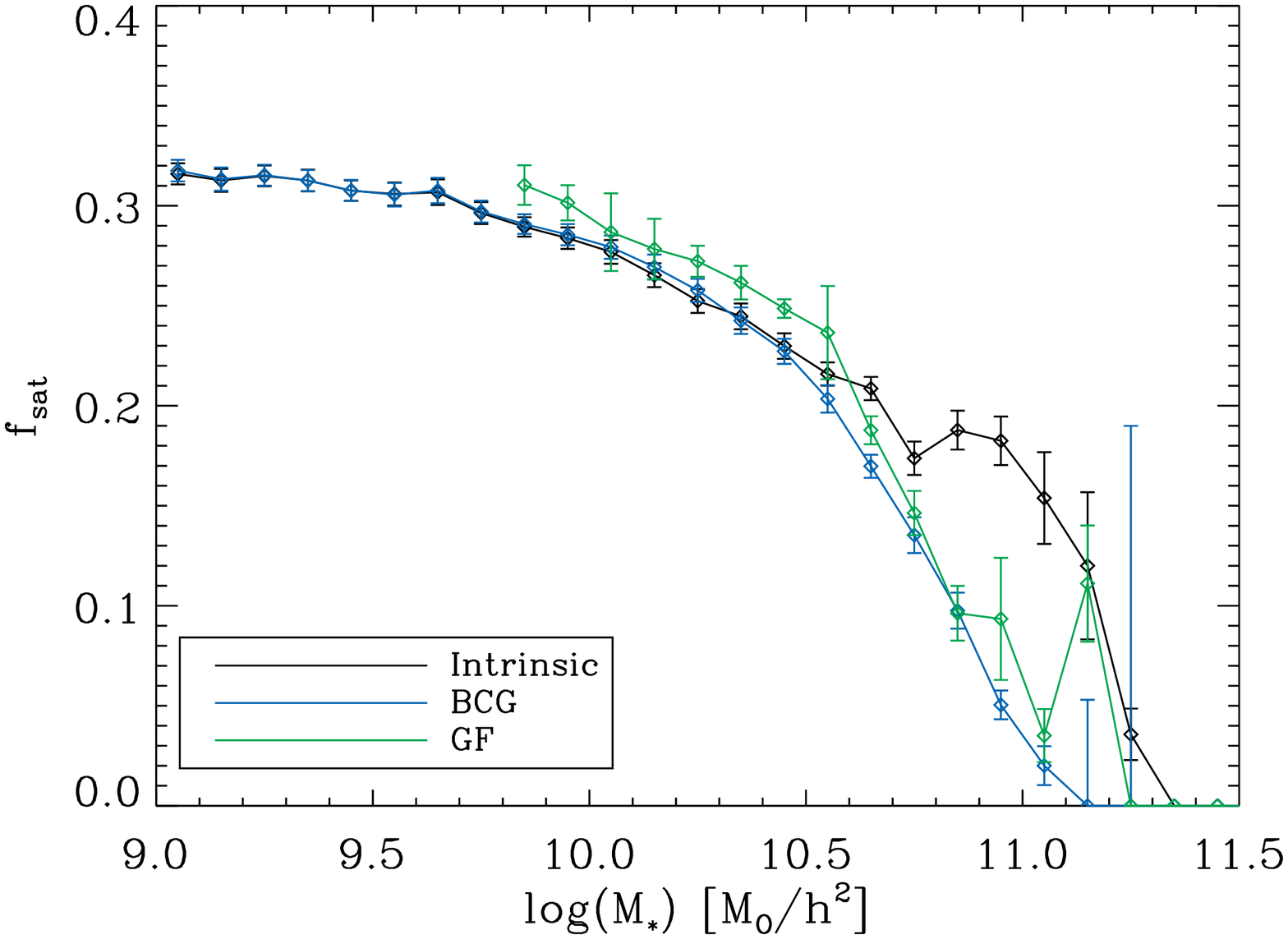}
\includegraphics[angle=90, width=0.48\textwidth]{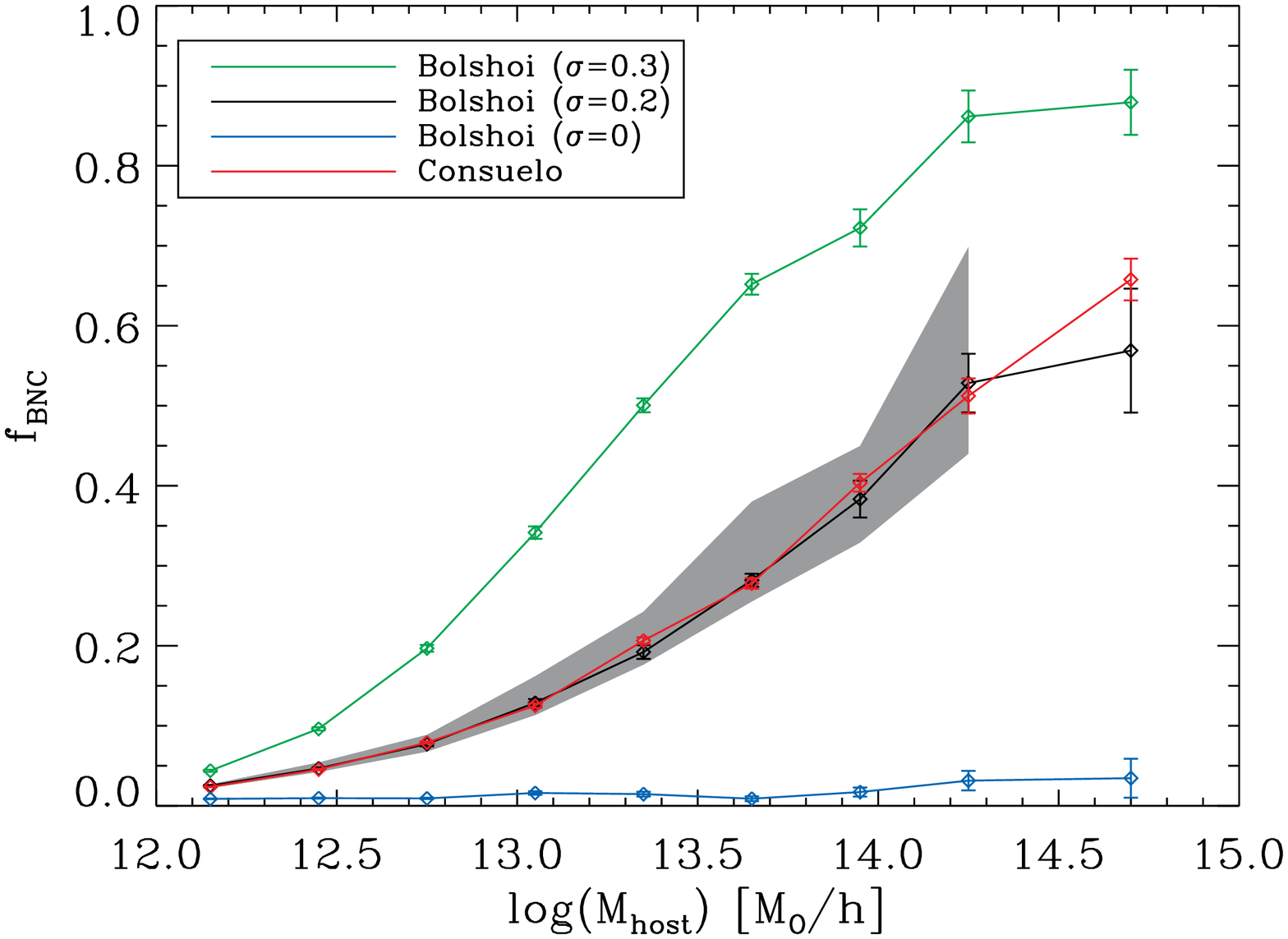}
\caption{{\em Left:} Effect of group finding on the satellite fraction.  The intrinsic satellite fraction in the model (black) is significantly higher than when reassigning the brightest cluster galaxy as the central (blue) in galaxies with high stellar masses.  This is because the nonzero scatter allows a significant number of true satellites to be scattered up in stellar mass, increasing the satellite fraction of massive galaxies.  This effect increases with scatter; in a zero-scatter model, the change is negligible.  This is also the primary difference between the intrinsic satellite fraction and that obtained via the group finder (green).  All lines are for the $\vpeak$, $\mucut$=0, scatter=0.20 dex model.
\em Right:\em~Fraction of central galaxies where at least one satellite in the same halo has higher stellar mass.  The result is shown on the mocks for two different simulation, the Bolshoi simulation (black) and the Consuelo simulation (red) which is lower resolution.  These both use a model with stellar mass, $\vpeak$, $\mucut=0.03$, and scatter of 0.20 dex.  Error bars show statistical jackknife errors.  The gray band gives the resulting range in the $f_{\rm{BNC}}$ fraction given the $1\sigma$ range in scatter for the fitted Bolshoi model.  This probability is also shown for two other values of scatter (0.30 dex and zero) in Bolshoi, which are ruled out by the data.}
\label{fig:fsat-gf}
\end{figure}

The group finder itself has a significant impact on our various measurements.  As discussed in the main text, the two primary systematic effects of the group finder are the artificial reduction of scatter in central galaxy stellar mass for low halo masses, and the assumption that the most massive galaxy in a group must be the central.  A clear demonstration of this may be seen in Fig.~\ref{fig:fsat-gf}.  Here, we show the difference in the model satellite fraction between using the intrinsic central galaxies, and assuming that the most massive galaxy is the central, both using the intrinsic group assignment.  As expected, this significantly reduces the satellite fraction of massive galaxies, since in large clusters it is not unlikely for at least one satellite to be more assigned a higher stellar mass than the central.  (This can be seen in the intrinsic CSMF in Fig.~\ref{fig:fit-clf}.)  This is the primary reason for the difference in satellite fraction between the intrinsic satellite fraction and that obtained from the group finder.  Furthermore, this effect becomes stronger in models with increased scatter, because non-central galaxies are more likely to be scattered up in stellar mass than the intrinsic central, and is almost negligible in models with zero scatter.

The fraction of central galaxies that do not have the most stellar mass (or are not the brightest) increases with host halo mass, as can be seen in the right-hand plot of Fig.~\ref{fig:fsat-gf}.  It also increases with intrinsic scatter, but is not strongly dependent on the resolution of the dark matter simulation.  The values we find for moderate scatter are in general agreement with the study of \citet{Ski2011}.  The recent weak lensing study of \cite{Geo2012} tests multiple different center definitions for groups with a range in $\mhost$ of $10^{13}-10^{14}~\Msun$.  They find that $\sim20-30\%$ of these groups have "ambiguous" centers, where multiple center definitions are in significant disagreement.  This is also in good 
agreement with the fractions we measure in Fig.~\ref{fig:fsat-gf}.

This effect of group finding can also be seen in a comparison between the intrinsic CSMF (Fig.~\ref{fig:fit-clf}) and that obtained after the use of the group finder (Fig.~\ref{fig:vp-bestfit}).  Note that although the distribution of galaxies in massive halos is not strongly changed, the central distribution in the low-mass halos sharpens considerably after group finding, lowering the inferred scatter due to correlations between central properties and group properties.  

\section{Resolution Requirements}\label{app:res}

\begin{figure*}[p]
\centering
\includegraphics[angle=90, width=0.8\textwidth]{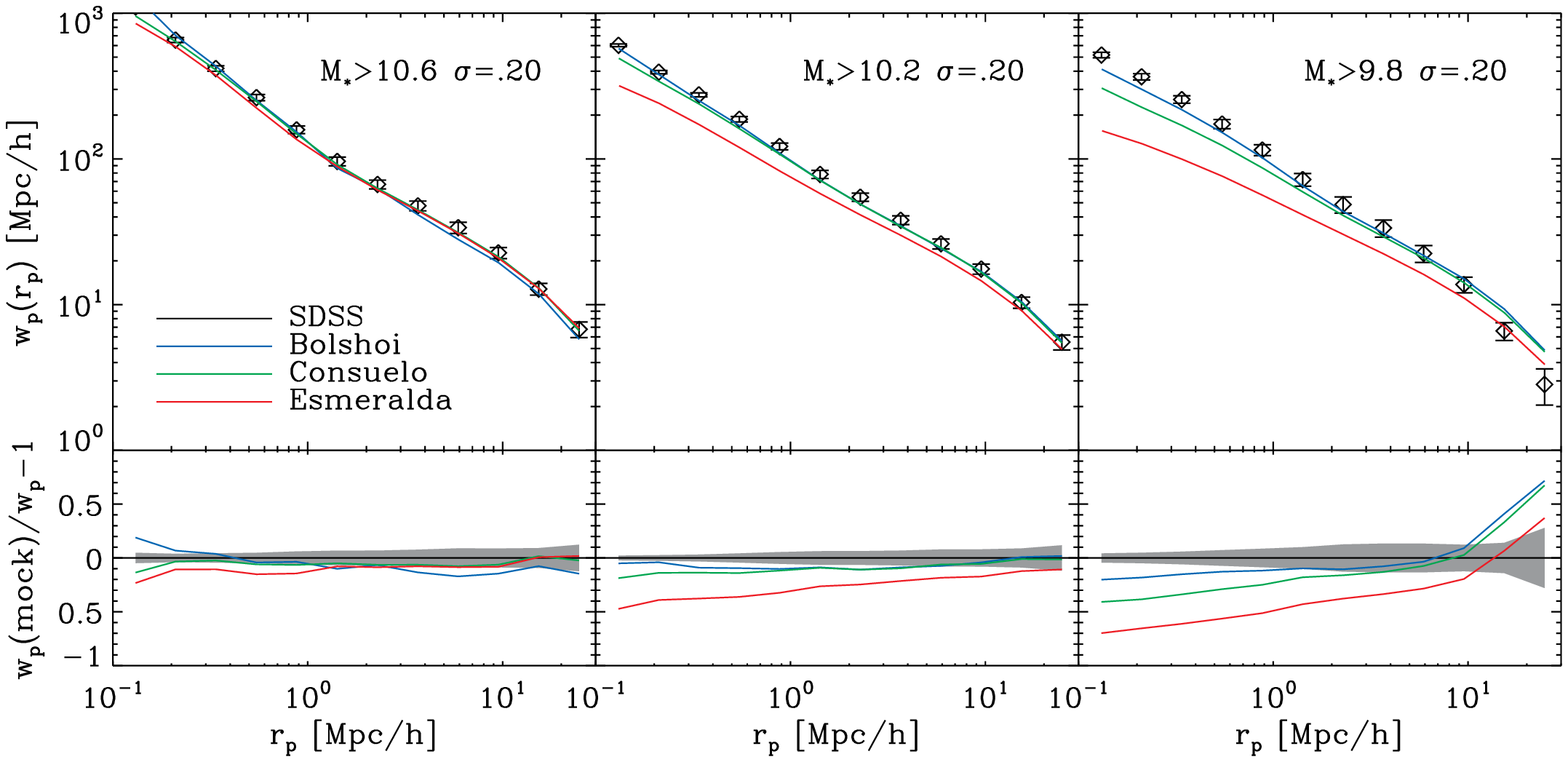}
\includegraphics[angle=90, width=0.8\textwidth]{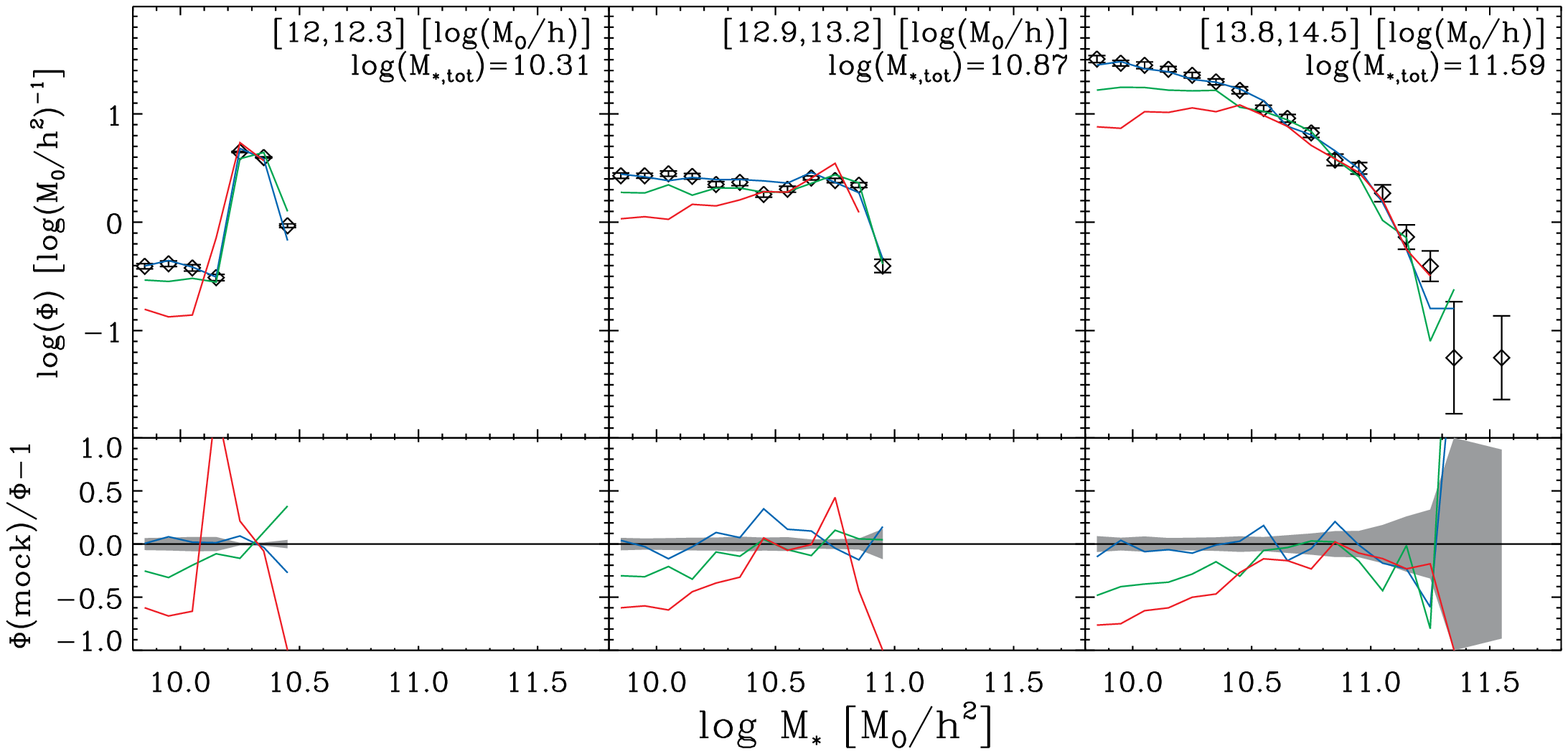}
\hspace{-0.25 cm}\includegraphics[angle=90, width=0.9\textwidth]{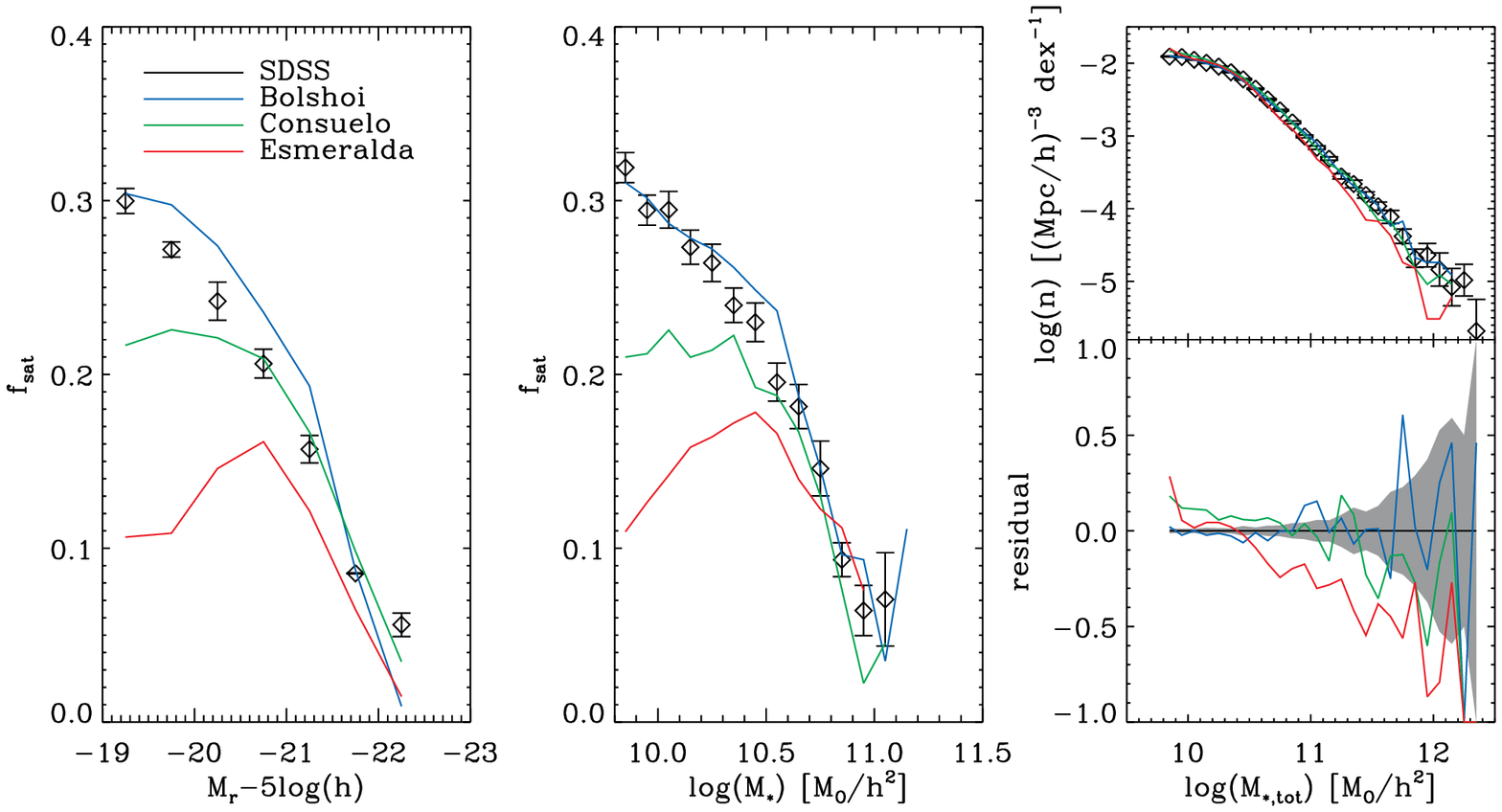}\hspace{-0.25 cm}
\caption{Impact of simulation resolution on statistics of resolved subhalos.  Figure shows the $\vpeak$ model with $\mucut$=0 and $\sigma=0.2$, applied to the Bolshoi (blue), Consuelo (green), and Esmeralda (red) simulations, with the measured values from the SDSS DR7 VAGC (black) shown for comparison.  The inability of lower resolution simulations to resolve all satellite halos results in a deficit of satellites and a drop in the small-scale clustering.  {\em Top:}  Correlation functions.  {\em Center:}  Conditional stellar mass functions.  Total stellar mass is given in $\log(\Msun/h^2)$.  {\em Bottom left:} Satellite fraction for the luminosity model with these parameters.  {\em Bottom center:} Satellite fraction in the stellar mass model.  {\em Bottom right:} Group total stellar mass function.  Based on the results from the satellite fraction,
the Bolshoi, Consuelo, and Esmeralda simulations are roughly complete for satellite galaxies at stellar masses of $\log(M_*/(\Msun/h^2))$ = 10.0, 10.5, and 10.8, respectively, or at luminosities of $M_r-5\log(h)<$-19.5, -20.5, and -21.5.
}
\label{fig:rescomp}
\end{figure*}

The use of a high-resolution simulation such as Bolshoi is essential to this work.  A simulation with more massive particles or a larger softening length would not be able to resolve as many subhalos, particularly those near the center of massive clusters (see \citealt{BWW2013} and \citealt{Oni2012} for related subhalo information, and \citealt{Wu2012} for a more detailed discussion) which tend to be victims of "overmerging" or otherwise become prematurely disrupted.  Fig.~\ref{fig:rescomp} shows the difference between using Bolshoi, and the Consuelo and Esmeralda simulations from the LasDamas suite \citep{McB2012}.  Consuelo \citep[see also][]{BWW2013, LTB2011} uses $1400^3$ particles in a volume of $(420~\hmpc)^3$ (with a particle mass of $1.9 \times 10^{9}$, while Esmeralda has $1250^3$ particles in $(640~\hmpc)^3$ (with a particle mass of $9.3\times 10^{9}$).  Bolshoi, Consuelo and Esmeralda have (physical) force resolution of 1, 8 and 15 ${\rm kpc}/h$, respectively.

The same abundance matching model was applied to all three simulations.  As can be seen in the figure, the model applied to Consuelo (with the same parameters) has a significant deficit of satellites with $M_*>10.5$, while the loss of satellites in Esmeralda is even more severe.  Because smaller subhalos are more easily disrupted, there are fewer of them.  Thus, for a selection at a fixed stellar mass to have the appropriate number density from abundance matching, a mixture of smaller halos (and sometimes subhalos) will be given a greater stellar mass than they would be assigned if the prematurely disrupted subhalos had not been lost.  Most of these halos will be isolated halos, reducing the satellite fraction.  This also reduces the clustering, particularly at the small scales where satellites contribute strongly.

Furthermore, this effect is worsened when using a property other than $\vmax$ or $\mnow$ for abundance matching.  In particular, when using $\vpeak$ as the abundance matching parameter as shown in the figure, there will be numerous relatively smaller subhalos at the present time which had a much higher $\vmax$ in the past, but are now lost to the simulation.  The additional force resolution of the Bolshoi simulation does a better job of capturing these satellites that have experienced significant stripping of their dark matter mass, allowing them to be tracked substantially longer than they can be tracked in the lower resolution Consuelo or Esmeralda simulations.

\section{Using Luminosity}\label{app:lum}

\begin{figure*}[p]
 \centering
\includegraphics[angle=90, width=0.8\textwidth]{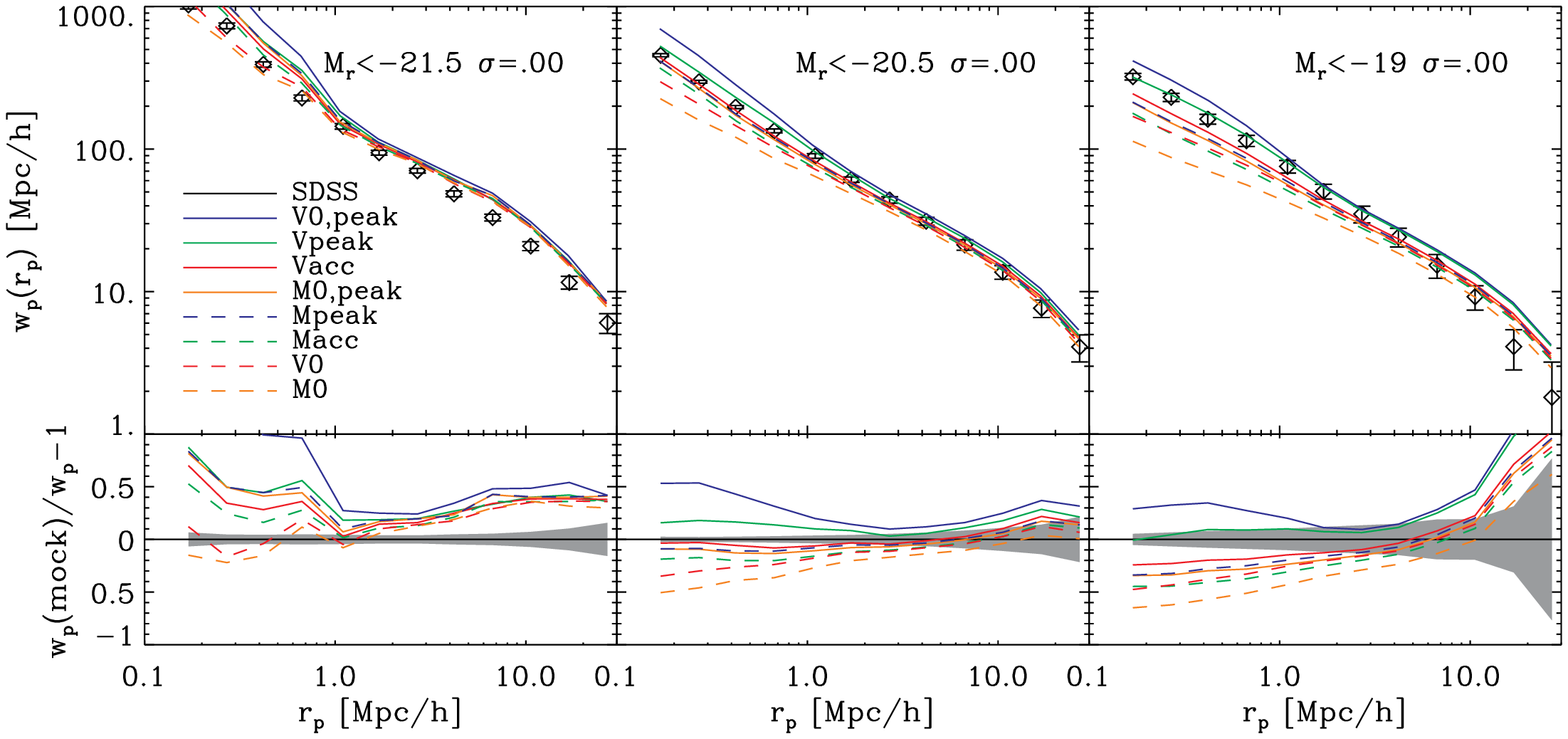}
\includegraphics[angle=90, width=0.8\textwidth]{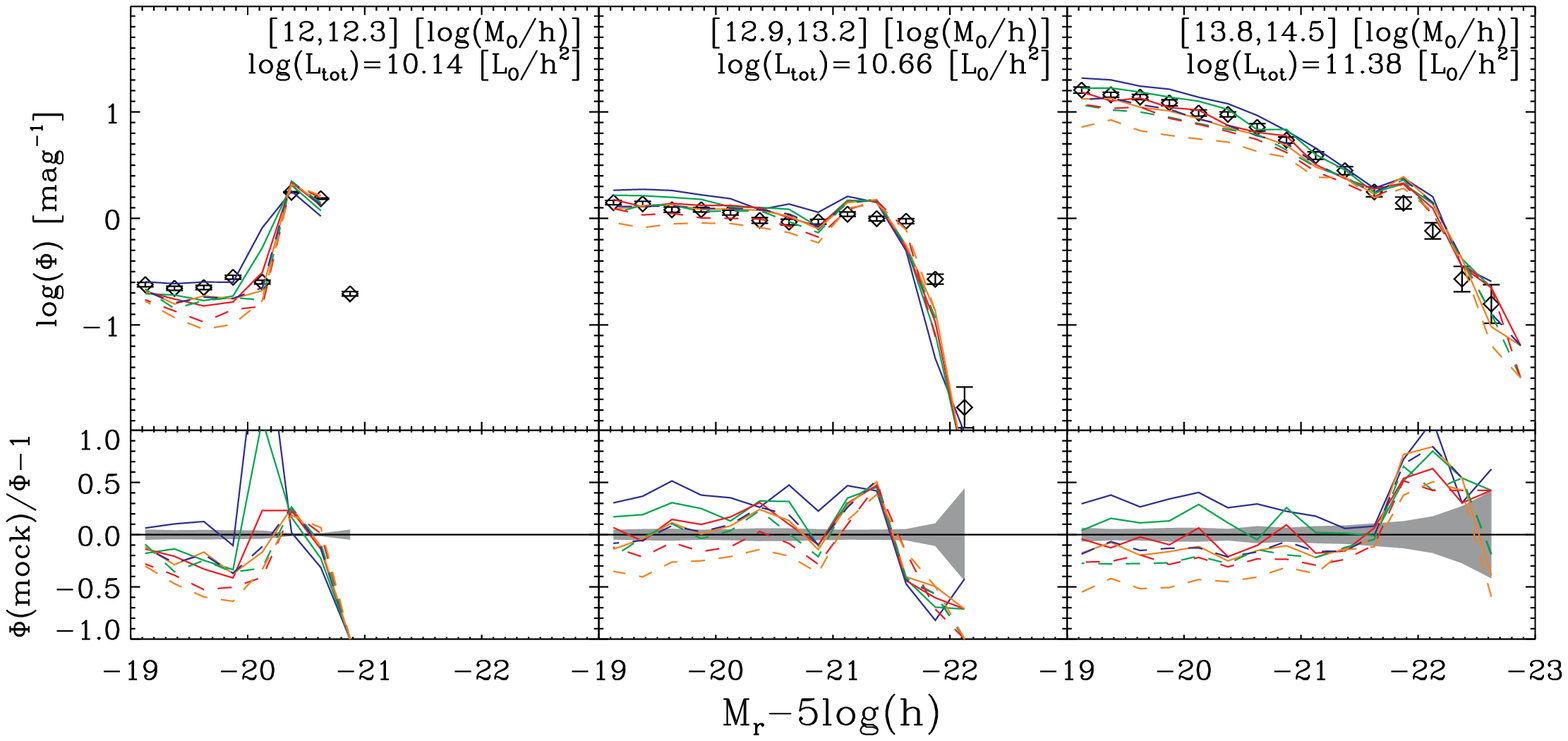}
\includegraphics[angle=90, width=0.9\textwidth]{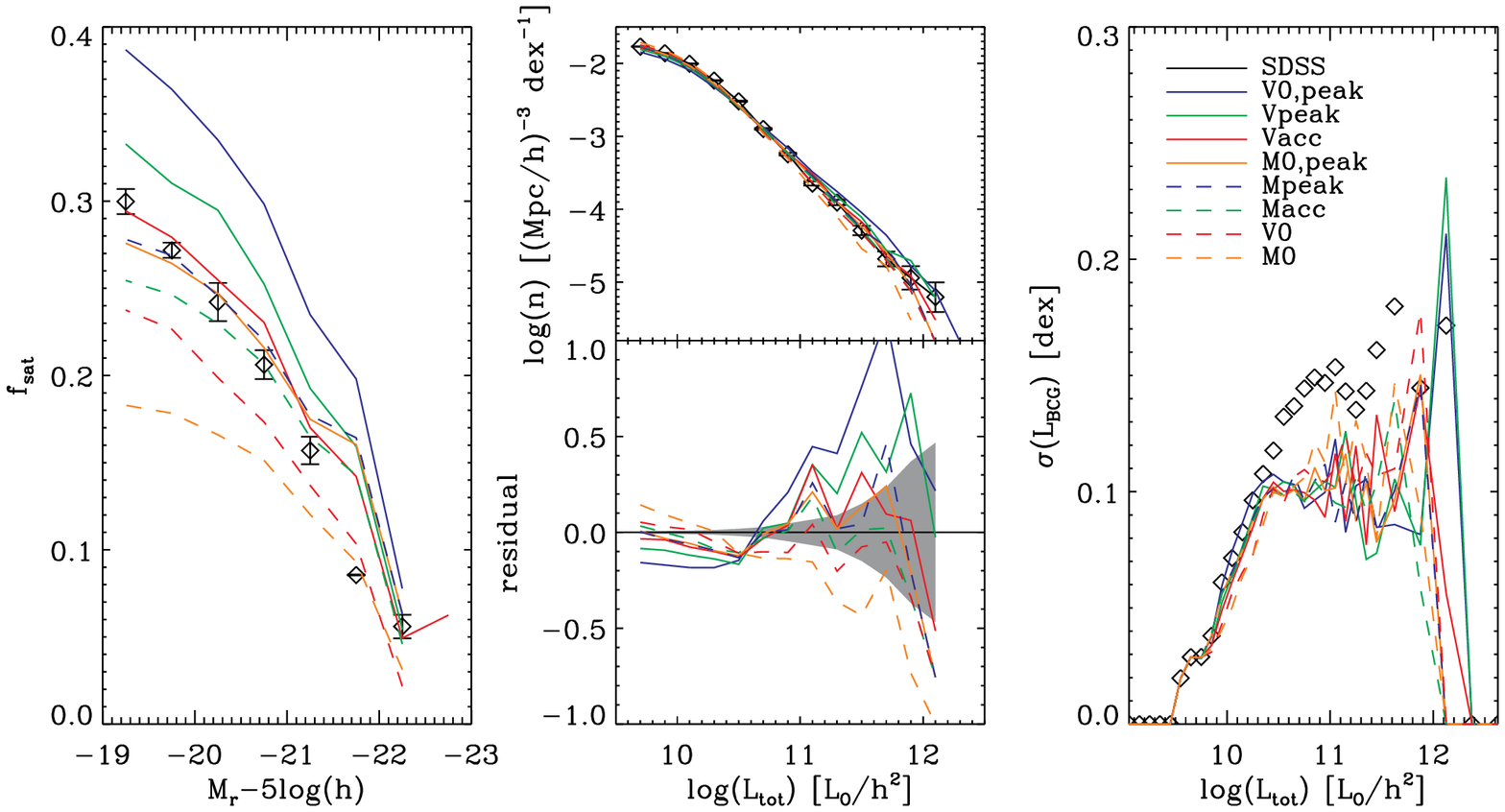}
\caption{Abundance matching results matching galaxy luminosity to different halo properties.  All shown here have zero scatter and $\mucut$=0.  \emph{Top:} Projected two-point correlation function.  Labels denote the luminosity thresholds.  Changes in model here are generally most noticeable in the one-halo term.  Because increases in scatter or $\mucut$ can only decrease the clustering, it follows that any model which falls significantly below the measured clustering (black) must be excluded.  \emph{Center:}  Conditional luminosity function (CLF).  Labels indicate the range in $\log( M_{vir})$ for each plot.  Non-zero scatter broadens this part of the distribution.  \emph{Bottom left:}  Satellite fraction as a function of luminosity.  As should be expected, models with higher satellite fraction correlate with stronger one-halo clustering and more satellites in the CLF.  \emph{Bottom center:}  Group luminosity function.  \emph{Bottom right:}  Standard deviation (scatter) in stellar mass of central as a function of total group stellar mass.  Error bars on the models are suppressed for clarity.}
\label{fig:comp-vtype-lum}
\end{figure*}

\begin{figure*}
\includegraphics[angle=90, width=\textwidth]{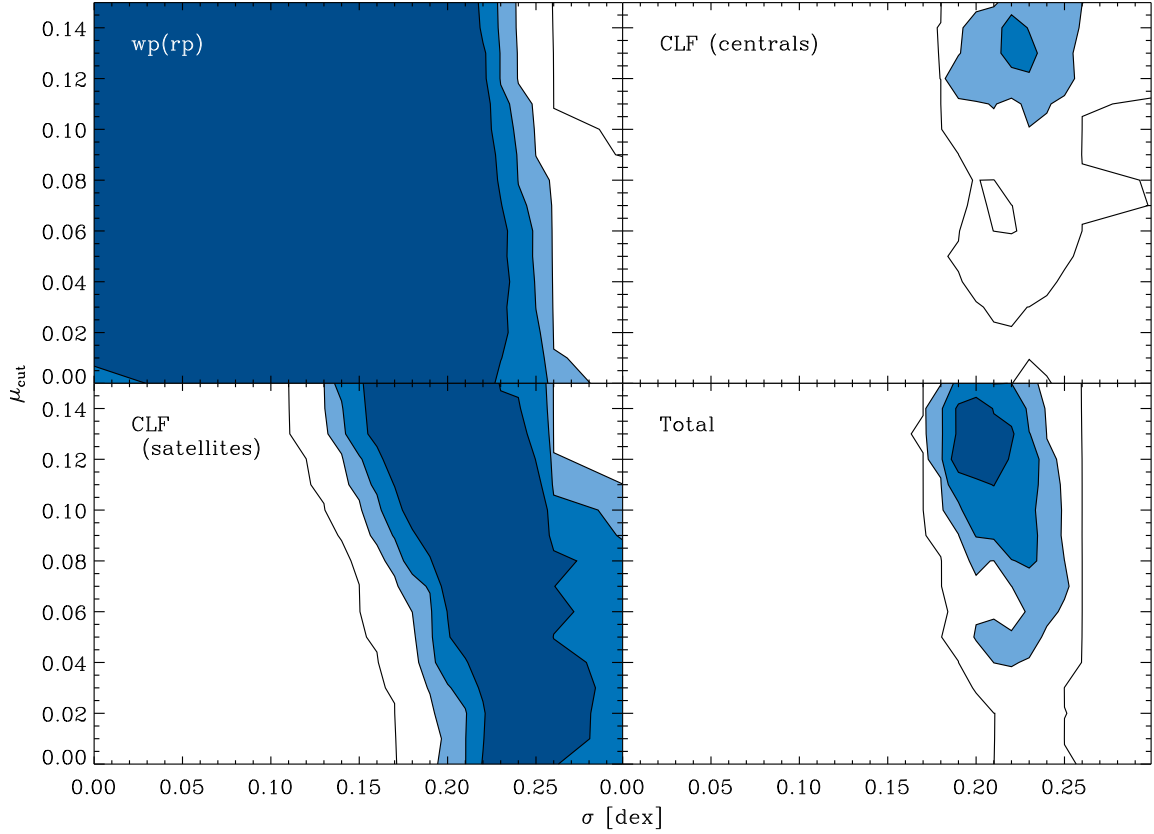}
\vspace{-1.5 cm}
\caption{Constraint on the scatter and $\mucut$ when using $\vpeak$. Levels give $P(>\chi^2)$, corresponding to 1, 2, 3, and 5-$\sigma$ contours.  \emph{Top left:}  Constraint from clustering only.  \emph{Top right:} Constraint from central part of CLF only.   \emph{Lower left:} Constraint from satellite part of CLF only.   \emph{Lower right:}  Constraint from all measures combined.}
\label{fig:constr-lum}
\end{figure*}

\begin{figure}
\includegraphics[angle=90, width=0.5\textwidth]{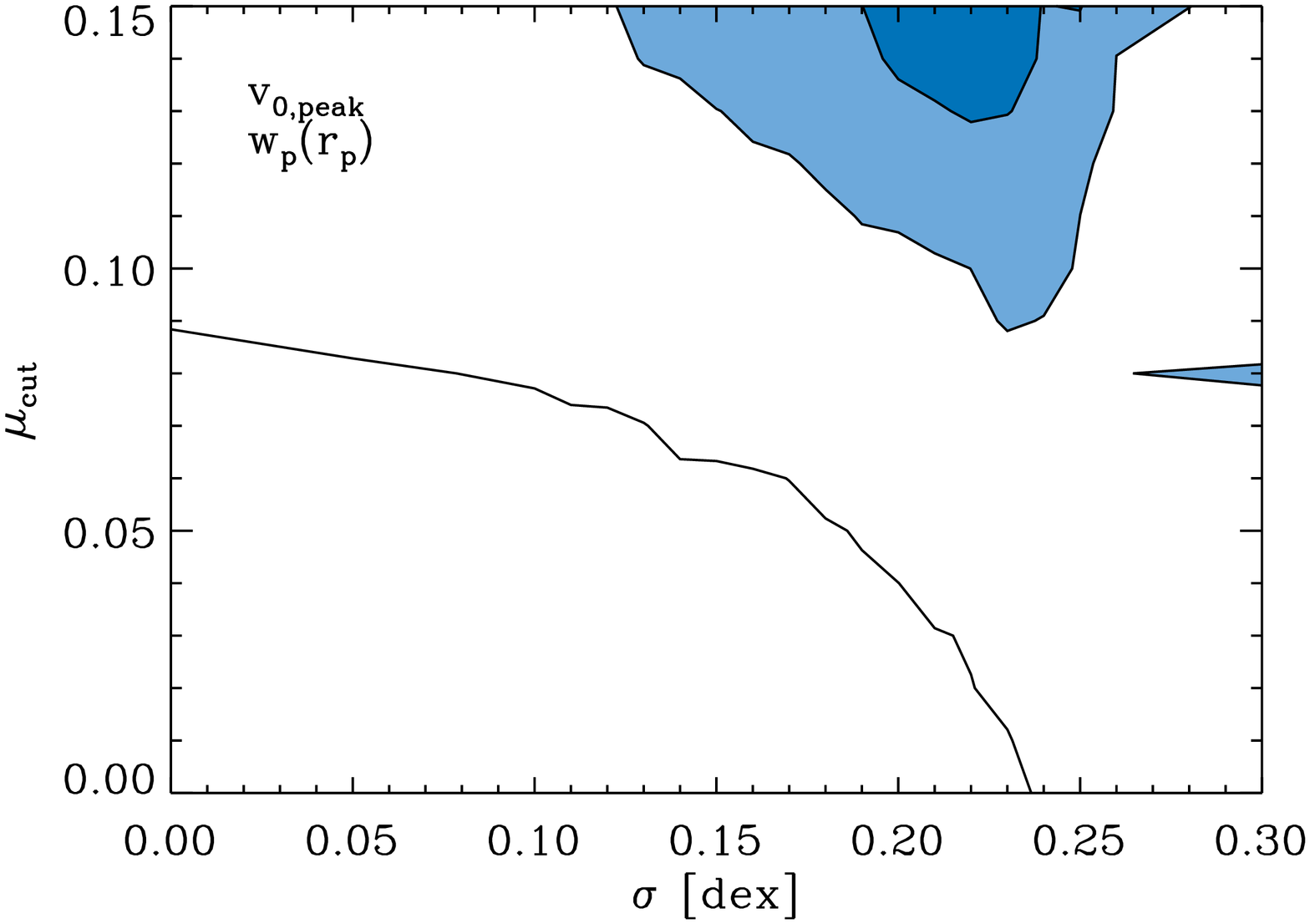}
\caption{Same as Fig. \ref{fig:constr-lum}, but using $\vnpeak$.  Constraints on the scatter and $\mucut$.  Levels give $P(>\chi^2)$, corresponding to 1, 2, 3, and 5-$\sigma$ contours, though here only the upper right corner with the 5-sigma contour appears.  The central and satellite CLF, and overall fit are everywhere more than 5-$\sigma$ deviations, and therefore omitted.}
\label{fig:vnpconstr-lum}
\end{figure}
We have repeated the entire study using luminosity in the SDSS $r$-band.  The global luminosity function from the SDSS \citep{Bla2003}, while having more information on dimmer galaxies, is not precisely the same as the luminosity function in our sample.  Therefore, for consistency with the group catalog, we use the luminosity function of galaxies in the corresponding volume-limited sample to perform the abundance matching, as was done when using stellar mass.  For comparisons of the two-point correlation function, we use the measurements of \citet{Zeh2011} defined with luminosity thresholds.

The same general trends apply for luminosity as for stellar mass, with a few complications.  First, while we use the same volume-limited sample as for the stellar mass-based comparison, the luminosity completeness limit is at $M_r<-19$.  We therefore have more galaxies present in a sample of the same volume in the luminosity sample.  Additionally, here we correct for changes in inferred absolute magnitude due to changes in inferred redshift due to fiber collisions, using the $k$-corrections to the $r$-band from \citet{BlRo2007}.  

Constraints are calculated including all correlations functions shown, and the central and satellite parts of the CLF.  The best-fit results are again for $\vpeak$, but this time with $\mucut$=0.12 and scatter of 0.21 dex.  (When not using the local averaging procedure, the best fit lies at $\mucut$=0.13 and scatter of 0.22 dex.)  Marginalizing over $\mucut$, we obtain limits of $\sigma=0.210^{+0.01}_{-0.02}$ dex (68\%) and 
$\sigma=0.21^{+0.02}_{-0.03}$ dex (95\%).  Marginalizing over scatter, the $\mucut$ limits are $\mucut$=$0.12^{+0.02}_{-0.01}$ (68\%) and $\mucut$>$0.09$ (95\% limit).

While the scatter agrees with our results for stellar mass, the $\mucut$ value is significantly higher.  This is favored by the parts of the CLF, which contribute most of the $\chi^2$, but not by the clustering alone, as can been seen with the low clustering in the brightest sample.  The $\vpeak$ model fits the satellite CLF somewhat well, but the group LF is low for small groups, and there is some offset in the central part of the CLF.

\begin{figure*}[p]
\centering
\includegraphics[angle=90, width=0.8\textwidth]{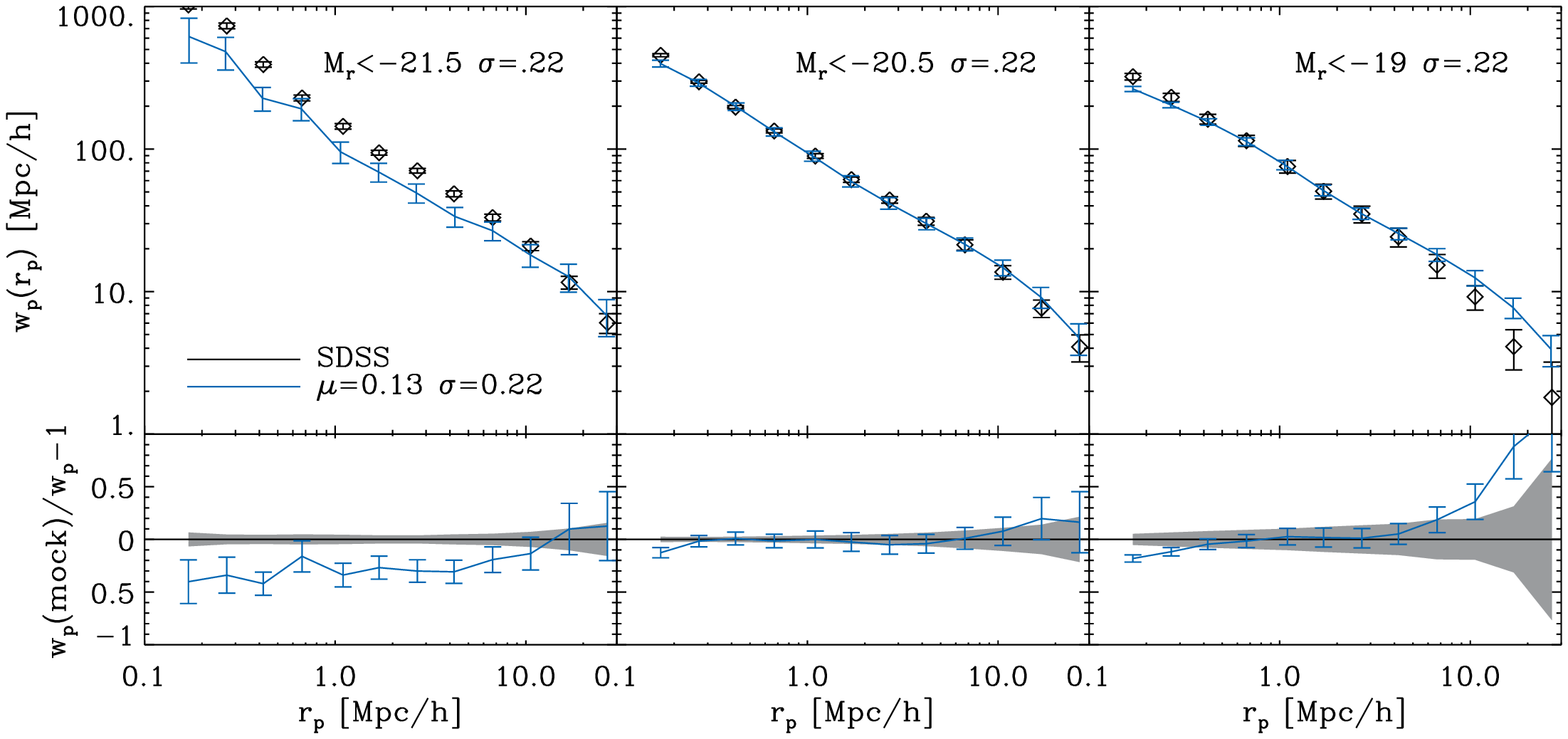}
\includegraphics[angle=90, width=0.8\textwidth]{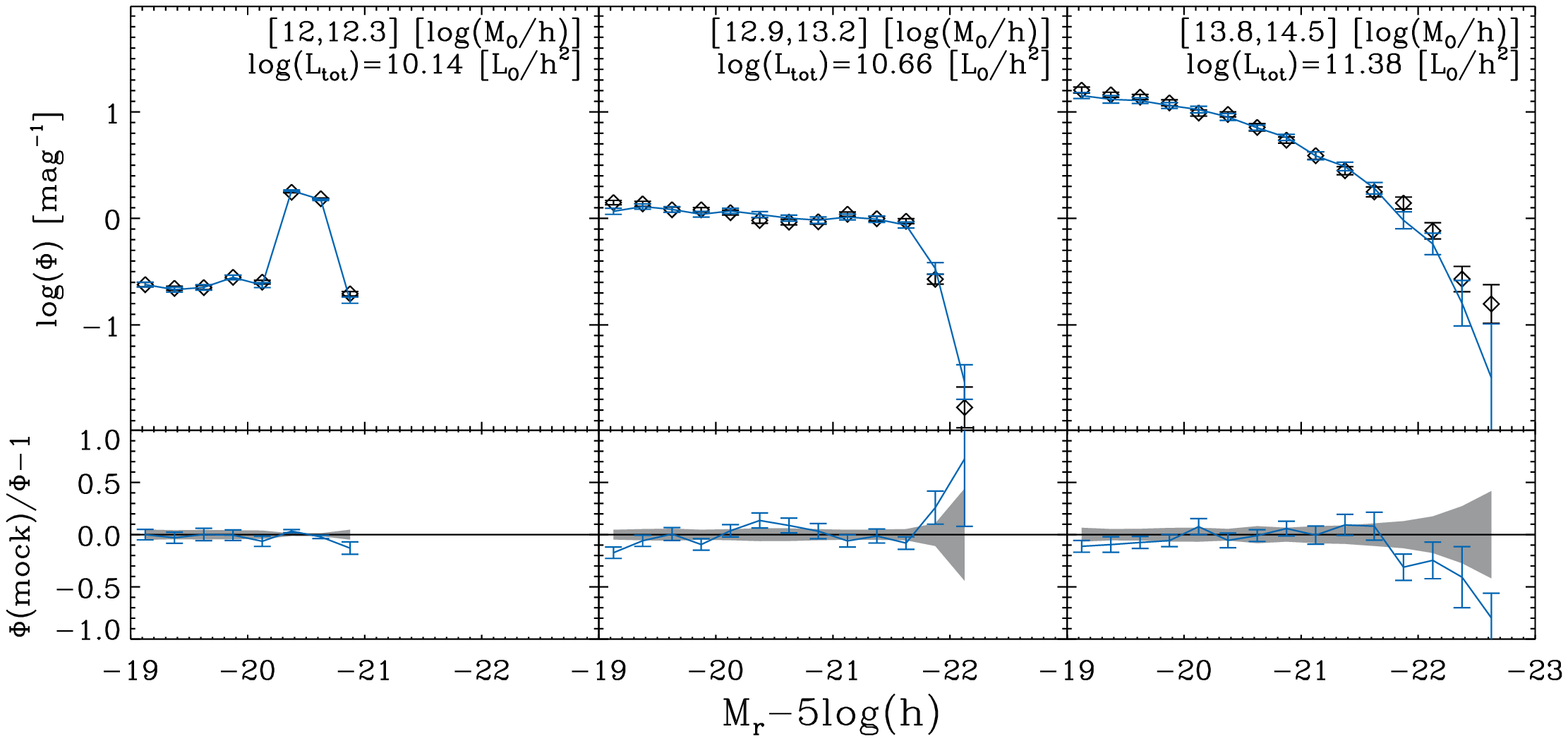}
\includegraphics[angle=90, width=0.9\textwidth]{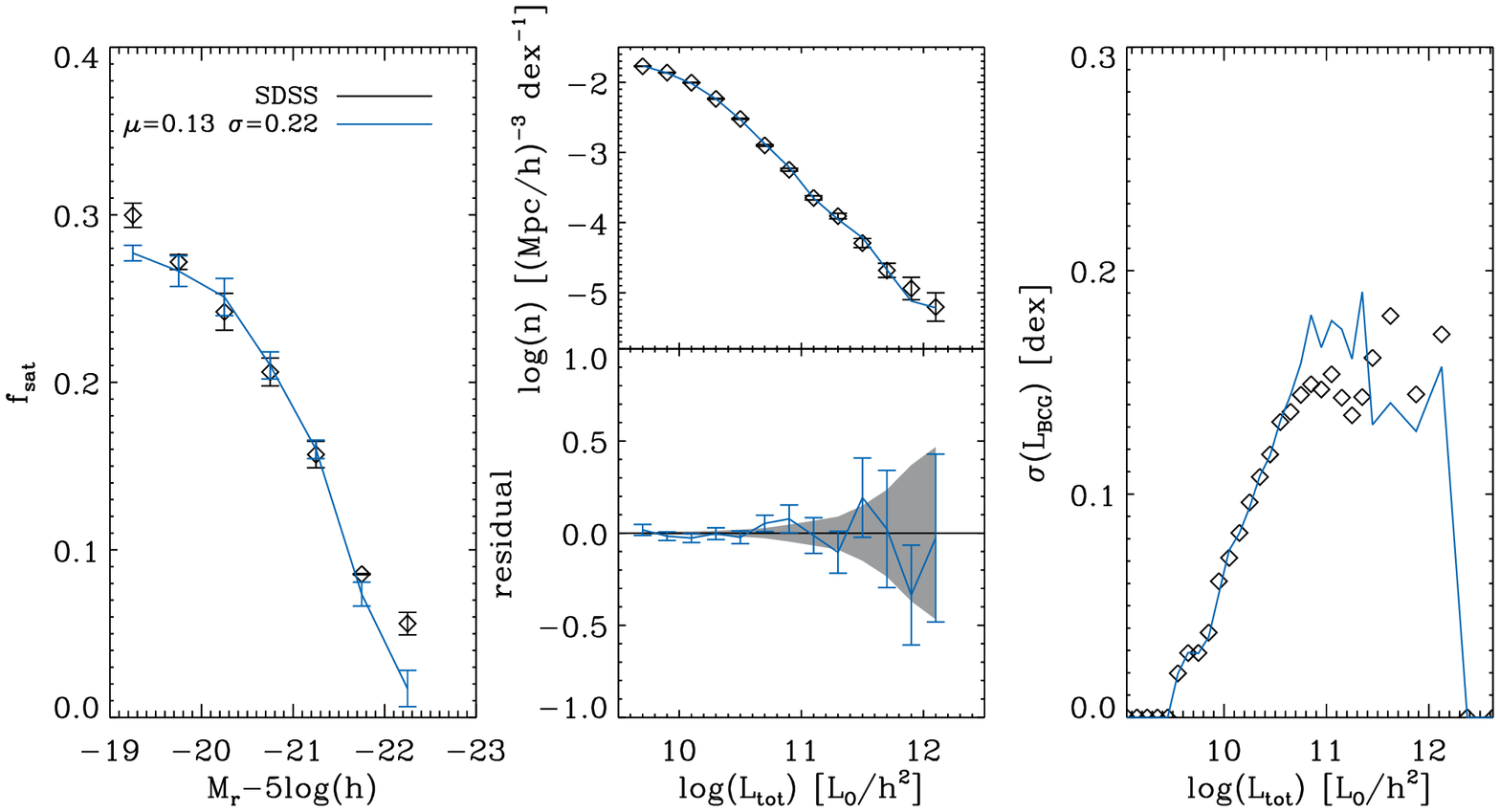}
\caption{Best-fit model when using $\vpeak$, with $\mucut$=0.13, scatter=0.22 dex.  Plots are the same as described in Fig.~\ref{fig:comp-vtype-lum}.  The low clustering of the $M_r$<-21.5 threshold is likely due to the high $\mucut$ value, but this does not have a large impact on the fit due to the large errors and correlations between data points.}
\label{fig:vp-bestfit-lum}
\end{figure*}

It remains true that $\vnpeak$ fits badly on all counts, being overclustered and having too many satellite galaxies.  (See Fig.~\ref{fig:comp-vtype-lum} for the comparison of different matching parameters with luminosity.)  Neither $\vpeak$ or $\vnpeak$ provides a good fit to the central part of the CLF, due primarily to an offset in the mean.  Even the best fit $\vpeak$ produces centrals that are too dim in low halo masses. and $\vnpeak$ centrals are too dim at low masses and somewhat too bright at higher halo masses.  The constraints are shown in Figs.~\ref{fig:constr-lum}, \ref{fig:vnpconstr-lum}, with the best-fit results in Fig.~\ref{fig:vp-bestfit-lum}.  The CLF fit parameters are given in Table~\ref{tab:clf-lum}, and the HOD fit is given in Table~\ref{tab:hod-lum}.  Note that the $C_{cen}$ value is an additional multiplicative factor applied to the central HOD, to account for the number of centrals not reaching unity for some luminosity thresholds.

\begin{centering}
\begin{table*}
	\centering
	\caption{Intrinsic CLF Luminosity Fit Parameters for Best-Fit Model}
	\begin{tabular}{l c c c c c r}
	\hline \hline
	$\mhost$ 	&	$\log(L_c)$ 	&	$\sigma_c$  &	$\phi_*$	&	$\alpha$	&	$\log(L_*)$	&	No. of hosts \\
	$[\log(M_{vir})]$	&	[$\log(L_\odot/h^2$)]		&	[$\log(L_\odot/h^2$)]		&	$[\log(L_\odot/h^2)^{-1}]$	&	&	[$\log(L_\odot/h^2)$] \\
	\hline
	12.0-12.3	&	$10.024\pm0.001$	&	$0.2338\pm0.0008$		&	$1.16\pm0.06$		&	$-0.93\pm0.08$	&	$9.77\pm0.02$		&	27948	\\
	12.3-12.6	&	$10.150\pm0.002$	&	$0.227\pm0.001$		&	$2.34\pm0.08$		&	$-0.684\pm0.060$	&	$9.842\pm0.018$	& 	14983	\\
	12.6-12.9	&	$10.238\pm0.003$	&	$0.224\pm0.001$		&	$4.36\pm0.16$		&	$-0.738\pm0.050$	&	$9.923\pm0.016$	&	7814		\\
	12.9-13.2 &	$10.284\pm0.004$	&	$0.228\pm0.002$		&	$7.54\pm0.31$		&	$-0.820\pm0.046$	&	$10.008\pm0.017$	&	4000		\\
	13.2-13.8 &	$10.332\pm0.004$	&	$0.230\pm0.002$		&	$18.0\pm0.6$		&	$-0.893\pm0.033$	&	$10.054\pm0.013$	&	2896		\\
	13.8-14.5 &	$10.381\pm0.009$	&	$0.217\pm0.004$		&	$66.2\pm3.1$		&	$-0.995\pm0.042$	&	$10.091\pm0.015$	&	595		\\
	\hline
	\end{tabular}
	\label{tab:clf-lum}
\end{table*}

\begin{table*}
	\centering
	\caption{Intrinsic HOD Luminosity Fit Parameters for Best-Fit Model}
	\begin{tabular}{l c c c c c c r}
	\hline \hline
	$M_r$ threshold	&	$M_{\rm{min}}$		&	$\sigma_m$ 	&	$C_{cen}$	&	$M_1$	&	$M_{\rm{cut}}$	&	$\alpha_{\rm HOD}$	&	No. of galaxies \\
	&	[$\log(\Msun/h)$]	&	[$\log(\Msun/h)$]	&	&	[$\Msun/h$]	&	[$\Msun/h$]	&	& \\
	\hline
	-21.5		&	$12.83\pm0.03$		&	$1.53\pm0.07$			&	$0.239\pm0.011$	& 	$14.33\pm0.02$		&	$12.2\pm0.6$		&	$1.06\pm0.07$		&	4437		\\
	-21.0		&	$12.49\pm0.01$		&	$1.26\pm0.02$			&	$0.497\pm0.007$	& 	$13.72\pm0.01$		&	$12.51\pm0.08$	&	$0.948\pm0.023$	&	16062	\\
	-20.5		&	$12.217\pm0.003$		&	$1.108\pm0.008$		&	$0.784\pm0.003$	& 	$13.27\pm0.01$		&	$12.37\pm0.04$	&	$0.948\pm0.013$	&	49718	\\
	-20.0		&	$11.936\pm0.002$		&	$0.959\pm0.005$		&	$0.936\pm0.002$	& 	$12.954\pm0.007$		&	$12.16\pm0.02$	&	$0.949\pm0.008$	&	103906	\\
	-19.5		&	$11.701\pm0.001$		&	$0.812\pm0.003$		&	$0.9854\pm0.0005$	& 	$12.736\pm0.005$		&	$11.97\pm0.02$	&	$0.960\pm0.005$	&	174937	\\
	-19.0		&	$11.503\pm0.001$		&	$0.723\pm0.002$		&	$0.9975\pm0.0002$	& 	$12.567\pm0.004$		&	$11.81\pm0.01$	&	$0.966\pm0.004$	&	261921	\\
	\hline
	\end{tabular}
	\label{tab:hod-lum}
\end{table*}

\begin{table*}
	\centering
	\caption{Luminosity HOD Parameters for Zehavi Fit}
	\begin{tabular}{l c c c c c c r}
	\hline \hline
	$M_r$ threshold	&	$\log M_{\rm{min}}$	&	$\sigma_{\log M}$	&	$\log M_0$	&	$\log M'_1$	&	$\alpha_{\rm HOD}$	&	No. of galaxies \\
	&	[$\log(\Msun/h)$]	&	[$\log(\Msun/h)$]	&	 $[\log \Msun/h]$	&	$[\log \Msun/h]$	&	&	\\
	\hline
	-21.5		&	$13.75\pm0.03$	&	$1.13\pm0.03$		& 	$13.75\pm0.38$	&	$14.35\pm0.12$	&	$1.33\pm0.47$		&	4437		\\
	-21.0		&	$12.83\pm0.01$	&	$0.731\pm0.009$	& 	$13.26\pm0.07$	&	$13.80\pm0.03$	&	$1.06\pm0.06$		&	18062	\\
	-20.5		&	$12.293\pm0.003$	&	$0.514\pm0.004$	& 	$12.73\pm0.02$	&	$13.29\pm0.01$	&	$0.965\pm0.014$	&	49715	\\
	-20.0		&	$11.919\pm0.002$	&	$0.392\pm0.003$	& 	$12.31\pm0.01$	&	$12.947\pm0.005$	&	$0.945\pm0.007$	&	103904	\\
	-19.5		&	$11.682\pm0.001$	&	$0.321\pm0.002$	& 	$11.682\pm0.007$	&	$12.729\pm0.004$	&	$0.953\pm0.004$	&	174932	\\
	-19.0		&	$11.491\pm0.001$	&	$0.295\pm0.001$	& 	$11.491\pm0.008$	&	$12.580\pm0.003$	&	$0.977\pm0.003$	&	261915	\\
	\hline
	\end{tabular}
	\label{tab:hod-zeh}
\end{table*}

\end{centering}

\section{Luminosity HOD Comparison to SDSS}\label{app:hod}

To perform a more exact comparison with the HOD of \cite{Zeh2011}, we use the best-fit luminosity-based abundance matching model.  This model has parameters $\mucut$=0.13 and scatter of 0.22 dex, and well-reproduces the SDSS clustering of \cite{Zeh2011}, as shown in Appendix~\ref{app:lum}.
We measure the HOD directly from the model, then perform a fit to the total HOD using the fitting function of \cite{Zeh2011}:

\be\label{eq:hod-zeh}
<N> = \frac{1}{2} \left[1+\mathrm{erf} \left( \frac{\log M_h-\log M_{\rm{min}}}{\sigma_{\log M}}\right)\right] \cdot \left[1+\left(\frac{M_h-M_0}{M'_1}\right)^{\alpha_{\rm HOD}}\right]
\ee

The final term gives the central and satellite parts, with the power law-like satellite part being set to zero when $M_h<M_0$.

\begin{figure*}
\centering
\includegraphics[angle=90, width=0.9\textwidth]{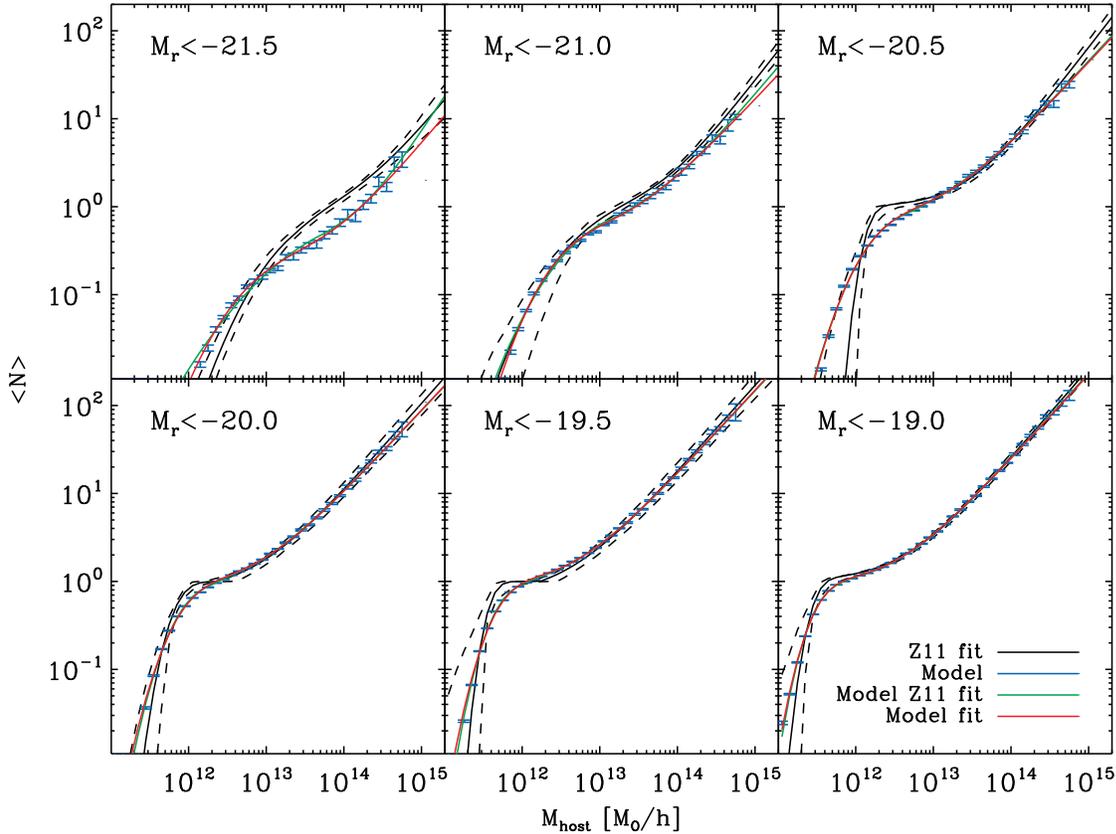}
\caption{Comparison of the best-fit model (abundance matched to luminosity) with \citet{Zeh2011} HOD derived from a fit to SDSS clustering measurements.  Solid black lines show the \cite{Zeh2011} HOD, with dashed lines showing the $1\sigma$ bounds based on the parameters they provide for their fit, assuming no correlation among parameters.  Blue error bars are the model results.  The green line is the fit to the model results using the \citet{Zeh2011} parameterization from Eq.~\ref{eq:hod-zeh}, while the red line shows our parameterization from Eq.~\ref{eq:hod-cen} and \ref{eq:hod-sat}, and modified as described in Appendix \ref{app:lum}.  The primary difference between the two lies in the location and width of the central host mass cutoff, which are somewhat degenerate when fitting to clustering measurements.  While this form provides a good fit to the overall
HOD, it does not well describe the central and satellite parts of the HOD separately.}
\label{fig:hod-lum-zeh}
\end{figure*}

The results of this fit, along with comparison to the results of \citet{Zeh2011} and our parameterization of the HOD are shown in Fig.~\ref{fig:hod-lum-zeh}.  The parameters for the luminosity model using this fitting function are given in Table~\ref{tab:hod-zeh}.  Both this figure and a comparison of the parameters indicate nearly the same behavior as described for the HODs in the stellar mass model.  Our model implies a higher and broader central mass cutoff then seen in \citet{Zeh2011}.  The fit for the satellite part is generally consistent between the two cases.  However, due to the high $\mucut$\ and scatter, the central part of the HOD never reaches unity for the brightest luminosity thresholds.  While the overall HOD can be well-fit with Eq.~\ref{eq:hod-zeh}, the centrals and satellites separately are not, particularly at the brighter thresholds.  This serves as additional motivation for our explicit separation of the central and satellite parts of the HOD.  For the luminosity case, we multiply Eq.~\ref{eq:hod-cen} by an additional overall normalization parameter to account for the reduced maximum number of central galaxies.  The closeness of the fits in general makes it difficult to claim a significant difference between the \citet{Zeh2011} results and our fits.  Further, in the highest luminosity thresholds where the differences are largest, the clustering produced by our model is also somewhat low.  This is in agreement with the shift of the brightest luminosity HOD to somewhat lower host halo masses, and thus, lower bias, which also obscures the comparison.

\section{Radial Profiles}\label{app:rp}

\begin{figure*}
\centering
\includegraphics[angle=90, width=1\textwidth]{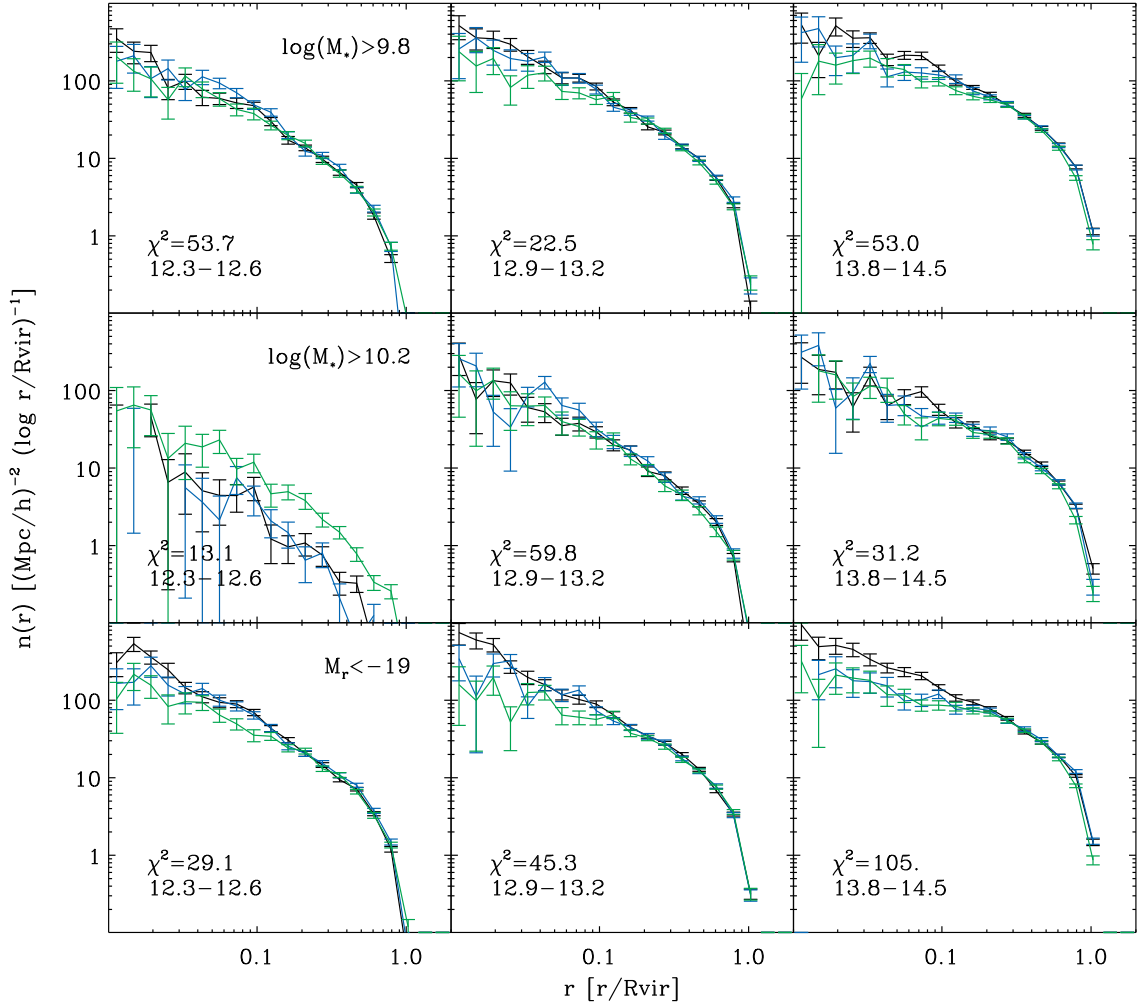}\hspace{-3 cm}
\caption{Projected radial profiles of galaxies in halos, for different cuts in stellar mass or luminosity.  {\em Top:}   Radial profiles for stellar masses with $\log(M_*)>9.8$.  {\em Center:}  Stellar masses with $\log(M_*)>10.2$.  {\em Bottom:}  Luminosity cut at $M_r<-19$.  In all plots, black is SDSS; blue is the best-fit model as it would be observed, which is $\vpeak$, $\mucut$=0.03, scatter=0.20 dex for stellar mass, and $\mucut=0.13$ and scatter=0.22 dex for luminosity.  Green is the intrinsic projected radial profile (without group finding).  $\chi^2$ values indicate the quality of the fit at $r/R_{vir}>0.1$ (nine data points).  While the fit in that range is quite good, it tends to fail at smaller radii, particularly for the more massive groups.}
\label{fig:rpr-sm}
\end{figure*}

Projected radial profiles are presented, as a further test of the input catalog and the group finding algorithm.  These show the satellites assigned to groups for each host halo mass, and give their projected number density at distances from the group center.  The group center is determined by the location of the central, and distances are given as a fraction of the virial radius.  Fig.~\ref{fig:rpr-sm} shows the profiles in the stellar mass best-fit case for two different cuts in stellar mass, and the same result for one cut in the best-fit luminosity model.

The larger differences in the profiles in the luminosity case may help explain why the luminosity model fits more poorly overall.  The higher $\mucut$ preferentially removes satellites near the centers of clusters which have already been significantly stripped.  This impacts the CSMF, but the change in radial profile shape also impacts the one-halo term in the clustering. Further discussion of satellite incompleteness and its dependence on galaxy luminosity and simulation specifications will be given in \citet{Wu2012}.

\end{appendices}

\end{document}